\documentclass[12pt]{iopart}

\usepackage{iopams}
\usepackage{xcolor}
\usepackage{hyperref}
\usepackage{setstack}
\usepackage{tikz}
\usepackage{cite}
\usepackage{subcaption}
\usepackage{slashed}
\usepackage{stmaryrd}

\newcommand{\funcdiff}[2]{\frac{\delta #1}{\delta #2}}
\renewcommand{\S}{\mathcal{S}}
\renewcommand{\P}{\mathcal{P}}
\renewcommand{\L}{\mathcal{L}}

\newcommand{\Pdf}{\mathcal{P}_{\mathrm{DF}}}
\newcommand{\Pdfb}{\cev{\mathcal{P}}_{\mathrm{DF}}}

\renewcommand{\Re}{\mathrm{Re}}
\renewcommand{\Im}{\mathrm{Im}}
\newcommand{\A}{\mathcal{A}}
\newcommand{\T}{\mathcal{T}}

\newcommand{\D}{\mathcal{D}}
\newcommand{\V}{\mathcal{V}}

\newcommand{\R}{\mathbb{R}}
\newcommand{\dS}{\dot{\S}}

\newcommand{\adf}{\A_{\mathrm{DF}}}
\newcommand{\adfb}{\cev{\A}_{\mathrm{DF}}}

\usepackage{graphicx,accents}
\makeatletter
\DeclareRobustCommand{\cev}[1]{%
  \mathpalette\do@cev{#1}%
}
\newcommand{\do@cev}[2]{%
  \fix@cev{#1}{+}%
  \reflectbox{$\m@th#1\vec{\reflectbox{$\fix@cev{#1}{-}\m@th#1#2\fix@cev{#1}{+}$}}$}%
  \fix@cev{#1}{-}%
}
\newcommand{\fix@cev}[2]{%
  \ifx#1\displaystyle
    \mkern#23mu
  \else
    \ifx#1\textstyle
      \mkern#23mu
    \else
      \ifx#1\scriptstyle
        \mkern#22mu
      \else
        \mkern#22mu
      \fi
    \fi
  \fi
}

\makeatother

\begin{document}
\title[Time-reversal symmetry violations and entropy production in polar active matter]{Time-reversal symmetry violations and entropy production in field theories of polar active matter}

\author{\O yvind L. Borthne$^1$, \'Etienne Fodor$^2$, Michael E. Cates$^1$}

\address{$^1$ DAMTP, Centre for Mathematical Sciences, University of Cambridge, Wilberforce Road, Cambridge CB3 0WA, UK}

\address{$^2$ Department of Physics and Materials Science, University of Luxembourg, L-1511 Luxembourg}

\ead{olb23@cam.ac.uk}

\begin{abstract}
{We investigate the steady-state entropy production rate (EPR) in the Hydrodynamic Vicsek Model (HVM) and Diffusive Flocking Model (DFM). Both models display a transition from an isotropic gas to a polar liquid (flocking) phase, in addition to traveling polar clusters and microphase-separation in the miscibility gap. The phase diagram of the DFM, which may be considered an extension of the HVM, contains additional structure at low densities where we find a novel crystal phase in which a stationary hexagonal lattice of high-density ridges surround low density valleys. From an assessment of the scaling of the EPR at low noise, we uncover that the dynamics in this limit may be organised into three main classes based on the dominant contribution. Truly nonequilibrium dynamics is characterised by a divergent EPR in this limit, and sustains global time-reversal symmetry (TRS) violating currents at zero noise. On the other hand, marginally nonequilibrium and effectively equilibrium dynamics have a finite EPR in this limit, and TRS is broken only at the level of fluctuations. For the latter of these two cases, detailed balance is restored in the small noise limit and we recover effective Boltzmann statistics to lowest nontrivial order. We further demonstrate that the scaling of the EPR may change depending on the dynamical variables that are tracked when it is computed, and the protocol chosen for time-reversal. Results acquired from numerical simulations of the dynamics confirm both the asymptotic scaling relations we derive and our quantitative predictions.}
\end{abstract}

\newpage

\section{Introduction}
\label{sec:intro}
Nonequilibrium statistical physics deals with fluctuating dynamics that violate detailed balance, or equivalently time-reversal symmetry (TRS), driving the system away from the classical Boltzmann statistics \cite{Sekimoto2010,Kubo1991,Zwanzig2001,Lebowitz1998,Seifert2012}. In recent decades considerable attention in this field has been granted to models of living systems, incorporating traits such as motility \cite{Wittkowski2014,Tailleur2008,Solon2015,Bertin2006,Peshkov2014,Chate2008,Mahault2018,Fodor2020,Chate2020,Fodor2016,Toner1998,Toner2005,Toner2012,Marchetti2013,Thompson2011,Cates2015,Barre2014,Vicsek1995,Nemoto2019,Ramaswamy2010}, birth and death \cite{Toner2012,Li2020,Grafke2017}, and quorum sensing \cite{Tailleur2008,Solon2015,Farrell2012}, although many also have important counterparts in inanimate systems \cite{Deseigne2010,Kumar2014,Deblais2018,Palacci2013}. Common to all such systems is the fact that they rely on a steady transfer of energy either via reservoirs or some external drive, engendering currents that violate TRS. Active matter forms an important subclass of nonequilibrium systems as motility is generated on the basis of a sustained \textit{local} exchange of energy, via e.g. consumption of chemical fuel \cite{Palacci2013} or conversion of vibrational energy \cite{Kumar2014,Deseigne2010} and subsequent dissipation, thus breaking TRS \cite{Nardini2017,Ganguly2013,Dadhichi2018,Shim2016,Fodor2016,Crosato2019,Dabelow2019}. By developing our understanding of this inherent irreversibility of active motion in simple models of active matter, we hope to provide key insight into the nonequilibrium nature of real living systems.

The surge of interest in active matter can also be attributed to discoveries of a vast range of novel phenomena associated with motility. Particularly prominent among these are flocking and motility-induced phase-separation (MIPS). The former arises from the combined effects of activity and alignment in response to e.g. pairwise collisions in suspensions of rod-shaped particles, hydrodynamic interactions or a sensory based steering in living systems \cite{Marchetti2013,Chate2020,Ramaswamy2010}. In MIPS, spherically symmetric colloidal particles interact via steric repulsion to form dense segregated clusters against a vapor background at sufficiently high average densities and long persistence times \cite{Cates2015}. We will be primarily concerned with field theoretic formulations of dry polar flocking, where `dry' refers to the fact that we neglect hydrodynamic interactions with the solvent fluid \cite{Marchetti2013}. In addition to a local conserved density, we describe the dynamics of such flocks by a local polar order parameter. This specifies either a head-to-tail orientation of the active particles or a local swimming velocity, and breaks rotational symmetry by attaining a nonzero global value in the flocking (or polar liquid) phase. Nonetheless, we believe that many of the principles we discuss are more widely applicable -- also to systems that display MIPS -- and so we will view them in light of previous work that has been conducted on similar themes.

Precise identification of TRS violations from the large scale dynamics of active matter is not always trivial, as the microscopic motion does not necessarily generate global net currents. For example, in field theories of MIPS such as Active Model B (AMB), the absence of steady mass currents renders the steady-state deceivingly similar to equilibrium phase-separation \cite{Cates2015,Wittkowski2014}. More specifically, for phase-separating dynamics of AMB type, the local density only provides information about the underlying irreversibility through fluctuations \cite{Nardini2017}. One might ask to what extent this qualitative notion of `looking like equilibrium' is reflected quantitatively by the entropy production rate (EPR), measuring the extent of TRS violation by the stochastic dynamics. To address this question we investigate the dominant contribution to the EPR for polar systems at low noise, allowing us to distinguish between TRS violation at the levels of fluctuating and mean global dynamics. In particular, from this analysis we determine the properties of the mean global dynamics which causes TRS violation at zero noise. Note also that the viewpoint we assume in this paper treats the EPR solely as a measure of irreversibility, although several interesting questions relate to the connections between this measure and the energetics of active systems on the hydrodynamic scale \cite{Markovich2020}.

In this paper we observe that dynamics in the small noise regime may be organised into three main classes based on the scaling of the EPR with the strength of local noise fluctuations. Within this scheme, truly nonequilibrium dynamics is characterised by a diverging EPR in the limit of small noise. The dominant divergent contribution stems from the ground-state dynamics at zero noise, signifying the presence of steady TRS-violating currents that persist in this limit. In fact, we will show that it \textit{is} possible for the dynamics to be truly nonequilibrium even when steady homogeneous mass currents do not break detailed balance alone. In this case, the violation of TRS at ground-state level is an emergent collective phenomenon which does not have any counterpart for a single active particle. When the EPR is finite in the limit of small noise, we further classify dynamics as either marginally nonequilibrium or effectively equilibrium, where the latter corresponds to the case where the EPR vanishes in this limit. Note that the EPR can also vanish on approach to a critical point while maintaining nonequilibrium behaviour but we do not address such cases here \cite{Caballero2020}. For dynamics of marginal or effectively equilibrium type, the ground-state dynamics do not violate TRS and entropy is produced only at the level of fluctuations. However, for effectively equilibrium dynamics, detailed balance is restored at small noise where we recover Boltzmann statistics to lowest nontrivial order, while it is broken by a finite amount for any infinitesimal fluctuation in the marginal case.

In the truly nonequilibrium case, the ground-state dynamics at zero noise determines the coefficient of the leading order term. When the EPR remains finite in the limit of vanishing noise, we go beyond the deterministic setting and show that we may access the leading order coefficient by including fluctuations via a systematic expansion of the dynamics in the noise strength about the steady profile. Furthermore, we confirm our predictions by explicitly comparing them with results from numerical simulations of the dynamics. We also show that the scaling exponent of the EPR can be bounded from below by symmetry arguments, and that this agrees both with the linearisation as well as simulations.

The type of field theory we consider has been granted extensive attention elsewhere in the literature. An important first analysis was performed by Toner and Tu \cite{Toner1998,Toner2005} in their seminal approach to the hydrodynamics of flocking based on symmetry considerations of the Vicsek model. Subsequent developments included derivations of this theory via explicit coarse-graining from the Vicsek dynamics by Boltzmann-Ginzburg-Landau \cite{Bertin2006,Bertin2009,Peshkov2014} and Dean's equation \cite{Farrell2012,Barre2014,Dean1996} approaches. For the first part of this article, we study phenomenological equations akin to those proposed by Solon et al. in \cite{Solon2015a}, albeit with noise. To compare forward and time-reversed paths of this system, the local polar density must respect a discrete polar symmetry on time-reversal. By following Marchetti \cite{Marchetti2013} and Dadhichi \cite{Dadhichi2018}, who consider a more general constitutive equation for the current that advects the density, we free the polar density from this constraint.

We structure the article as follows. In \sref{sec:field-theory} we introduce the Hydrodynamic Vicsek Model (HVM), where the density current is locally proportional to the polar density. Our discussion is meant to highlight the familiar phase behaviour of the model, with particular emphasis on the location of phase boundaries with respect to the phenomenological parameters of the model (as opposed to any underlying microscopic parameters). Next, in \sref{sec:fluct-hydro-epr} we define the EPR via the difference between time-forward and reversed path probability weights and discuss the implications of the polar density changing sign on time-reversal. Here, we also introduce the asymptotic scaling relation for the EPR at small noise, and proceed to study its behaviour in the various phases of the model. \Sref{sec:density-flucts} introduces the model we refer to as the Diffusive Flocking Model (DFM), in which we consider a more general constitutive equation for the advective current that includes noise and depends nonlinearly on both density and the local polar order parameter. This allows us to consider both the case of a time-even and time-odd polar density, and we explore how this choice changes our results from \sref{sec:fluct-hydro-epr}. In addition, we demonstrate that in order to fully account for the entropy produced due to density currents in the flocking phase, we must also track this advective current. Finally, in \sref{sec:conclusion} we summarise our findings and present our concluding remarks and perspectives.

\section{The Hydrodynamic Vicsek Model}
\label{sec:field-theory}
Initially, the seminal numerical study by Vicsek et al. \cite{Vicsek1995} inspired a large body of research on the transition to collective `flock' motion in active particle systems with aligning interactions. Their original article considers a discrete-time, continuous-space automaton in $d=2$, where ferromagnetic spins travel in space at constant motility and align with their nearest neighbours. In essence, the transition to collective flock motion in the  Vicsek model occurs due to the coupling between the XY-type spin interaction and a \textit{time-dependent} connectivity matrix of spins, separating it from the classical Heisenberg model where true long range order cannot occur in $d=2$ \cite{Toner1998,Toner2005,Ginelli2016}. Since its inception, the model has been generalized and recast in various different forms; in continuous-time \cite{Farrell2012,Barre2014}, for spatial dimensions $d\neq2$ \cite{OLaighleis2018,Mahault2018}, to topological (rather than metric) interactions \cite{Castellana2016,Peshkov2012}, to systems with nematic symmetry \cite{Patelli2019,Peshkov2014,Mahault2018,Marchetti2013} as well as to include additional interactions such as hard-core central forces and density-dependent bare self-propulsion speeds\cite{Farrell2012,Barre2014,Marchetti2013}.

Of the many hydrodynamic theories that have been derived from the various microscopic models, most bear resemblance with that considered initially by Toner and Tu \cite{Toner1998,Toner2005} on phenomenological grounds, although important insights have been gained from explicit coarse-graining. For our present perspective, particularly important are those that relate the dependence of the coupling parameters in Toner and Tu's theory on the density and local polar order with the nonlinear dynamics. In particular, we adopt equations for the local particle and polarisation densities $\rho$ and $\bi{P}$ (respectively) that are motivated by explicit coarse-graining and include the familiar microphase-separated and polar cluster regimes of the microscopic Vicsek dynamics \cite{Solon2015a,Solon2015b}. Nonetheless, the approach we choose is a phenomenological one without specific reference to any underlying microscopic model.

In the following we consider a conserved density $\rho(\bi{x},t)$ of particles, where $\bi{x}\in\V$ and $\V\subset\R^2$ is a periodic domain in $d=2$, which changes locally due to the flow induced by the current $\bi{J}$ and thus obeys a continuity equation
\begin{equation}
\label{eq:density}
\partial_t\rho=-\nabla\cdot\bi{J}.
\end{equation}
Further to this, we assume that locally particles tend to move in a direction specified by the polarisation density $\bi{P}(\bi{x},t)=\rho(\bi{x},t)\bi{W}(\bi{x},t)$, where $\bi{W}$ is an order parameter for the local polar order. It is important to note that we associate with $\bi{P}$ either a local head-to-tail orientation of the particles or a direction of the \textit{bare} self-propulsive force. In general therefore, one might expect the constitutive equation for the current in \eref{eq:density} to be some complicated expression $\bi{J}\equiv\bi{J}(\rho,\bi{P},\nabla\rho,\nabla\bi{P},\ldots)$ of $\rho$ and $\bi{P}$ in addition to their spatial derivatives. Indeed, this will be the case for example if particles interact via steric repulsion, or if they undergo thermal Brownian motion in addition to self-propulsion due to interactions with an underlying substrate. For now we delay this issue, which we will revisit in \sref{sec:density-flucts}, and focus instead on the simplest case where the constitutive equation is given by
\begin{equation}
\label{eq:constitutive-eqn}
\bi{J}=w\bi{P}.
\end{equation}
This is consistent with the case where $\bi{W}$ is the local average direction of the velocity of particles and $w$ is the constant self-propulsion speed.

Symmetry considerations alone generally lead to a high-dimensional parameter space spanned by the coefficients appearing in the equation for the polarisation density, even when the hydrodynamic expansion is truncated at lowest non-trivial order \cite{Toner1998,Toner2005}. One of the achievements of explicit coarse-graining has therefore been to provide a more pragmatic approach to reducing the number of independent parameters by relating them to microscopic quantities. A particularly clever observation is that many of these computations lead to an equation of the form \cite{Marchetti2013,Dadhichi2018}
\begin{equation}
\label{eq:polarity}
\partial_t\bi{P}+\lambda \bi{P}\cdot\nabla\bi{P}=-\funcdiff{F}{\bi{P}}+\bfeta.
\end{equation}
Expressed in this way, the equation is reminiscent of a vectorial Model A (in the Halperin-Hohenberg classification \cite{Hohenberg1977}), with a self-advection piece, i.e. the $\lambda$-term, that explicitly violates TRS as it cannot be written as a functional derivative \cite{Nardini2017}. In fact, TRS-violation in this model is a slightly more involved issue due to the coupling between $\rho$ and $\bi{P}$, and we will return to this in \sref{sec:fluct-hydro-epr}. 

We further take the functional $F[\rho,\bi{P}]$ to be given by
\begin{equation}
\label{eq:free-energy}
F[\rho,\bi{P}]=\int_{\V}\rmd\bi{x}\,\left(f(\rho,\bi{P})+\frac{\nu}{2}(\nabla_{\alpha}P_{\beta})^2+\bi{P}\cdot\nabla\Phi(\rho,\bi{P})\right).
\end{equation}
Here, $F$ contains a local free-energy density $f(\rho,\bi{P})$ which is of standard quartic form, i.e.
\begin{equation}
f(\rho,\bi{P})=\frac{a}{2}(\rho_c-\rho)|\bi{P}|^2+\frac{b}{4}|\bi{P}|^4,
\end{equation}
with the notable exception that the coefficient of the quadratic term controlling the transition to the low-temperature phase depends explicitly on the local density $\rho$. As mentioned above, this dependence is a product of coarse-graining and is kept here in order to capture the inhomogeneous phases, while we assume all other parameters to be constant. In fact, many such procedures lead to other coefficients also carrying a nontrivial dependence on $\rho$, although this is the simplest dependence needed to make the isotropic-to-flock transition similar to a first order vapor-to-liquid transition. In this form, $f$ attains the characteristic bistable form when $\rho>\rho_c$ which marks the transition to the ordered phase. In addition, in \eref{eq:free-energy} we use the function $\Phi$ defined by
\begin{equation}
\label{eq:pressure}
\Phi(\rho,\bi{P})=w_1\rho-\frac{\kappa}{2}|\bi{P}|^2,
\end{equation}
which is often referred to as an effective pressure \cite{Marchetti2013,Bertin2006,Bertin2009,Gopinath2012}. The first term on the right-hand side is the ideal gas pressure contribution. \Eref{eq:pressure} also implies that pressure may be reduced locally by increasing the polar order, an effect often associated with a tendency to splay in polar liquid type systems \cite{Gopinath2012}. In fact, this competing effect, in which $\bi{P}$ wants to align against gradients in $\rho$ and towards increasing $|\bi{P}|$, culminates in an instability at sufficiently large $\kappa$ leading to the formation of localized traveling polar clusters. Finally, fluctuations are accounted for in \eref{eq:polarity} via the mean-zero spatiotemporal Gaussian white noise field $\bfeta$ with covariance
\begin{equation}
\label{eq:eta-cov}
\langle\eta_{\alpha}(\bi{x},t)\eta_{\beta}(\bi{x}',t')\rangle=2D\,\delta_{\alpha\beta}\delta(\bi{x}-\bi{x}')\delta(t-t'),
\end{equation}
where the noise coefficient $D$ parameterises the strength of fluctuations. Importantly, the noise term is added completely \textit{ad-hoc}, and is not a direct result of coarse-graining. Several authors have addressed the effects of different noise statistics, including scalar versus vectorial noise, as well as additive versus multiplicative \cite{Solon2015a,Marchetti2013}, although all find that the phase diagram is reasonably stable against such modifications. For our purposes \eref{eq:eta-cov} will therefore suffice.

Even in this simplified model, henceforth referred to as the \textit{Hydrodynamic Vicsek Model} (HVM), we are still left with a rather large parameter space. Some reduction of this can be made by choosing suitable units; indeed we observe that under a rescaling of the time, space and the fields we may set $a$, $b$, $\nu$ and $\rho_c$ all equal to one, which we adopt in the following. Again motivated by explicit coarse-graining from microscopic Vicsek dynamics, we will also mostly be concerned with the special case in which $\kappa=\lambda$ and $w_1=w/2$, leaving us with two independent parameters $(\lambda,w)$ in addition to the noise coefficient $D$, system size $L$ (taking $\V=[-L,L]^2$ for simplicity) and the conserved average local density $\rho_0=\V^{-1}\int_{\V}\rmd\bi{x}\,\rho$. With these definitions, writing out explicitly the equation for $\bi{P}$ in \eref{eq:polarity} we obtain
\begin{equation}
\label{eq:polarity-full}
\fl\partial_t\bi{P}+\lambda \bi{P}\cdot\nabla\bi{P}=\left(\rho-1-|\bi{P}|^2\right)\bi{P}+\nabla^2\bi{P}+\frac{\kappa}{2}\nabla|\bi{P}|^2-\kappa\bi{P}\nabla\cdot\bi{P}-w_1\nabla\rho+\bfeta.
\end{equation}
\begin{figure}[t]
\centering
\begin{subfigure}[b]{0.42\linewidth}
\includegraphics[scale=0.48]{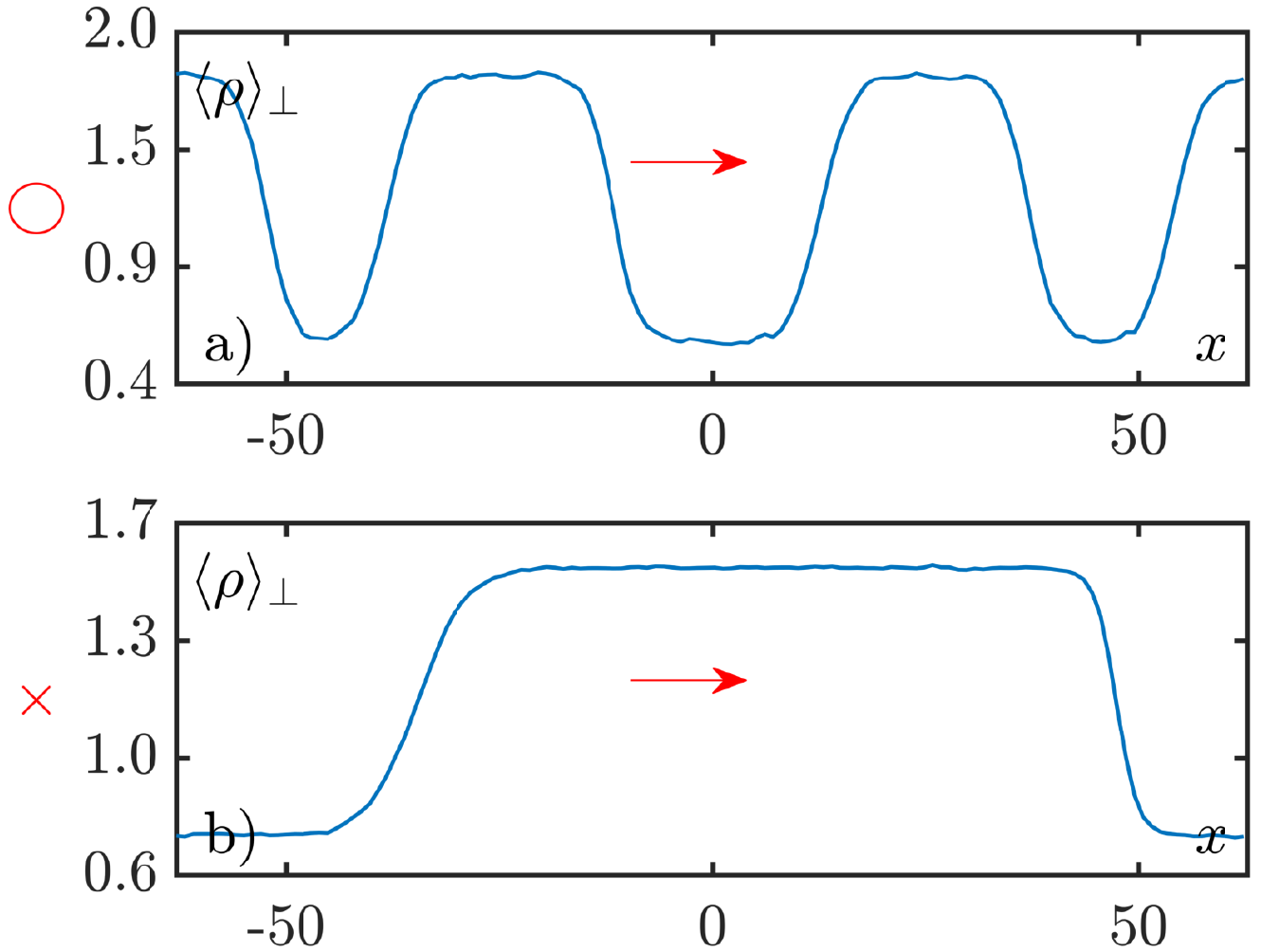}
\vspace{0.5cm}
\end{subfigure}
\begin{subfigure}[b]{0.47\linewidth}
\includegraphics[scale=0.75]{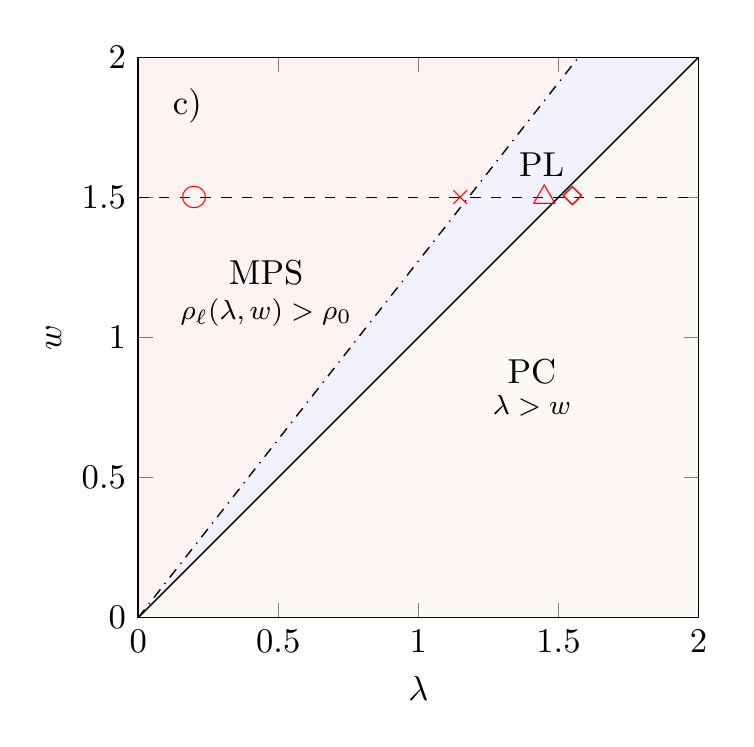}
\end{subfigure}

\begin{subfigure}[b]{0.86\linewidth}
\includegraphics[scale=0.22]{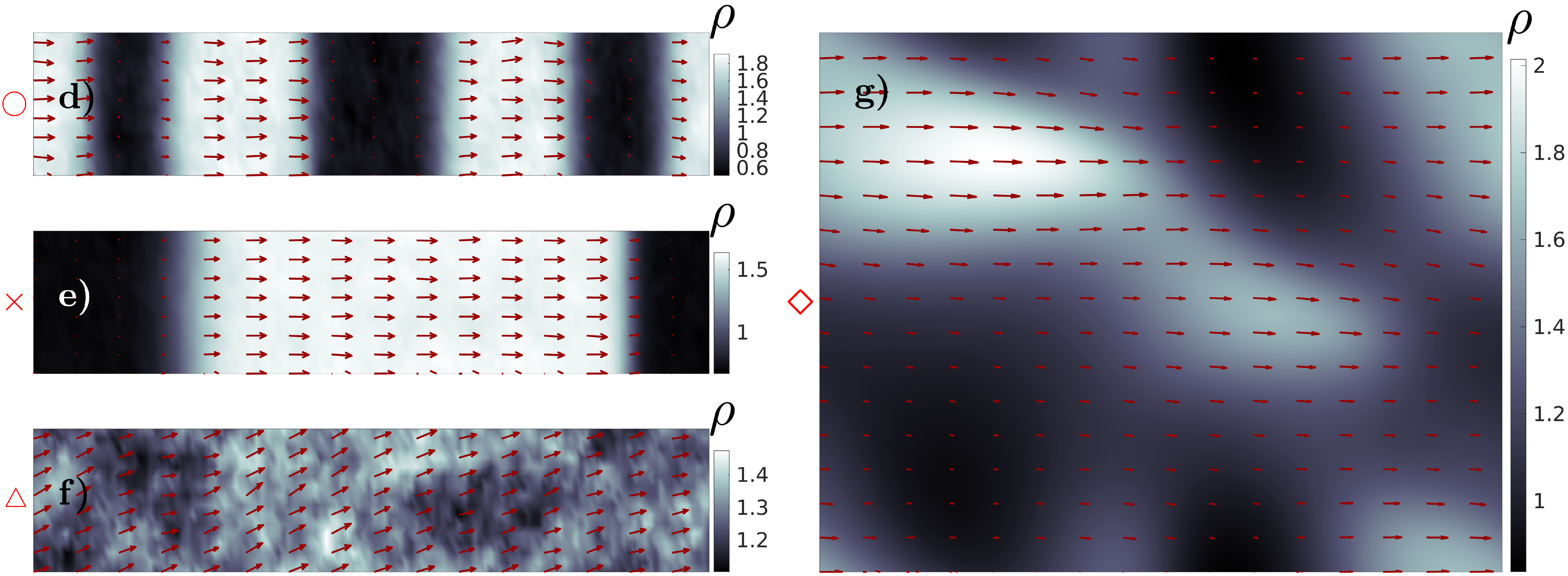}
\end{subfigure}
\caption{Plots in a) - b) and d) - g) illustrate the ordered phases of the HVM in \eref{eq:density}-\eref{eq:constitutive-eqn} and \eref{eq:polarity-full}: a) - b) show two microphase-separated profiles of the system, where in a) the traveling bands form a smectic arrangement and in b) a single solitonic band travels against an isotropic background. In both plots, $\langle\rho\rangle_{\perp}$ denotes an instantaneous (in time) average over the direction perpendicular to the motion indicated by red arrows ({\color{red}$\rightarrow$}). c) Phase diagram of the model at fixed $\rho_0=1.28$, $\kappa=\lambda$ and $w_1=w/2$, with data points ({\color{red}$\opencircle$}, {\color{red}$\times$}, {\color{red}$\opentriangle$}, {\color{red}$\diamond$}) corresponding to figures a) - b) and d) - g) (in order of increasing $\lambda$). Solid (\full) and dash-dotted (\chain) lines correspond to the phase boundaries $w=\lambda$ and $\rho_{\ell}(\lambda,w)=\rho_0$ respectively, determined from the linear stability analysis. Figures a) - b) and d) - e) display microphase-separation (MPS), where the number of bands is seen to increase as $\lambda$ is decreased. In figure f) the system is homogeneously polarized, while in g) where $\lambda>w$, both the MPS and homogeneous polar liquid (PL) phases are unstable and localized polar clusters (PC) form.}   
\label{fig:phase-diagram} 
\end{figure}

It is well known that the HVM in \eref{eq:density}-\eref{eq:polarity} displays both an isotropic and a polar liquid (or flocking) phase, in addition to a microphase-separated regime with both solitonic and smectic arrangements of polar bands traveling against an isotropic background \cite{Marchetti2013}. As mentioned, we also find a regime in which the polar liquid state becomes unstable and traveling polar clusters emerge. In \fref{fig:phase-diagram} we show typical realisations of the nonlinear steady-states and polar liquid in the HVM.  

A standard linear stability analysis provides us with some insight into the nature of the phase diagram of the model, including the nonlinear phases, although we don't expect to fully classify it by such means. We provide the details of this calculation in \ref{app:lin-analysis} for completeness. In essence, one finds that the only constant and homogeneous solutions $\rho_0$, $\bi{P}_0$ to \eref{eq:density}-\eref{eq:polarity} at zero noise are the mean field isotropic and polarly ordered solutions. More specifically, when $\rho_0<1$ the only solution is the isotropic one for which $|\bi{P}_0|=0$. When $\rho_0>1$, one additionally finds polarly ordered solutions with
\begin{equation}
|\bi{P}_0|=\sqrt{\rho_0-1}.
\end{equation}

The isotropic phase is linearly stable only when $\rho_0<1$ beyond which global polar order emerges. However, in the miscibility gap $\rho_0\in(1,\rho_{\ell})$ where
\begin{equation}
\label{eq:rho-ell}
\rho_{\ell}(\lambda,w,w_1)=1+\frac{1}{2}\frac{w}{\lambda+2w_1},
\end{equation}
the homogeneous polar liquid state is linearly unstable to perturbations due to the coupling between fluctuations of $\rho$ and of $\bi{P}$ for fluctuations that are parallel to the direction of broken symmetry. In this region we observe both spatially inhomogeneous phases reported above, i.e. microphase-separation and traveling polar clusters, separated by a phase boundary that appears at sufficiently large $\kappa$. Insight into this is again provided by the linear stability analysis of the polar liquid phase, which requires that
\begin{equation}
\label{eq:pc-boundary}
\kappa<2w_1
\end{equation}
for stability. From simulations we observe that polar clusters are formed when $\kappa>2w_1$, both within the miscibility gap and for larger $\rho_0$. We also find numerically that within the region where microphase-separation occurs, the number of bands increases with decreasing $\lambda$. Finally, note that when $\kappa=\lambda$, $w_1=w/2$, and the inequality \eref{eq:pc-boundary} holds, we have that $\rho_{\ell}\in(5/4,3/2)$. In particular, when $\rho_0$ is within this region, all three polarly ordered phases can be realized by varying $\lambda$ and $w$. This is illustrated in \fref{fig:phase-diagram}, where we plot the resulting phase diagram for fixed $\rho_0\in(5/4,3/2)$ in the $(\lambda,w)$-plane. All simulations were performed using a Fourier-Galerkin pseudospectral scheme with semi-implicit time stepping \cite{Kloeden1992,Canuto2006}, initiated with both a homogeneous isotropic and polar liquid state and allowed to relax to the steady-state at several choices for $D$ to ensure stability.

There are still many aspects of the phenomenology of the HVM that deserve deeper investigation, including the exact nature of the various transitions. However, for our purposes the available knowledge is sufficient, as our analysis will only treat the stable regimes of the isotropic, polar liquid, microphase-separated and polar cluster phases. Our aim in the following will first be to investigate the EPR in the HVM, and in particular its scaling in the limit $D\rightarrow0$, the physical significance of which will become more readily apparent in the next section.

\section{Entropy production at the fluctuating hydrodynamic level}
\label{sec:fluct-hydro-epr}
Although a large body of research in statistical physics has been devoted to the study of nonequilibrium systems, arguably few general principles have emerged. Among the more important successes are the discoveries of fluctuation theorems \cite{Seifert2012}. Within the framework of stochastic thermodynamics, these have formalised the connection between entropy production and time-reversal at the level of fluctuating trajectories \cite{Lebowitz1998}. Informally, fluctuation theorems capture the idea that the EPR is a measure of the probabilistic disparity between observing a time-forward trajectory (or history) of a system and its time-reversal under \textit{the same} ensemble. Because of this, fluctuations are essential in order to allow the time-reversed trajectory to be realisable under the time-forward dynamics. Despite this, a well-defined limit of vanishing noise strength can often be established \cite{Nardini2017,Dadhichi2018}. 

Previous studies have investigated this limit of vanishing noise in field theories of active matter, e.g. for AMB describing MIPS on the hydrodynamic scale \cite{Nardini2017}. Here it was found that the scaling of the steady-state EPR at small noise depends on the phase of the system. For an isotropic system, the EPR in AMB is $\Or(D)$, while it is $\Or(D^0)$ when phase-separation has occurred. On the other hand, Dadhichi et al. noted in \cite{Dadhichi2018} that in their model of flocking the EPR scales as $\Or(D^0)$ in both the homogeneous isotropic and polar liquid phases. Here we aim to provide some further details on the physics behind these results, and to organise them within a few unifying principles.

We construct the steady-state EPR from the Freidlin-Wentzell probability measure of trajectories on the time interval $[-\tau,\tau]$ \cite{Touchette2009}. For the system in \eref{eq:density}-\eref{eq:polarity}, this is defined via the action $\A$, where
\begin{equation}
\label{eq:fw-action}
\fl\A[\rho,\bi{P}]=
\frac{1}{4}\int_{-\tau}^{\tau}\rmd t\int_{\V}\rmd\bi{x}\,\left|\partial_t\bi{P}+\lambda\bi{P}\cdot\nabla\bi{P}+\funcdiff{F}{\bi{P}}\right|^2\quad\mbox{if}\quad\partial_t\rho+w\nabla\cdot\bi{P}=0,
\end{equation}
and $\A=\infty$ otherwise. The transition probability measure $\P[\rho,\bi{P}]$ of a trajectory $(\rho(t),\bi{P}(t))_{t\in[-\tau,\tau]}$ is then constructed in the standard way, i.e. by setting
\begin{equation}
\label{eq:path-prob}
\P[\rho,\bi{P}]\propto\exp\left(-\A[\rho,\bi{P}]/D\right).
\end{equation}
It is important to note that in this formulation of the stochastic dynamics, equation \eref{eq:density} acts as a constraint which limits the space of observable trajectories (i.e. those with $\P>0$). In order for all observable trajectories to have an observable time-reversal under $\P$, it is necessary therefore that the protocol $\mathcal{T}$ we choose for time-reversal involves a polarity flip, i.e.
\begin{equation}
\label{eq:trs-odd}
\T=\left\{\begin{array}{lll}
\rho(\bi{x},t) & \mapsto & \rho(\bi{x},-t), \\
\bi{P}(\bi{x},t) & \mapsto & -\bi{P}(\bi{x},-t).
\end{array}\right.
\end{equation}
Indeed, this ensures that $\A<\infty$ if and only if the composition $\A\circ\T<\infty$, which can be seen directly from \eref{eq:fw-action}. We may thus define a time-conjugate ensemble to \eref{eq:path-prob} by setting
\begin{equation}
\label{eq:rev-path-prob}
\cev{\P}[\rho,\bi{P}] = (\P\circ\T)[\rho,\bi{P}],
\end{equation}
which is supported on the same constrained space of trajectories as $\P$. 

Interestingly, one observes that the functional $F[\rho,\bi{P}]$ defined in \eref{eq:free-energy} is not invariant under $\T$. In particular, under this protocol $F$ decomposes into $\T$-symmetric and $\T$-antisymmetric contributions $F^S$ and $F^A$ respectively, where
\begin{eqnarray}
\label{eq:f-antisym}
F^A[\rho,\bi{P}]&=\frac{1}{2}\left(F[\rho,\bi{P}]-F[\rho,-\bi{P}]\right) \nonumber \\
&=\int_{\V}\rmd\bi{x}\,\bi{P}\cdot\nabla\Phi(\rho,\bi{P}),
\end{eqnarray}
is the part of $F$ that is odd in $\bi{P}$, and $F=F^S+F^A$. In fact, because of this we cannot interpret $F$ as a true free-energy since it would clearly have to remain invariant under time-reversal. Moreover, this also implies that the system in \eref{eq:density}-\eref{eq:polarity} is out of equilibrium even when $\lambda=0$, meaning that the self-advective contribution is not the only explicitly TRS violating component in the equations of motion.

Following standard treatments of stochastic thermodynamics, we formally define the steady-state entropy production rate $\dS$ as the (log) ratio between the forward and time-reversed ensembles $\P$ and $\cev{\P}$ \cite{Nardini2017,Lebowitz1998,Seifert2012}
\begin{equation}
\label{eq:epr-defn}
\dS\equiv\lim_{\tau\rightarrow\infty}\frac{1}{2\tau}\log\frac{\P[\rho,\bi{P}]}{\cev{\P}[\rho,\bi{P}]}.
\end{equation}
We assume that \eref{eq:epr-defn} holds in the almost sure sense. That is, we assume that $\dS=\langle\dS\rangle$, where $\langle\cdot\rangle$ denotes a steady-state expectation, for almost all realisations of the noise under the distribution $\P$. This assumption of ergodicity implies that we may replace noise averages by temporal averages and vice versa when computing $\dS$. Furthermore, definition \eref{eq:epr-defn} allows us to consider the EPR pathwise, i.e. as a functional of a trajectory $\dS\equiv\dS[\rho,\bi{P}]$. By construction, this functional satisfies the symmetry $\dS\circ\T=-\dS$ as can be readily observed from \eref{eq:epr-defn}, a fact closely related with the much celebrated Gallavotti-Cohen symmetry \cite{Lebowitz1998}.

In \ref{app:epr-calculation} we show from \eref{eq:fw-action}, \eref{eq:path-prob} and \eref{eq:rev-path-prob} that $\dS$ can be expressed in integral form as
\begin{equation}
\label{eq:epr}
\dS = D^{-1}\int_{\V}\rmd\bi{x}\,\left\langle\frac{w}{2}|\bi{P}|^2(\nabla\cdot\bi{P})-\left(\lambda \bi{P}\cdot\nabla\bi{P}+\funcdiff{F^A}{\bi{P}}\right)\cdot\funcdiff{F^S}{\bi{P}}\right\rangle.
\end{equation}
We will view $\dS\equiv\dS(D)$ as a function of the noise coefficient $D$ and look to determine the asymptotic scaling
\begin{equation}
\label{eq:asymp-scaling}
\dS(D)\sim D^{\chi},\qquad D\ll 1.
\end{equation}
In \ref{app:epr-half-order} we show that $\chi\in\{-1,0,1,\ldots\}$ can only take integer values. As we will argue, this result also makes sense physically. We will see that when $\chi=-1$ the steady ground-state dynamics at $D=0$ violates detailed balance in a pathwise sense, meaning that macroscopic irreversible currents are inherent to the dynamics and are not solely observable at the level of fluctuations. On the other hand, when $\chi=0$ the system is in a sense `marginally nonequilibrium'. In particular, the steady $D=0$ dynamics does not violate detailed balance, yet it is broken by a finite amount for any infinitesimal fluctuation and is never recovered as we send $D\rightarrow0$. In contrast, when $\chi\geq1$, the small noise limit is effectively an equilibrium regime where detailed balance is restored. Indeed, we will show that in the isotropic phase where $\chi=1$ an expansion of the fields in small $D$ allows the lowest order contribution beyond the steady $D=0$ solution to be mapped onto an equilibrium system of decoupled underdamped harmonic oscillators.

In the following two subsections we investigate analytically as well as numerically the scaling \eref{eq:asymp-scaling} in the various phases of the model in \eref{eq:density}-\eref{eq:polarity}. We begin by studying the homogeneous isotropic and polar liquid phases in addition to the polar cluster phase, where some analytical progress can be made at the fluctuating level. Subsequently, in \sref{sec:mps-analysis} we look at the micophase-separated and polar cluster regimes.

\subsection{Constant homogeneous ground-states}
\label{sec:iso-polar-analysis}
For the computations we present here, we will assume that the steady-state dynamics relaxes onto a `ground-state' trajectory $(\rho_0(\bi{x},t),\bi{P}_0(\bi{x},t))_{t\in(-\infty,\infty)}$ \footnote{From here on we omit the subscript notation in $(\rho_0(\bi{x},t),\bi{P}_0(\bi{x},t))_{t\in(-\infty,\infty)}$ when talking about a trajectory.} in the limit $D\rightarrow0$. That is, we assume that the probability distribution over trajectories concentrates on a single path as $D\rightarrow0$, and express this by
\begin{equation}
\label{eq:ground-state-limit}
(\rho,\bi{P})\overset{D\rightarrow0}{\rightarrow}(\rho_0,\bi{P}_0),
\end{equation}
where the limit is understood in the almost sure sense. We also assume that $(\rho_0,\bi{P}_0)$ solves \eref{eq:density}-\eref{eq:polarity} at $D=0$ and that this limit is unique up to possible degeneracies arising from rotational invariance. Firstly our aim will be to classify the ground-states that satisfy the \textit{pathwise} equilibrium condition
\begin{equation}
\label{eq:pathwise-eq}
\dS[\rho_0,\bi{P}_0]=0.
\end{equation} 
In particular, if both \eref{eq:ground-state-limit} and \eref{eq:pathwise-eq} hold, the dynamics must have $\chi>-1$. Clearly, the pathwise equilibrium ground-states include those that are invariant under $\T$ in \eref{eq:trs-odd}, meaning that they satisfy
\begin{equation}
(\rho_0(\bi{x},t),\bi{P}_0(\bi{x},t))=(\rho_0(\bi{x},-t),-\bi{P}_0(\bi{x},-t)),
\end{equation}
which follows from the fact that $\dS\circ\T=-\dS$. The constant homogeneous isotropic state with $\rho_0=\mathrm{const.}$ and $\bi{P}_0=0$ provides an example of such a state. On the other hand, the polar liquid state with $\rho_0>1$ and $|\bi{P}_0|=\sqrt{\rho_0-1}$ clearly violates $\T$ alone. This is where rotational invariance arises as an important symmetry principle, because it implies that $\P$ (and thus $\dS$) must be invariant under the parity transformation
\begin{equation}
\label{eq:parity}
\mathrm{P}=\left\{\begin{array}{lll}
\rho(\bi{x},t) & \mapsto & \rho(-\bi{x},t), \\
\bi{P}(\bi{x},t) & \mapsto & -\bi{P}(-\bi{x},t),
\end{array}\right.
\end{equation}
which translates to the statement that a flock is equally likely to travel to the left as to the right. Now, if the ground-state trajectory $(\rho_0,\bi{P}_0)$ is P$\T$-symmetric, i.e. it satisfies
\begin{equation}
\label{eq:pt-symmetry}
(\rho_0(\bi{x},t),\bi{P}_0(\bi{x},t))=(\rho_0(-\bi{x},-t),\bi{P}_0(-\bi{x},-t)),
\end{equation}
then it follows that it is pathwise equilibrium. Perhaps surprisingly then, one realises that the constant homogeneous polar liquid state in fact is pathwise equilibrium since it satisfies P$\T$. However, this is to be expected: a charged particle gyrating at constant frequency in the plane perpendicular to an imposed magnetic field is certainly in equilibrium (although here $\T$ should be replaced by C$\T$ to include charge conjugation). Interestingly, these observations also imply that if rotational symmetry is broken a priori, for example by driving the system with an external electric field, then $\P$ would no longer be P-invariant and the polar liquid state would have $\chi=-1$. We also note that the fact that $\chi>-1$ in both the homogeneous isotropic and polar liquid states also follows directly from \eref{eq:epr} by evaluating the integral at constant $(\rho_0,\bi{P}_0)$.

In order to go beyond the $D=0$ dynamics, we must take account of fluctuations. We do so by assuming that the fluctuating dynamics admit an expansion in small $\sqrt{D}$, following \cite{Nardini2017,Dadhichi2018}, so that
\begin{eqnarray}
\label{eq:density-expansion}
\rho = \rho_0+\rho_1D^{1/2}+\Or(D), \\
\label{eq:polarity-expansion}
\bi{P}=\bi{P}_0+\bi{P}_1D^{1/2}+\Or(D).
\end{eqnarray}
Furthermore, we restrict here to the case where $(\rho_0,\bi{P}_0)$ is constant and homogeneous. By substituting \eref{eq:density-expansion}, \eref{eq:polarity-expansion} into the equations of motion in \eref{eq:density}-\eref{eq:polarity} and collecting terms, we obtain at order $D^{1/2}$
\begin{eqnarray}
\label{eq:density-lin}
\partial_t\rho_1=-w\nabla\cdot\bi{P}_1, \\
\label{eq:polarity-lin}
\partial_t\bi{P}_1+\lambda\bi{P}_0\cdot\nabla\bi{P}_1=-\funcdiff{F_L}{\bi{P}}[\rho_1,\bi{P}_1]+\bfeta_1.
\end{eqnarray}
Here, $F_L$ is the quadratic functional
\begin{equation}
\label{eq:f-lin}
F_L[\rho,\bi{P}]=\int_{\V}\rmd\bi{x}\left(f_L(\rho,\bi{P})+\frac{1}{2}(\nabla_{\alpha}P_{\beta})^2+\bi{P}\cdot\nabla\Phi_L(\rho,\bi{P})\right),
\end{equation}
where we have defined the local free energy $f_L$ by
\begin{equation}
\label{eq:f-den-lin}
f_L(\rho,\bi{P})=\frac{a_0}{2}|\bi{P}|^2-\rho\bi{P}_0\cdot\bi{P}+(\bi{P}_0\cdot\bi{P})^2,
\end{equation}
in addition to the linearised effective pressure
\begin{equation}
\label{eq:pressure-lin}
\Phi_L=w_1\rho-\kappa\bi{P}_0\cdot\bi{P}.
\end{equation}
Moreover, $a_0=1-\rho_0+|\bi{P}_0|^2$ and $\bi{P}_0$ satisfies $a_0\bi{P}_0=0$, while $\bfeta_1$ is a mean-zero Gaussian white noise with
\begin{equation}
\left\langle\eta_{1\alpha}(\bi{x},t)\eta_{1\beta}(\bi{x}',t')\right\rangle=2\delta_{\alpha\beta}\delta(\bi{x}-\bi{x}')\delta(t-t').
\end{equation}

\begin{figure}
\centering
\begin{minipage}{0.5\textwidth}
\includegraphics[scale=0.95]{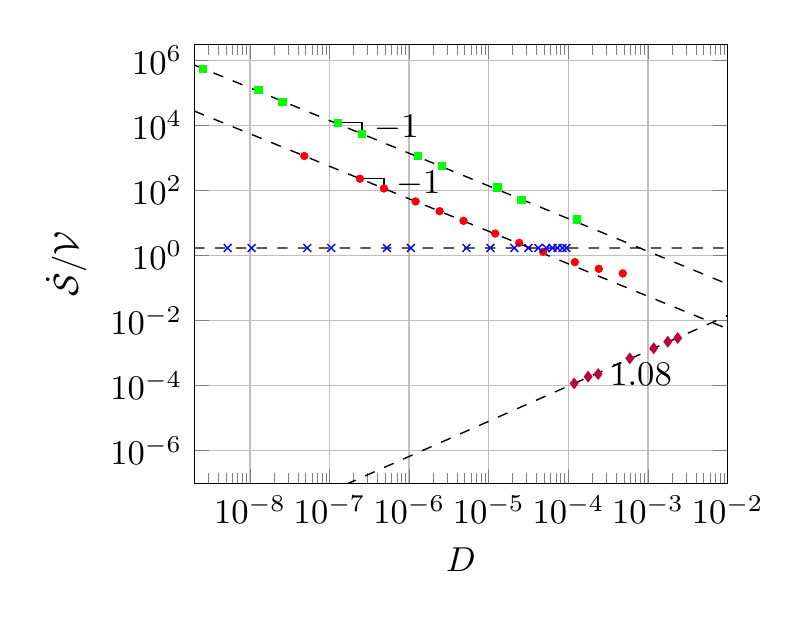}
\end{minipage}%
\begin{minipage}{0.5\textwidth}
\hspace{0.5cm}
{\footnotesize\begin{tabular}{l|l@{}}
{\tiny\color{purple}$\fulldiamond$} & Isotropic \\
 & $\rho_0=0.9$, $\lambda=1.1$, $w=1.2$ \\
\mr
{\color{blue}$\times$} & Polar liquid \\
 & $\rho_0=1.33$, $\lambda=1.1$, $w=1.2$ \\
 & Intercept: $1.6984$\\
 & $\dS_0(\Lambda)/\V\approx1.6992$ \\
\mr
{\color{red}$\bullet$} & Microphase-separation \\
 & $\rho_0=1.04$, $\lambda=1.1$, $w=1.2$ \\
\mr
{\tiny\color{green}$\fullsquare$} & Polar cluster \\
 & $\rho_0=1.28$, $\lambda=1.53$, $w=1.5$ \\
\end{tabular}}
\vspace{0.5cm}
\end{minipage}
\caption{Scaling of the EPR $\dS$ (normalised by volume $\V$) with the noise coefficient $D$ in the isotropic, polar liquid, microphase-separated and polar cluster regimes. Dashed lines (\dashed) represent the best linear fit to the data from simulations (marked by {\footnotesize\color{purple}$\fulldiamond$}, {\color{blue}$\times$},  {\color{red}$\bullet$}, {\tiny\color{green}$\fullsquare$}), with the associated intercept (best estimate of $\lim_{D\rightarrow0}\dS(D)$ from simulation data) reported in the legend for the polar liquid. The intercept is compared with the numerically evaluated analytical result in \eref{eq:epr-zero-fourier} for $\dS_0(\Lambda)/\V$, where $\Lambda=2\pi N/L$, $L=14\pi$ and $N=96$.} 
\label{fig:scaling-homogeneous-phases} 
\end{figure}

We may perform a similar procedure in order to obtain an expansion of $\dS$ from \eref{eq:epr} in small $D$ of the form
\begin{equation}
\label{eq:epr-expansion}
\dS(D)=\dS_{-1}D^{-1}+\dS_0+\dS_1D+\Or(D^2).
\end{equation}
In \ref{app:epr-half-order} we show these are the only possible terms that could enter in the expansion of $\dS$, i.e. that there are no terms of half-integer order in $D$. Also, from the asymptotic scaling relation \eref{eq:asymp-scaling} and the positivity of the EPR, we know that $\dS_k=0$ for $k<\chi$ and that the leading order coefficient $\dS_{\chi}>0$. By explicitly computing this expansion of $\dS$ to order $D^0$, it is straightforward to show that
\begin{equation}
\label{eq:hom-scaling}
\chi\quad\left\{\begin{array}{ll}
\geq1,&\mbox{isotropic}, \\
=0,&\mbox{polar liquid},
\end{array}\right.
\end{equation}
and so $\chi>-1$ in the homogeneous phases as argued for above. For the explicit calculation of \eref{eq:hom-scaling}, we refer to \ref{app:epr-calculation} for the details. From simulations we further find that in fact $\chi=1$ in the isotropic phase, as shown in \fref{fig:scaling-homogeneous-phases}, although we do not explicitly compute $\dS_1$. In the polar liquid case, we obtain an explicit expression for $\dS_0$ given by
\begin{eqnarray}
\label{eq:epr-zero}
\fl\dS_0=P_0^2(2w_1-\kappa+\lambda)\int_{\V}\rmd\bi{x}\,\left\langle\rho_1\partial_{\parallel}P_{\parallel}\right\rangle+P_0(w-2P_0^2\kappa)\int_{\V}\rmd\bi{x}\,\left\langle P_{\parallel}\partial_{\perp}P_{\perp}\right\rangle \nonumber\\
+2P_0\kappa\int_{\V}\rmd\bi{x}\,\left\langle(\partial_{\perp}P_{\perp})(\nabla^2P_{\parallel})\right\rangle.
\end{eqnarray}
In \eref{eq:epr-zero} we use subscripts $\parallel$ and $\perp$ to denote components of $\bi{P}_1$ and $\nabla$ that are parallel and perpendicular to $\bi{P}_0$ respectively. 

Consistently with \eref{eq:hom-scaling}, equation \eref{eq:epr-zero} implies that $\dS_0\sim P_0=|\bi{P}_0|$ for $P_0\ll1$. In fact, this could have been predicted without explicitly performing the systematic expansion of $\dS$ in small $D$. Indeed, if we were to imagine expanding the integral expression for $\dS$ in \eref{eq:epr} using the series representations \eref{eq:density-expansion}, \eref{eq:polarity-expansion} at $\bi{P}_0=0$, we see from simple power counting that the only combinations of fields that could possibly appear within the integrand at order $D^0$ are of the form
\begin{equation}
\label{eq:combinations-of-fields}
\langle\rho_2\rangle,\qquad\langle\nabla\cdot\bi{P}_2\rangle,\qquad\langle\rho_1^2\rangle,\qquad\langle|\bi{P}_1|^2\rangle,\qquad\langle\rho_1\nabla\cdot\bi{P}_1\rangle,\ldots
\end{equation}
Now, the symmetry $\dS\circ\T=-\dS$ excludes all of $\rho_2$, $\rho_1^2$ and $|\bi{P}_1|^2$ from entering, while $\nabla\cdot\bi{P}_2$ would just integrate to zero over $\V$. For the final average in \eref{eq:combinations-of-fields}, observe that \eref{eq:density-lin} implies
\begin{equation}
\label{eq:density-polarity-correlator}
\langle\rho_1\nabla\cdot\bi{P}_1\rangle=-w^{-1}\langle\rho_1\partial_t\rho_1\rangle=\frac{-1}{2w}\partial_t\langle\rho_1^2\rangle=0.
\end{equation}
Hence, there are in fact no nontrivial contributions that could enter in the expansion of $\dS$ at order $D^0$ when $\bi{P}_0=0$, so we must have $\chi\geq1$ in the isotropic phase.

Since $\dS$ is $\Or(D)$ in the isotropic phase, we in fact recover effective equilibrium in the limit $D\rightarrow0$. To see this, we transform the linearised equations of motion \eref{eq:density-lin}, \eref{eq:polarity-lin} to Fourier space. Throughout we use the convention that the Fourier coefficients $\hat{h}(\bi{q})$ of a function $h(\bi{x})$ are given by
\begin{equation}
\label{eq:fourier-conv}
\hat{h}(\bi{q})=\V^{-1}\int_{\V}\rmd\bi{x}\,h(\bi{x})\exp(-i\bi{x}\cdot\bi{q}),
\end{equation}
where we with slight abuse of notation denote by $\V=\mathrm{vol}(\V)$. We thus obtain the set of equations
\begin{eqnarray}
\label{eq:density-lin-fourier}
\partial_t\hat{\rho}_1=-iw\bi{q}\cdot\hat{\bi{P}}_1, \\
\label{eq:polarity-lin-fourier}
\partial_t\hat{\bi{P}}_1=-\Gamma(q)\hat{\bi{P}}_1-iw_1\bi{q}\hat{\rho}_1+\hat{\bfeta}_1,
\end{eqnarray}
where we have defined the damping coefficient $\Gamma(q)=a_0+q^2$ and the noise term $\hat{\bfeta}_1(\bi{q},t)$ is mean zero, Gaussian and white with autocovariance
\begin{equation}
\langle\hat{\eta}_{1\alpha}(\bi{q},t)\hat{\eta}^*_{1\beta}(\bi{q}',t')\rangle=2\V^{-1}\delta_{\alpha\beta}\delta_{\bi{q},\bi{q}'}\delta(t-t').
\end{equation} 
Using the mapping
\begin{eqnarray}
\hat{\rho}_1(\bi{q},t)=x(\bi{q},t)+iy(\bi{q},t), \\
\partial_t\hat{\rho}_1(\bi{q},t)=v_x(\bi{q},t) + iv_y(\bi{q},t),
\end{eqnarray} 
and setting $\bi{X}=(x,y)$, $\bi{V}=(v_x,v_y)$ we immediately see that these follow standard equilibrium Langevin equations for an underdamped particle in a harmonic potential \cite{Kubo1991},
\begin{eqnarray}
\dot{\bi{X}}=\bi{V}, \\
\label{eq:velocity}
\dot{\bi{V}}=-\Gamma \bi{V}-\nabla_{X}U+\sqrt{2\Gamma T}\bzeta,
\end{eqnarray}
where the potential $U=ww_1q^2|\bi{X}|^2/2$. The final degrees of freedom in \eref{eq:density-lin-fourier}, \eref{eq:polarity-lin-fourier} are captured by the transverse component $\bi{V}_T=(v_{Tx},v_{Ty})$ of $\hat{\bi{P}}_1$ with respect to $\bi{q}$, i.e. 
\begin{equation}
v_{Tx}(\bi{q},t) + iv_{Ty}(\bi{q},t)=-iw\bi{q}_{\perp}\cdot\hat{\bi{P}}_1(\bi{q},t),
\end{equation}
and $\bi{q}_{\perp}$ is perpendicular to $\bi{q}$ with $|\bi{q}_{\perp}|=q$. Again this follows an equilibrium Langevin equation,
\begin{equation}
\label{eq:v-trans}
\dot{\bi{V}}_T=-\Gamma \bi{V}_T+\sqrt{2\Gamma T}\bzeta_T.
\end{equation}
In \eref{eq:velocity}, \eref{eq:v-trans} the noise terms $\bzeta$, $\bzeta_T$ are mean zero unit variance Gaussian white noises, and interestingly the effective temperature $T$ is defined by
\begin{equation}
T(q)=\frac{1}{2\V}\frac{w^2q^2}{a_0+q^2}.
\end{equation}
Since the modes $\bi{V}$, $\bi{V}_T$ are independent for all $\bi{q}$, the dependence of the effective temperature $T$ on $q$ does not lead to any current in phase space. At higher order in $D$, however, modes $\hat{\rho}_1(\bi{q},t)$ and $\hat{\bi{P}}_1(\bi{q},t)$ are coupled at different wavevectors $\bi{q}$ via the nonlinear terms in the equation for the polar density in \eref{eq:polarity}. In particular, these terms couple heat baths at different temperatures $T(q)$, driving the dynamics at the next order away from equilibrium.

Linear theory also allows us to make quantitative predictions about $\dS_0$ from \eref{eq:epr-zero} in the polar liquid phase. Indeed, transforming this to Fourier space we obtain
\begin{equation}
\label{eq:epr-zero-fourier}
\dS_0(\Lambda)/\V=\sum_{|\bi{q}|\leq\Lambda}\left\langle\left(\hat{\bi{u}}^{\mathrm{pl}}\right)^{\dagger}\dot{\sigma}^{\mathrm{pl}}\hat{\bi{u}}^{\mathrm{pl}}\right\rangle=\sum_{|\bi{q}|\leq\Lambda}\Tr\left(\dot{\sigma}^{\mathrm{pl}}\mathcal{C}^{\mathrm{pl}}\right),
\end{equation}
where the Hermitian matrix $\dot{\sigma}^{\mathrm{pl}}$ is given by
\begin{equation}
\label{eq:sigma-pl}
\fl\dot{\sigma}^{\mathrm{pl}}=\frac{iP_0}{2}\left(\begin{array}{ccc}
0 & P_0(2w_1-\kappa+\lambda)q_{\parallel} & 0 \\
P_0(\kappa-2w_1-\lambda)q_{\parallel} & 0 & (w-2(P_0^2+q^2)\kappa)q_{\perp} \\
0 & (2(P_0^2+q^2)\kappa-w)q_{\perp} & 0
\end{array}\right).
\end{equation}
In addition, in \eref{eq:epr-zero-fourier} we have defined the vector
\begin{equation}
\label{eq:u-vec}
\hat{\bi{u}}^{\mathrm{pl}}=\left(\hat{\rho}_1,\hat{P}_{\parallel},\hat{P}_{\perp}\right)^T
\end{equation}
of Fourier modes, as well as the matrix $\mathcal{C}^{\mathrm{pl}}\equiv(\mathcal{C}^{\mathrm{pl}}_{ij})$ of equal-time correlators by
\begin{equation}
\label{eq:corr-mat-pl}
\mathcal{C}^{\mathrm{pl}}_{ij}(q)\delta_{\bi{q},\bi{q}'}=\left\langle\hat{u}^{\mathrm{pl}}_i(\bi{q},t)\left(\hat{u}^{\mathrm{pl}}_j(\bi{q}',t)\right)^*\right\rangle.
\end{equation}
The sum in \eref{eq:epr-zero-fourier} runs over modes with wavenumbers smaller than the ultraviolet cutoff $\Lambda$, which is introduced since the sum is divergent with $\Lambda\rightarrow\infty$. This is sometimes seen in field theories of this kind, since they are often derived based on the assumption that they are only valid down to a certain length scale. In \ref{app:epr-calculation} we show from this that $\dS_0(\Lambda)$ as predicted by the linear theory diverges in the ultraviolet as $\Lambda^2$. Although the closed form expressions for the correlators entering in \eref{eq:epr-zero-fourier} are too algebraically involved to report explicitly, they may be calculated straightforwardly by numerical methods. By doing this, we may calculate the corresponding sum in \eref{eq:epr-zero-fourier} and quantitatively compare the results with measurements of $\dS$ from simulations. In \fref{fig:scaling-homogeneous-phases} we plot the results obtained from simulations, which show good agreement with the analytical results. Plot a) in \fref{fig:scaling-homogeneous-phases} demonstrates that the EPR is $\Or(D)$ in the isotropic phase, as well as the predicted $\Or(D^0)$ scaling in the polar liquid phase. In particular, both remain finite at fixed $\Lambda$ as $D\rightarrow0$, with $\dS\rightarrow0$ in the isotropic phase.

\subsection{Nonlinear ground-states: Polar clusters and microphase-separation}
\label{sec:mps-analysis}
Previous studies have investigated the nonlinear solutions to \eref{eq:density}-\eref{eq:polarity} at $D=0$, and particularly interesting to our present context are the seminal contributions by Solon et al. \cite{Solon2015a,Solon2015b} on the structure of the banded profiles. These are effectively one-dimensional traveling wave solutions that are invariant along the direction perpendicular to the motion. We thus write $\bi{P}_0=(P_0,0)$ without loss of generality, and look for solutions of the form
\begin{eqnarray}
\label{eq:density-wave-ansatz}
\rho_0(\bi{x},t)\equiv\tilde{\rho}_0(x-ct), \\
\label{eq:polar-wave-ansatz}
P_0(\bi{x},t)\equiv\tilde{P}_0(x-ct).
\end{eqnarray}
Direct substitution then allows us to deduce a set of equations for $\tilde{\rho}_0$ and $\tilde{P}_0$ in terms of the variable $z=x-ct$ given explicitly by
\begin{eqnarray}
\tilde{\rho}_0=\rho_g+\frac{w}{c}\tilde{P}_0,\\
\label{eq:polar-wave}
\tilde{P}_0''=-\left(c-\frac{ww_1}{c}-\lambda\tilde{P}_0\right)\tilde{P}_0'-\left(\rho_g-1+\frac{w}{c}\tilde{P}_0-\tilde{P}_0^2\right)\tilde{P}_0,
\end{eqnarray}
where primes denote differentiation with respect to $z$. \Eref{eq:polar-wave} can be mapped onto a Newton problem for a particle in a potential under the influence of a nonlinear drag, and all stable orbits in the $(\tilde{P}_0,\tilde{P}_0')$ plane with $\tilde{P}_0\geq0$ can be uniquely identified with a pair $(c,\rho_g)$. In terms of the stochastic equations in \eref{eq:density}-\eref{eq:polarity}, it is assumed that the noise selects the stable steady-state profile (of which there are infinitely many \cite{Solon2015a,Solon2015b}).
\begin{figure}
\centering
\includegraphics[scale=0.65]{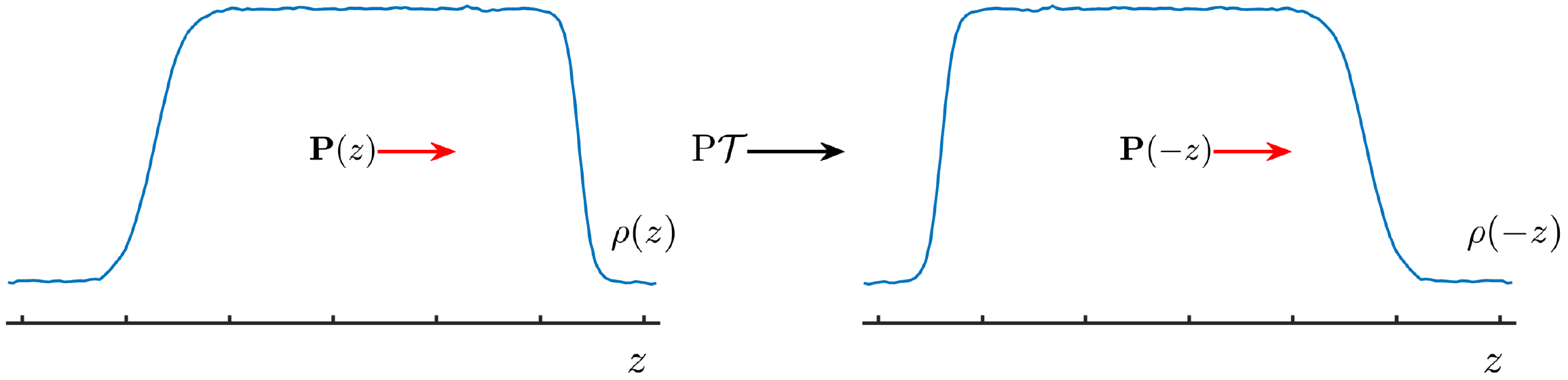}
\caption{Illustration showing a cross section of a banded profile traveling in the positive $x$ direction (left) and its image under the map P$\T$ (right).}   
\label{fig:scaling-banded-phase} 
\end{figure}

Importantly, these solutions to \eref{eq:polar-wave-ansatz} break both $\T$ and P$\T$-symmetry. Thus we expect the microphase-separated steady-state to have $\chi=-1$. Indeed, using the traveling wave ansatz in \eref{eq:density-wave-ansatz} and \eref{eq:polar-wave-ansatz} we deduce two expressions for $\dS_{-1}=\lim_{D\rightarrow0}D\dS(D)$, that are
\begin{eqnarray}
\label{eq:dissipation-positive}
\dS_{-1}/\V&=\frac{1}{2L}\int_{-L}^{L}\rmd z\left(\left(1-\tilde{\rho}_0\right)\tilde{P}_0+\tilde{P}_0^3-\tilde{P}_0''\right)^2\\
\label{eq:dissipation-odd}
&=-\frac{\lambda}{2L}\int_{-L}^{L}\rmd z\,(\tilde{P}_0')^3.
\end{eqnarray}
The first of these is most straightforwardly derived from \eref{eq:epr-1} in \ref{app:epr-calculation} by using the ODE \eref{eq:polar-wave} for $\tilde{P}_0$, and verifies that $\dS_{-1}\geq0$ as must be the case. The latter, i.e. \eref{eq:dissipation-odd}, follows immediately from \eref{eq:epr} after integrating out total derivatives. Now, from the final expression in \eref{eq:dissipation-odd} one observes that $\dS_{-1}$ vanishes identically for even distributions, i.e. those that satisfy $\tilde{P}_0(z_0+z)=\tilde{P}_0(z_0-z)$ for some $z_0$, which is exactly the P$\T$-symmetry in \eref{eq:pt-symmetry}. However, the traveling wave profiles are clearly asymmetric with a steeper front than tail, leading in general to the observed $\dS_{-1}\neq0$. In \fref{fig:scaling-banded-phase} we illustrate this, and in particular how the banded profile transforms under P$\T$. Finally, a sanity check also verifies that both expressions \eref{eq:dissipation-positive} and \eref{eq:dissipation-odd} are invariant under P alone ($\tilde{P}_0(z)\rightarrow-\tilde{P}_0(-z)$) as they should be since $\dS$ is, as remarked previously, oblivious to whether the wave is moving left or right (in \eref{eq:polar-wave} this must be complemented by $c\rightarrow-c$).

Interestingly, this mode of TRS violation at $D=0$ is a collective emergent phenomenon and does not have any counterpart for a single active particle. On the microscopic scale, it is associated with different rates of promotion and demotion of spins at the head and tail of the nonlinear profile respectively, leading to a difference in the rate of dissipation from the active alignment at these edges -- a point which we will explore further in a separate paper. In the following section, this point will become more explicit when we see how TRS can also be violated at $D=0$ by explicitly tracking mass currents in addition to the local polar density. In particular, in this case there is an analogous mode of TRS violation on the level of a single active particle.

We include in \fref{fig:scaling-homogeneous-phases} the scaling of the EPR $\dS$ in both the microphase-separated and polar cluster regimes. As shown, both are truly nonequilibrium within our classification scheme, with $\dS\sim D^{-1}$. Although we do not possess explicit polar cluster solutions to the dynamics at $D=0$, it is straightforward to argue heuristically that this is what one should expect due to the highly inhomogeneous nature of the phase. For future work, we aim to investigate this in more detail.

\subsection{Summary}
So far we have seen that the EPR of the HVM at small noise satisfies the asymptotic scaling relation $\dS\sim D^{\chi}$ for $D\ll1$, where the exponent $\chi\in\{-1,0,1,\ldots\}$. By performing a small noise expansion, we may systematically investigate the coefficients that appear at each order to determine the lowest order nontrivial contribution, and thus $\chi$. We also find that symmetries effectively bound $\chi$ from below; when the ground-state dynamics is pathwise equilibrium, the contribution $\dS_{-1}/D=\dS[\rho_0,\bi{P}_0]$ is locked out and $\chi>-1$. In the isotropic case, the fact that $\langle\rho_1\nabla\cdot\bi{P}_1\rangle=0$ further constrains $\chi>0$, and the small noise limit becomes an effective equilibrium regime. In \sref{sec:density-flucts} we look at the entropy production in a generalised model, where the constitutive equation for the density current in \eref{eq:constitutive-eqn} is modified to include a diffusive contribution.

\section{TRS violations in the generalised Diffusive Flocking Model}
\label{sec:density-flucts}
Above we found that for the HVM, pathwise violation of detailed balance at ground-state level is the direct result of a P$\T$-symmetry breaking by asymmetric steady $D=0$ profiles, and that a steady current of density was not sufficient to cause this alone. Here we aim to show that by changing the model so as to allow independent density current fluctuations, and by tracking this current explicitly, this may no longer hold. In this case therefore, we find that the EPR in fact does diverge as $D^{-1}$ due to the presence of circulating homogeneous currents of mass. Moreover, we recover an explicit expression for the pathwise EPR of a constant homogeneous polar state in this case.

\subsection{The Diffusive Flocking Model}
In the following, we add a diffusive contribution and noise to the constitutive equation for the density current $\bi{J}$. Specifically, we consider $\bi{J}=\bi{J}_d+\bxi$ as in \cite{Marchetti2013}, where the noise $\bxi$ is mean zero, Gaussian and white with covariance
\begin{equation}
\langle\xi_{\alpha}(\bi{x},t)\xi_{\beta}(\bi{x}',t')\rangle=2D_{\rho}\delta_{\alpha\beta}\delta(\bi{x}-\bi{x}')\delta(t-t').
\end{equation}
Furthermore, we take the deterministic part $\bi{J}_d$ of the current $\bi{J}$ to be of the form
\begin{equation}
\label{eq:jd}
\bi{J}_d=w\bi{P}-\gamma^{-1}\nabla\mu,
\end{equation}
where $\gamma$ is a constant friction coefficient. Here, $\mu$ serves an analogous purpose with the chemical potential known from equilibrium diffusive systems. However, since TRS is broken in this model, there is a priori no reason that it should be the functional variation of a free-energy. Notwithstanding, we will for simplicity ignore this issue and assume that we may write
\begin{equation}
\mu=\funcdiff{F}{\rho},
\end{equation}
with the same functional $F$ as that which appears in the equation for $\bi{P}$. We only make minor changes to $F$ for stability purposes by modifying the local free-energy $f(\rho,\bi{P})$ to include a quadratic term in $\rho$, so that now
\begin{equation}
f(\rho,\bi{P}) = \frac{a_{\rho}}{2}\rho^2+\frac{1}{2}(1-\rho)|\bi{P}|^2+\frac{1}{4}|\bi{P}|^4.
\end{equation}
In addition we add a square gradient contribution, giving us
\begin{equation}
\label{eq:dfm-free-energy}
F[\rho,\bi{P}]=\int_{\V}\rmd\bi{x}\,\left(f(\rho,\bi{P})+\frac{\nu_{\rho}}{2}|\nabla\rho|^2+\frac{1}{2}(\nabla_{\alpha}P_{\beta})^2+\bi{P}\cdot\nabla\Phi(\rho,\bi{P})\right).
\end{equation}
Observe, however, that these new terms do not change the equation for $\bi{P}$ in \eref{eq:polarity-full}, since they both drop out when considering the functional variation of $F$ with respect to $\bi{P}$. With this choice, we have that
\begin{equation}
\mu = a_{\rho}\rho-\nu_{\rho}\nabla^2\rho-\frac{1}{2}|\bi{P}|^2-w_1\nabla\cdot\bi{P}.
\end{equation}

Again, \eref{eq:jd} is motivated by coarse graining, and the diffusive contribution arises for example in cases where interactions such as steric repulsion are included in the microscopic model \cite{Marchetti2013}. Notably, there is a kind of paradigm shift when breaking the local linear relation $\bi{J}\propto\bi{P}$, which implies that $\bi{W}=\bi{P}/\rho$ should no longer be considered the local average direction of the velocity of particles. Physically, this reflects a situation on the microscopic scale where the bare self-propulsion may be thwarted by e.g. repulsive forces, so that mass currents may move against the local polar order. More importantly in the context of entropy production, this means that a trajectory of the system in which $\bi{J}$ and $\bi{P}$ do \textit{not} point in the same direction is realisable in the forward time ensemble, since fluctuations alone can now reverse $\bi{J}$ at fixed $\bi{P}$ even if highly unlikely. 
\begin{figure}[t]
\centering
\begin{subfigure}[b]{.3\linewidth}
\includegraphics[scale=0.6]{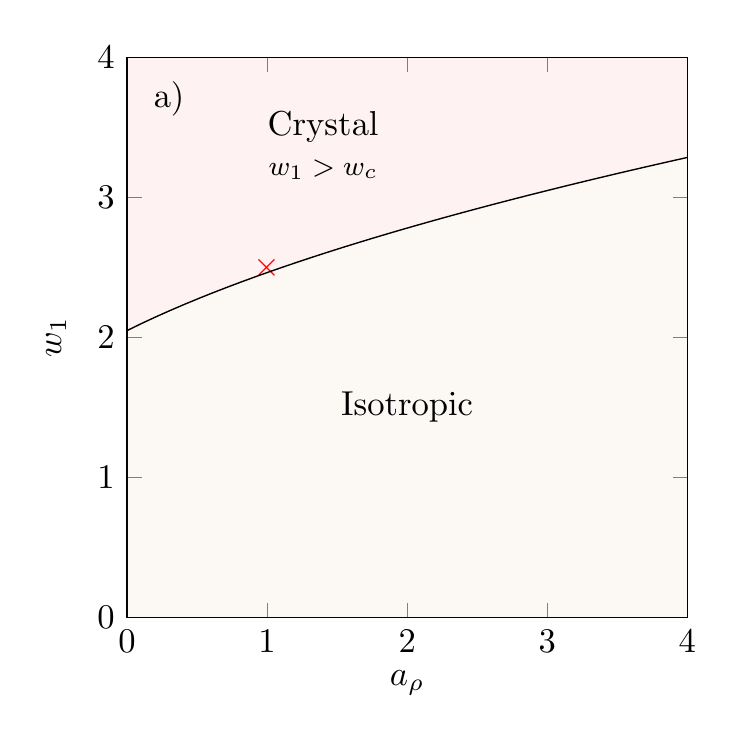}
\end{subfigure}
\begin{subfigure}[b]{.3\linewidth}
\includegraphics[scale=0.35]{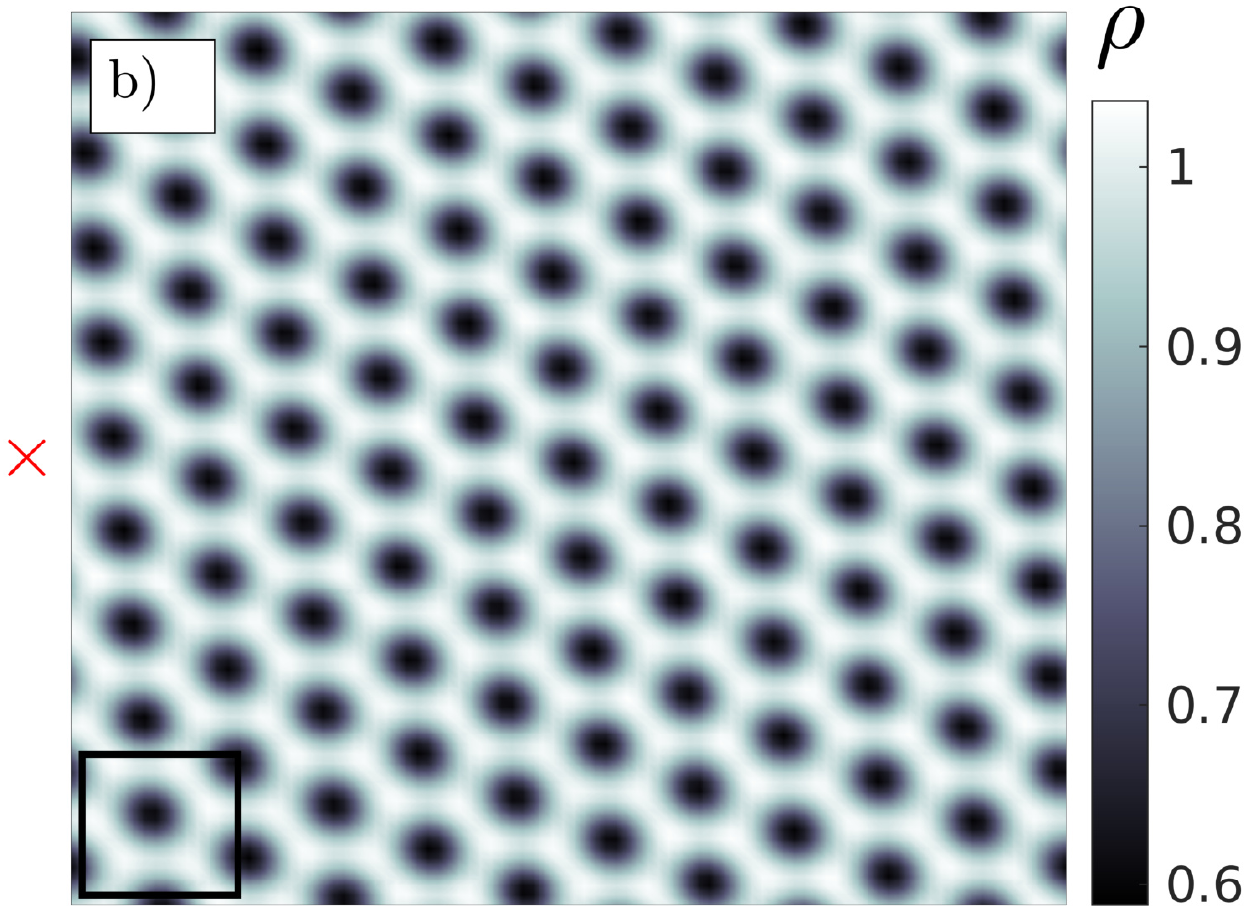}
\vspace{0.75cm}
\end{subfigure}
\begin{subfigure}[b]{.3\linewidth}
\includegraphics[scale=0.35]{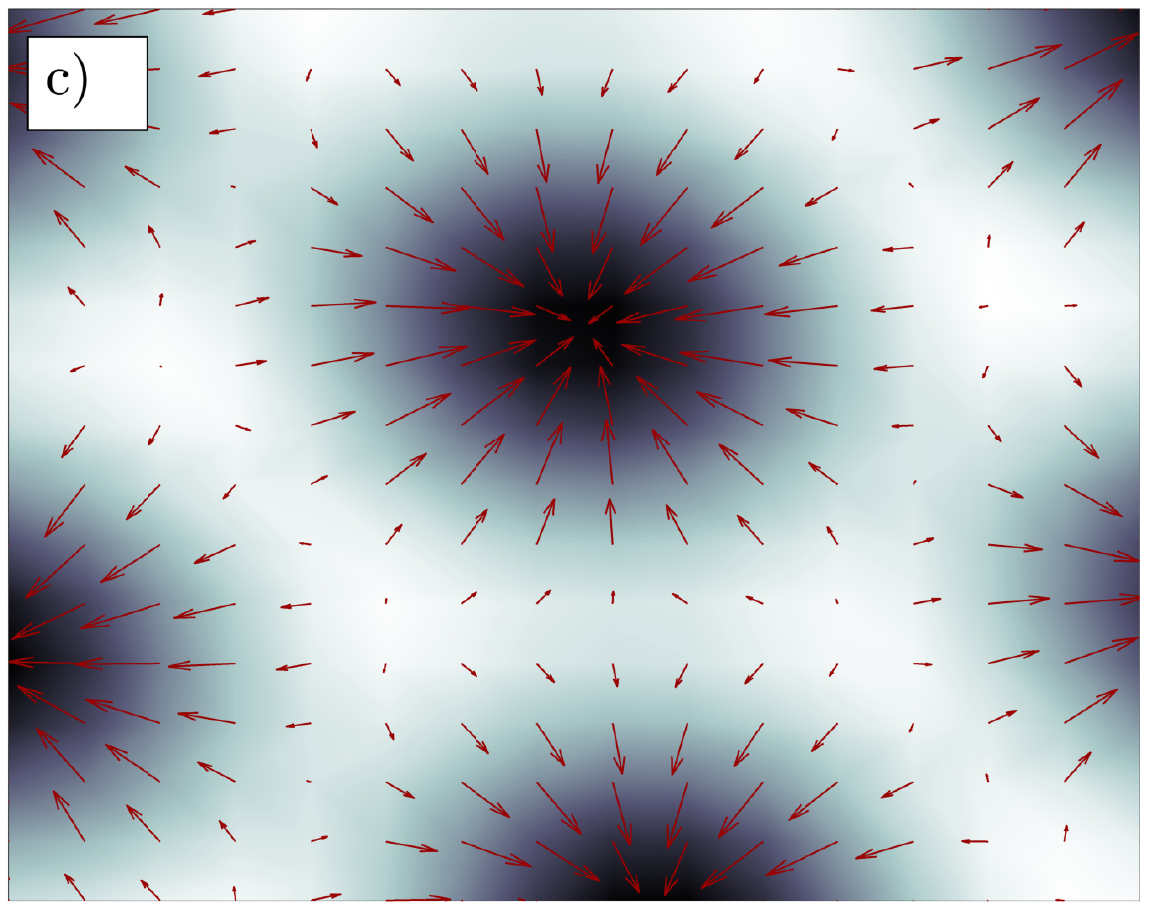}
\vspace{0.75cm}
\end{subfigure}
\caption{a) Phase diagram of the DFM at fixed $\rho_0=0.9$, $\nu_{\rho}=1$, $\gamma=0.5$ and $w_1=w/2$. When the condition \eref{eq:c-stability} is met, there is a finite range $q\in[q_-,q_+]$ of modes that are unstable to perturbations away from the isotropic state even when $\rho_0<1$. The resulting steady-state is a type of polar crystal in which a hexagonal lattice formed by high-density ridges enclose low density valleys, as illlustrated in b). Figure c) shows an enlarged part of the plot in b), indicated by a black square, including also the local polar density plotted with red arrows ({\color{red}$\rightarrow$}). The simulation parameters used in b) are given by $w_1=2.5$ and $a_{\rho}=1$, corresponding to the data point ({\color{red}$\times$}) in a), and in addition $\lambda=\kappa=1.1$, $D=10^{-4}$ and $L=7\pi$.}   
\label{fig:diff-flock-phases} 
\end{figure}

Numerical integration of the dynamics with $\bi{J}=\bi{J}_d+\bxi$, hereafter referred to as the Diffusive Flocking Model (DFM), allows us to investigate the resulting phase diagram as in \sref{sec:field-theory}. On the other hand, achieving analytical progress to a comparable extent as with the HVM is more difficult. Notably, however, from a linear analysis we do in fact find a finite wavelength instability in the region where $\rho_0<1$, in which the coarsening dynamics develop a polar crystalline structure as illustrated in \fref{fig:diff-flock-phases}. Our analysis also provides us with the isotropic-to-crystal phase-boundary, and we find that in the case $w_1=w/2$ the system is unstable to perturbations when
\begin{equation}
\label{eq:c-stability}
w_1^2>w_c^2\equiv a_{\rho}+\nu_{\rho}(1-\rho_0)+4\gamma\nu_{\rho}+2\sqrt{\nu_{\rho}(1-\rho_0+2\gamma)(a_{\rho}+2\gamma\nu_{\rho})}.
\end{equation}
More specifically, when the inequality \eref{eq:c-stability} holds there is a finite range $q\in[q_-,q_+]$ of modes that are unstable, where the exact expressions for $q_{\pm}$ are provided in \ref{app:lin-analysis}. In contrast, we do not observe significant changes to the phase diagram in the region where $\rho_0>1$. We explain this behaviour by observing that when $\rho_0<1$, and the local polar order is weak, the diffusive dynamics is significant while it is overpowered by advective transport when the polar order is strong \cite{Marchetti2013}. At $D=0$ the state is stationary and has both $\partial_t\rho=0$ and $|\partial_t\bi{P}|=0$. In fact, from simulations we observe that the stronger condition $|\langle\bi{J}_d\rangle|=0$ is met, meaning that at $D=0$ we expect 
\begin{equation}
\label{eq:zero-jd}
w\bi{P}=\gamma^{-1}\nabla\mu.
\end{equation}
From simulations we observe that the local polar order is directed such that it points \textit{in} towards low density, as illustrated in \fref{fig:diff-flock-phases}. \Eref{eq:zero-jd} then tells us that the advective transport induced by $\bi{P}$ is compensated by a reversed `diffusion' running \textit{up} gradients in $\rho$.

In the following we also restrict to the case where $D_{\rho}=D/\gamma$ for simplicity \cite{Marchetti2013}, in which case the Freidlin-Wentzell action for the DFM takes the form
\begin{equation}
\label{eq:diff-action}
\fl\adf[\rho,\bi{P}]=\frac{1}{4}\int_{-\tau}^{\tau}\rmd t\int_{\V}\rmd\bi{x}\,\left[\gamma\left|\nabla^{-1}\left(\partial_t\rho+\nabla\cdot\bi{J}_d\right)\right|^2+\left|\partial_t\bi{P}+\lambda\bi{P}\cdot\nabla\bi{P}+\funcdiff{F}{\bi{P}}\right|^2\right],
\end{equation}
and the path transition density $\Pdf[\rho,\bi{P}]$ is constructed as before by setting $\Pdf\propto\exp(-\adf/D)$. Note that in \eref{eq:diff-action}, we have defined the inverse gradient operator $\nabla^{-1}=\nabla^{-2}\nabla$, i.e. with gauge choice $|\nabla\times\nabla^{-1}h(\bi{x})|=0$ \cite{Nardini2017}. Crucially, with the added density current fluctuations, we are now free to define time-reversal without the polarity flip used in \eref{eq:trs-odd}. Specifically, we let
\begin{equation}
\label{eq:trs-diff}
\T^{\pm}=\left\{\begin{array}{lll}
\rho(\bi{x},t) & \mapsto & \rho(\bi{x},-t) \\
\bi{P}(\bi{x},t) & \mapsto & \pm\bi{P}(\bi{x},-t)
\end{array}\right.
\end{equation}
and observe that both compositions $\adf\circ\T^{\pm}$ are now well defined on the full space of trajectories. As before, this means that when we define the two time-reversed ensembles to $\Pdf$ by setting $\Pdfb^{\pm}=\Pdf\circ\T^{\pm}$, all trajectories that are observable under $\Pdf$ are also observable under $\Pdfb^{\pm}$. Now, comparing the time-forward ensemble with each of these gives rise to two different definitions of the entropy production rate, given by
\begin{equation}
\label{eq:epr-pm-defn}
\dS^{\pm}\equiv\lim_{\tau\rightarrow\infty}\frac{1}{2\tau}\log\frac{\Pdf[\rho,\bi{P}]}{\Pdfb^{\pm}[\rho,\bi{P}]}.
\end{equation}
In \sref{sec:epr-dfm} we will attempt to understand how this choice of polar signature may alter the scaling of the EPR at low noise \cite{Shankar2018,Dadhichi2018}.

Analogously with our treatment in the previous section, we observe that when $\bi{P}$ is odd under time-reversal, the functional $F$ splits into even and odd pieces $F^S$ and $F^A$ respectively. However, in our current setting this has further consequences as well, since it also implies that we should not consider $\mu$ a chemical potential like quantity either. Indeed, we see that $\mu$ splits into contributions
\begin{equation}
\mu=\mu^S+\funcdiff{F^A}{\rho}.
\end{equation}
Furthermore, the deterministic part of the current, $\bi{J}_d$, also splits into a $\bi{P}$-like odd piece under time-reversal and a $\nabla\mu^S$-like even piece. That is, we write $\bi{J}_d=\bi{J}^S_d+\bi{J}^A_d$, where we define
\begin{eqnarray}
\bi{J}^S_d=-\gamma^{-1}\nabla\mu^S, \\
\bi{J}^A_d=w\bi{P}-\gamma^{-1}\nabla\funcdiff{F^A}{\rho}.
\end{eqnarray}
Consequently, since $F$ does not remain invariant under time-reversal, and therefore neither $\mu$ nor $\bi{J}_d$ either, it could not feature in an equilibrium theory and violates TRS.

With these definitions, we may again deduce explicit integral expressions for the EPRs $\dS^{\pm}$ by using \eref{eq:diff-action}-\eref{eq:epr-pm-defn} and the definitions of $\Pdf$, $\Pdfb^{\pm}$, and we refer to \ref{app:epr-calculation} for the details. There we show that
\begin{equation}
\label{eq:eprp}
\fl\dS^+=D^{-1}\int_{\V}\rmd\bi{x}\,\left\langle \gamma w^2|K\bi{P}|^2-w\bi{P}\cdot\nabla\mu+\lambda\left[\left(\bi{P}\cdot\nabla\right)\bi{P}\right]\cdot\left(\lambda\left(\bi{P}\cdot\nabla\right)\bi{P}+\funcdiff{F}{\bi{P}}\right)\right\rangle 
\end{equation}
and
\begin{equation}
\label{eq:eprm}
\dS^-=D^{-1}\int_{\V}\rmd\bi{x}\,\left\langle\bi{J}^A_d\cdot\nabla\mu^S-\left(\lambda(\bi{P}\cdot\nabla)\bi{P}+\funcdiff{F^A}{\bi{P}}\right)\cdot\funcdiff{F^S}{\bi{P}}\right\rangle
\end{equation}
where $K$ is a matrix operator with entries $K_{\alpha\beta}=\nabla^{-1}_{\alpha}\nabla_{\beta}=\nabla^{-2}\nabla_{\alpha}\nabla_{\beta}$. Note that
in \eref{eq:eprp} we have performed an average over noise histories in order to obtain the given expression, which explains why the symmetry $\dS^+\circ\T=-\dS^+$ does not seem to hold pathwise any longer. However, it is recovered when writing the expression out as in e.g. \eref{eq:eprs-1} in \ref{app:epr-calculation}. As in \sref{sec:fluct-hydro-epr}, we proceed to analyse \eref{eq:eprp} and \eref{eq:eprm} in the low $D$ limit both analytically and numerically. We also carry over the definitions we employed there, in particular defining the exponents $\chi^{\pm}$ via the asymptotic scaling relation $\dS^{\pm}(D)\sim D^{\chi^{\pm}}$ for $D\ll1$. As we will see, we find that similar considerations to those made before carry over in a straightforward manner, allowing us to predict the correct scaling in all cases.

\subsection{Entropy production in the DFM}
\label{sec:epr-dfm}
Continuing as in \sref{sec:fluct-hydro-epr}, we look for ground-state trajectories $(\rho_0,\bi{P}_0)$ that satisfy the pathwise equilibrium condition
\begin{equation}
\dS^{\pm}[\rho_0,\bi{P}_0]=0,
\end{equation}
in order to determine when we should expect $\chi^{\pm}=-1$. Again, it is clear that these include all states that are either $\T^{\pm}$ or P$\T^{\pm}$-symmetric, and so the situation remains unchanged when choosing $\T^-$. Indeed, in this case the isotropic $\bi{P}_0=0$ state satisfies both symmetries, while the polar liquid $|\bi{P}_0|=\sqrt{\rho_0-1}$ state is P$\T^-$-symmetric only. However, we should also expect a similar situation when choosing the $\T^+$-protocol for time-reversal; since $\bi{P}_0$ does not flip sign upon time-reversal, any constant trajectory satisfies $\T^+$ alone. Thus we expect $\chi^{\pm}>-1$ for both the isotropic gas and polar liquid, meaning that there is no clear distinction between the two protocols for the homogeneous phases at ground-state level.

On the other hand, the situation changes quite drastically once the density current dynamics are tracked explicitly. In particular, in doing so, we expect that TRS violation at the $D=0$ level should become visible from a misalignment of the density current and polar density. To see this, we promote $\bi{J}$ to a dynamical variable and consider the Freidlin-Wentzell action at this level. That is, we define
\begin{equation}
\label{eq:diff-action-j}
\fl\adf^J[\rho,\bi{J},\bi{P}]=\frac{1}{4}\int_{-\tau}^{\tau}\rmd t\int_{\V}\rmd\bi{x}\left[\gamma|\bi{J}-\bi{J}_d|^2+\left|\partial_t\bi{P}+\lambda\bi{P}\cdot\nabla\bi{P}+\funcdiff{F}{\bi{P}}\right|^2\right],
\end{equation}
if $\partial_t\rho+\nabla\cdot\bi{J}=0$ and $\adf^J=\infty$ otherwise. Importantly, $\bi{J}$ now takes the role that $w\bi{P}$ had previously in \sref{sec:fluct-hydro-epr}, in the sense that it must be odd on time-reversal. Again, this is due to the constraint imposed by the continuity equation which limits the space of observable trajectories under the action \eref{eq:diff-action-j}. Time-reversal is thus generalised accordingly by setting
\begin{equation}
\T^{\pm}_J=\left\{\begin{array}{lll}
\rho(\bi{x},t) & \mapsto & \rho(\bi{x},-t), \\
\bi{J}(\bi{x},t) & \mapsto & -\bi{J}(\bi{x},-t), \\
\bi{P}(\bi{x},t) & \mapsto & \pm\bi{P}(\bi{x},-t).
\end{array}\right.
\end{equation}
Pathwise there is now a clear distinction between the two protocols $\T^{\pm}_J$. Indeed, when $(\rho_0,\bi{J}_0,\bi{P}_0)$ is a constant trajectory with both $|\bi{J}_0|>0$ and $|\bi{P}_0|>0$, the protocol $\T^+_J$ introduces a discrepancy between $\bi{J}_0$ and $\bi{P}_0$ that cannot be transformed away by parity. On the other hand, since both $\bi{J}_0$ and $\bi{P}_0$ transform the same way under $\T^-_J$, the trajectory $(\rho_0,\bi{J}_0,\bi{P}_0)$ remains invariant under P$\T^-_J$ as before. This fact is reflected by the EPR computed at the level of the action in \eref{eq:diff-action-j}. To see this, we set $\Pdf^J\propto\exp\left(-\adf^J/D\right)$ and $\Pdfb^{J,\pm}=\Pdf^J\circ\T^{\pm}_J$ as before, and define $\dS^{\pm}_J$ via the usual definition as in \eref{eq:epr-pm-defn}. In \ref{app:epr-calculation} we show that this computation results in the expression
\begin{equation}
\label{eq:eprjp}
\fl\dS^+_J=D^{-1}\int_{\V}\rmd\bi{x}\left\langle \gamma w^2|\bi{P}|^2-w\bi{P}\cdot\nabla\mu-\lambda\left[\left(\bi{P}\cdot\nabla\right)\bi{P}\right]\cdot\left(\lambda\left(\bi{P}\cdot\nabla\right)\bi{P}+\funcdiff{F}{\bi{P}}\right)\right\rangle
\end{equation}
for $\dS^+_J$, which differs from \eref{eq:eprp} by omission of the factor $K$ inside the first term, while in fact $\dS^-_J=\dS^-$ remains unchanged from \eref{eq:eprm}. 

It is instructive to investigate the difference between the two expressions \eref{eq:eprp} for $\dS^+$ and \eref{eq:eprjp} for $\dS^+_J$, and in particular to show that indeed the operator $K\neq\mathrm{Id}$. To this end, we introduce the potential $\varphi$ as the solution to the Poisson's equation
\begin{equation}
\nabla^2\varphi=\nabla\cdot\bi{P},
\end{equation}
which is unique up to an additive constant (assuming periodic boundaries on $\V$). Defining $\bi{P}_T\equiv\bi{P}-\nabla\varphi$ it follows by construction that
\begin{equation}
\label{eq:p-decomp}
\bi{P}=\nabla\varphi+\bi{P}_T,
\end{equation}
where $\bi{P}_T$ is solenoidal, i.e. $\nabla\cdot\bi{P}_T=0$. Importantly, this construction is in general not the same as the standard Helmholtz decomposition, since $\bi{P}_T$ is not necessarily the curl of a vector potential. A specific example demonstrating that these are indeed different is provided via simplest case for which $\bi{P}=\bi{P}_0$ is constant and nonzero. Indeed, in this case, periodic boundaries forces $\varphi=\mathrm{const.}$ and $\bi{P}_T=\bi{P}_0$, and the latter cannot be written as the curl of a vector field that respects the periodic boundaries. The decomposition in \eref{eq:p-decomp} is, however, orthogonal in $L^2(\V)$, meaning that
\begin{equation}
\int_{\V}\rmd\bi{x}\,|\bi{P}|^2=\int_{\V}\rmd\bi{x}\,|\nabla\varphi|^2+\int_{\V}\rmd\bi{x}\,|\bi{P}_T|^2.
\end{equation}
Thus, observing that by definition we have $K\bi{P}=\nabla\varphi$, it follows that the difference between $\dS^+$ and $\dS^+_J$ is simply
\begin{equation}
\label{eq:sj-s-diff}
\dS^+_J-\dS^+=\gamma w^2D^{-1}\int_{\V}\rmd\bi{x}\,\left\langle|\bi{P}_T|^2\right\rangle.
\end{equation}
This is consistent with our observation that the polar liquid breaks $\T^+_J$ at ground-state level as remarked above, and in particular it follows that we have
\begin{equation}
\label{eq:d0-eprp}
\lim_{D\rightarrow0}D\dS^+_J/\V=\gamma w^2|\bi{P}_0|^2
\end{equation}
for the constant homogeneous ground-states. This mechanism by which TRS is broken at $D=0$ due to $\bi{J}$ and $\bi{P}$ having different polar signatures under time-reversal has an analogue for a single active particle on the microscopic level. To see this we make the identifications $\dot{\bi{x}}_t\,\delta(\bi{x}_t-\bi{x})\rightarrow\bi{J}$ and $\hat{\bi{n}}_t\,\delta(\bi{x}_t-\bi{x})\rightarrow\bi{P}$, where $\bi{x}_t$ is the position of the active particle and $\hat{\bi{n}}_t$ denotes its polar orientation and direction of self-propulsion. Similarly to the current $\bi{J}$, the particle velocity $\dot{\bi{x}}_t$ must change sign under time-reversal. The polar orientation $\hat{\bi{n}}_t$ need not, on the other hand, and thus generally leads to a bare entropy production associated with the motility.

Continuing as in \sref{sec:fluct-hydro-epr}, we include fluctuations by performing a systematic expansion of the equations of motion and the EPRs $\dS^{\pm}$ via the integral expressions in \eref{eq:eprp} and \eref{eq:eprm} in small $D$. Again, we start by assuming that the ground-state $(\rho_0,\bi{P}_0)$ is constant and homogeneous. Thus, substituting the expansions in \eref{eq:density-expansion} and \eref{eq:polarity-expansion} into the continuity equation \eref{eq:density} with $\bi{J}=\bi{J}_d+\bxi$ we obtain to lowest nontrivial order
\begin{equation}
\label{eq:lin-density-dfm}
\partial_t\rho_1=-\nabla\cdot\left(w\bi{P}_1-\gamma^{-1}\nabla\funcdiff{F_L}{\rho}[\rho_1,\bi{P}_1]+\bxi_1\right),
\end{equation}
where now with slight abuse of notation
\begin{equation}
\label{eq:f-dfm-lin}
\fl F_L[\rho,\bi{P}]=\int_{\V}\rmd\bi{x}\left(f_L(\rho,\bi{P})+\frac{\nu_{\rho}}{2}|\nabla\rho|^2+\frac{1}{2}(\nabla_{\alpha}P_{\beta})^2+\bi{P}\cdot\nabla\Phi_L(\rho,\bi{P})\right).
\end{equation}
In addition, the local free energy $f_L$ is given by
\begin{equation}
\label{eq:f-den-dfm-lin}
f_L(\rho,\bi{P})=\frac{a_{\rho}}{2}\rho^2+\frac{a_0}{2}|\bi{P}|^2-\rho\bi{P}_0\cdot\bi{P}+(\bi{P}_0\cdot\bi{P})^2,
\end{equation}
while $\Phi_L$ remains the same as in \eref{eq:pressure-lin}. The noise term $\bxi_1$ is a mean zero Gaussian white noise process with covariance
\begin{equation}
\langle\xi_{1\alpha}(\bi{x},t)\xi_{1\beta}(\bi{x}',t')\rangle=2\gamma^{-1}\delta_{\alpha\beta}\delta(\bi{x}-\bi{x}')\delta(t-t'),
\end{equation}
so that the linearised equation \eref{eq:lin-density-dfm} is indeed independent of $D$. Note also that there is no change to the linearised equation for the polar density since this is the same for both models under consideration. 

Similarly, an expansion of the EPRs $\dS^{\pm}$ in small $D$ allows us to write
\begin{equation}
\label{eq:eprpm-expansion}
\dS^{\pm}(D)=\dS^{\pm}_{-1}D^{-1}+\dS^{\pm}_0+\dS^{\pm}_1D+\Or(D^2),
\end{equation}
where $\dS^{\pm}_k=0$ for $k<\chi^{\pm}$, and $\dS^{\pm}_{\chi^{\pm}}>0$ as before. By explicitly computing this expansion one finds that
\begin{equation}
\label{eq:hom-scaling-dfm}
\chi^{\pm}=\left\{\begin{array}{ll}
0,&\mbox{isotropic}, \\
0,&\mbox{polar liquid}.
\end{array}\right.
\end{equation}
In particular, since the linearised continuity equation \eref{eq:lin-density-dfm} of the DFM implies that the steady-state expectation $\langle\rho_1\nabla\cdot\bi{P}_1\rangle$ no longer vanishes identically as in \eref{eq:density-polarity-correlator}, we cannot any longer expect that $\chi^->0$ in the isotropic phase. Since $\chi^{\pm}=0$ in the isotropic and polar liquid phases, we classify both phases as being marginally nonequilibrium for the DFM. This means that the linearised dynamics of the DFM cannot be mapped onto an equilibrium dynamics for any choice of $\bi{P}_0$.
\begin{figure}
\centering
\begin{minipage}{0.5\textwidth}
\includegraphics[scale=0.98]{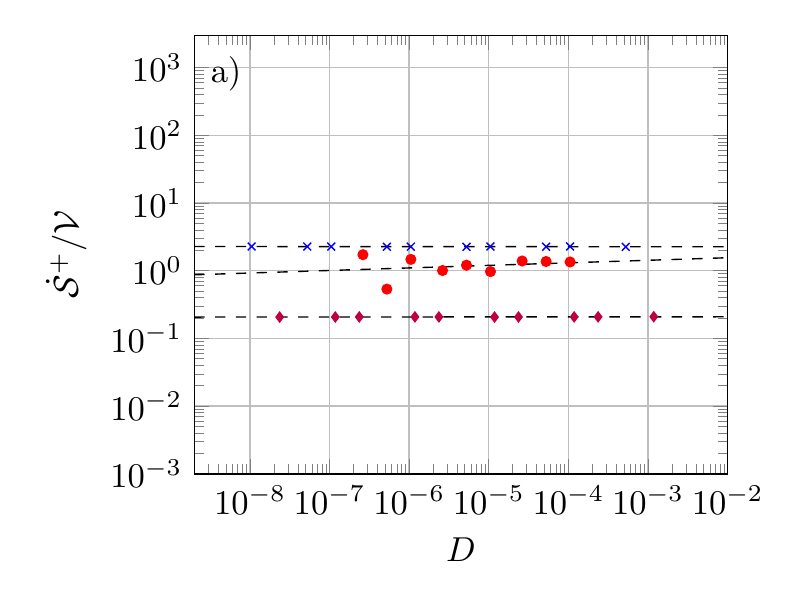}
\end{minipage}%
\begin{minipage}{0.5\textwidth}
\includegraphics[scale=0.98]{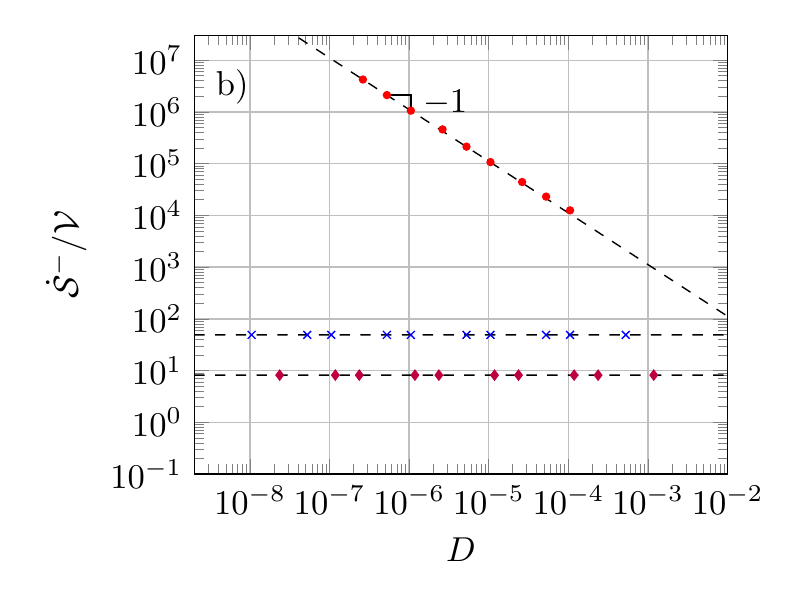}
\end{minipage}

{\scriptsize\begin{tabular}{@{}ll|ll|ll}
{\tiny\color{purple}$\fulldiamond$} & Isotropic & {\color{blue}$\times$} & Polar liquid & {\color{red}$\bullet$} & Crystal \\
\mr
 & $\rho_0=0.9$, $\lambda=1.1$, $w=1.2$ & & $\rho_0=1.33$, $\lambda=1.1$, $w=1.2$ & & $\rho_0=0.9$, $\lambda=1.1$, $w=4.95$ \\
 & $\gamma=1$, $a_{\rho}=1$, $\nu_{\rho}=1$ & & $\gamma=1$, $a_{\rho}=1$, $\nu_{\rho}=1$ & & $\gamma=0.5$, $a_{\rho}=1$, $\nu_{\rho}=1$ \\
  & Intercept: 0.2082(+), 8.1803(-) & & Intercept: 2.2743(+), 49.055(-) & & \\
  & $\dS^{\pm}_0(\Lambda)$: 0.21004(+), 8.1776(-) & & $\dS^{\pm}_0(\Lambda)$: 2.2677(+), 49.053(-) & & 
\end{tabular}}
\caption{Scaling of the EPRs $\dS^{\pm}$ (normalised by volume $\V$) with the noise coefficient $D$ in the isotropic, polar liquid, crystal phases. Dashed lines (\dashed) represent the best linear fit to the data from simulations (marked by {\footnotesize\color{purple}$\fulldiamond$}, {\color{blue}$\times$},  {\color{red}$\bullet$}), with the associated intercept (best estimate of $\lim_{D\rightarrow0}\dS(D)$ from simulation data) reported in the legend for the isotropic and polar liquid phases. The intercepts are compared with the numerically evaluated analytical results in \eref{eq:eprpm-iso-fourier} for $\dS_0(\Lambda)/\V$, where $\Lambda=2\pi N/L$, $L=14\pi$ and $N=96$.} 
\label{fig:scaling-dfm} 
\end{figure}

As in \sref{sec:fluct-hydro-epr}, we now present a more in-depth analysis of the leading order term in the expansion \eref{eq:eprpm-expansion} of $\dS^{\pm}$. Beginning with the isotropic phase where $|\bi{P}_0|=0$, we find after a Fourier transform that $\dS^{\pm}_0$ may be written in bilinear form as in \eref{eq:epr-zero-fourier}:
\begin{equation}
\label{eq:eprpm-iso-fourier}
\dS_0^{\pm}(\Lambda)/\V=\sum_{0<|\bi{q}|\leq\Lambda}\left\langle\left(\hat{\bi{u}}^{\mathrm{iso}}\right)^{\dagger}\dot{\sigma}^{\pm,\mathrm{iso}}\hat{\bi{u}}^{\mathrm{iso}}\right\rangle=\sum_{0<|\bi{q}|\leq\Lambda}\Tr\left(\dot{\sigma}^{\pm,\mathrm{iso}}\mathcal{C}^{\mathrm{iso}}\right).
\end{equation}
Here, the sum runs over wavevectors $\bi{q}$, and an explicit dependence in $\dS^{\pm}_0(\Lambda)$ on the ultraviolet cutoff $\Lambda$ is introduced in order to study the limit in which it is taken to infinity. We denote the Fourier modes of $\rho_1$ and $\bi{P}_1$ by $\hat{\rho}_1$ and $\hat{\bi{P}}_1$ respectively, and have defined
\begin{equation}
\hat{\bi{u}}^{\mathrm{iso}}=\left(\hat{\rho}_1,\hat{P}_L,\hat{P}_T\right)^T,
\end{equation}
where $\hat{P}_L=\hat{\bi{q}}\cdot\bi{P}_1$ and $\hat{P}_T=\hat{\bi{q}}_{\perp}\cdot\bi{P}_1$ are respectively the longitudinal and transverse components of $\bi{P}_1$ with respect to $\hat{\bi{q}}$, and $\hat{\bi{q}}_{\perp}$ is perpendicular to $\bi{q}$ and of unit length. The equal-time steady-state correlation matrix $\mathcal{C}^{\mathrm{iso}}\equiv(\mathcal{C}^{\mathrm{iso}}_{ij})$ is then defined by
\begin{equation}
\label{eq:corr-mat-iso}
\mathcal{C}^{\mathrm{iso}}_{ij}(q)\delta_{\bi{q},\bi{q}'}=\left\langle\hat{u}^{\mathrm{iso}}_i(\bi{q},t)\left(\hat{u}^{\mathrm{iso}}_j(\bi{q}',t)\right)^*\right\rangle.
\end{equation}
In addition, the Hermitian matrices $\dot{\sigma}^{\pm,\mathrm{iso}}$ in \eref{eq:eprpm-iso-fourier} are given by
\begin{equation}
\label{eq:sigmap-iso}
\sigma^{+,\mathrm{iso}}=\frac{1}{2}\left(\begin{array}{ccc}
0 & iwq\Gamma_{\rho} & 0 \\
-iwq\Gamma_{\rho} & 2\gamma w\tilde{w} & 0 \\
0 & 0 & 0 
\end{array}\right),
\end{equation}
and
\begin{equation}
\label{eq:sigmam-iso}
\sigma^{-,\mathrm{iso}}=\frac{1}{2}\left(\begin{array}{ccc}
0 & iq\left(w_1\Gamma-\tilde{w}\Gamma_{\rho}\right) & 0 \\
-iq\left(w_1\Gamma-\tilde{w}\Gamma_{\rho}\right) & 0 & 0 \\
0 & 0 & 0 
\end{array}\right),
\end{equation}
where we have defined $\tilde{w}=w(1-w_1q^2/\gamma w)$ as well as damping coefficients $\Gamma=1-\rho_0+q^2$ and $\Gamma_{\rho}=a_{\rho}+\nu_{\rho}q^2$. In fact, we may explicitly compute $\mathcal{C}^{\mathrm{iso}}$ from the linearised dynamics and we refer to \ref{app:lin-analysis} for the details. By substituting the result of this calculation back into \eref{eq:eprpm-iso-fourier}, we obtain
\begin{equation}
\label{eq:eprp-zero-fourier}
\dS_0^+(\Lambda)=\sum_{0<|\bi{q}|\leq\Lambda}\frac{\gamma w^2}{\Gamma+\gamma^{-1}q^2\Gamma_{\rho}},
\end{equation}
and
\begin{equation}
\label{eq:eprm-zero-fourier}
\dS_0^-(\Lambda)=\sum_{|\bi{q}|\leq\Lambda}\frac{\gamma^{-1}q^2(w_1\Gamma-\tilde{w}\Gamma_{\rho})^2}{(\Gamma+\gamma^{-1}q^2\Gamma_{\rho})(\tilde{w}w_1+\gamma^{-1}\Gamma_{\rho}\Gamma)}.
\end{equation}
Interestingly, we see from direct power counting that $\dS^+_0(\Lambda)$ converges as $\Lambda\rightarrow\infty$, while $\dS^-_0(\Lambda)\sim\log\Lambda$. In fact, expressions \eref{eq:eprp-zero-fourier} and \eref{eq:eprm-zero-fourier} generalise trivially to dimensions $d\neq2$, meaning that
\begin{equation}
\dS^+_0(\Lambda)\sim\left\{\begin{array}{ll}
1, & d<4 \\
\Lambda^{d-4}, &d\geq 4
\end{array}\right.
\end{equation}
and
\begin{equation}
\dS^-_0(\Lambda)\sim\left\{\begin{array}{ll}
1, & d<2 \\
\Lambda^{d-2}, &d\geq 2
\end{array}\right.,
\end{equation}
where we denote by $\Lambda^0$ a logarithmic divergence.
 
We may perform an identical procedure in the polar liquid case, although we leave the details of this calculation in \ref{app:epr-calculation} to simplify the presentation. We note, however, that in the polar liquid phase the scaling of $\dS^{\pm}_0(\Lambda)$ with the ultraviolet cutoff $\Lambda$ changes. For our present case where $d=2$, we find that $\dS^+_0\sim\Lambda^2$ while $\dS^-_0\sim\Lambda^4$. 

Similarly to our treatment in \sref{sec:fluct-hydro-epr}, we find good agreement between predictions and the results from simulations. In \fref{fig:scaling-dfm} we demonstrate this comparison for both of the homogeneous phases. Furthermore, all considerations extend straightforwardly to the level where we explicitly track the current $\bi{J}$; the scaling exponent $\chi^+_J$ of the EPR $\dS^+_J$ is at this level given by
\begin{equation}
\label{eq:hom-scaling-j}
\chi^+_J=\left\{\begin{array}{ll}
0,&\mbox{isotropic} \\
-1,&\mbox{polar liquid}
\end{array}\right.
\end{equation}
while $\chi^-_J=\chi^-$. Note, however, that in the isotropic phase the coefficient of the leading order term of $\dS^+_J$ changes by virtue of the equation \eref{eq:sj-s-diff}.

Finally, we consider inhomogeneous ground-states $(\rho_0,\bi{P}_0)$ (or $(\rho_0,\bi{J}_0,\bi{P}_0)$ at the level of $\bi{J}$), specifically the nonlinear polar cluster, microphase-separated and polar crystal states. Following the same reasoning as in \sref{sec:mps-analysis}, we conclude that $\chi^{\pm}_J=\chi^{\pm}=-1$ for both the banded profiles and polar clusters. Indeed, these profiles still break both $\T^{\pm}$ and P$\T^{\pm}$ as before and are therefore truly nonequilibrium at small noise. On the other hand, for the crystal state we find that $\chi^+=0$, while $\chi^-=-1$, as shown in \fref{fig:scaling-dfm}. We explain this by the observation that at $D=0$ the ground-state solution is stationary, i.e. it has both $\partial_t\rho_0=0$ and $\partial_t\bi{P}_0=0$. In particular, since $(\rho_0,\bi{P}_0)$ is independent of time, it is in fact also invariant under $\T^+$. This may also be confirmed by inspection of e.g. \eref{eq:eprs-1}, where it is apparent that the stationary condition implies we must have $\chi^+>-1$. Similarly, at the level of the current $\bi{J}$ we also find that $\chi_J^+=0$ for the crystal state. To see why this should be true, note that also $|\bi{J}_0|=0$ for the polar crystal at ground-state level, meaning that there is no difference between $(\rho_0,\bi{J}_0,\bi{P}_0)$ and its time-reversal under $\T_J^+$ for this phase. Coincidentally, this also shows that inhomogeneity is not necessarily sufficient alone to make the system truly non-equilibrium. On the other hand, the polar crystal state is clearly not invariant under $\T^-$ or P$\T^-$ (nor $\T^-_J$, P$\T^-_J$), and so we conclude that $\chi^-_J=\chi^-=-1$.

\section{Conclusion}
\label{sec:conclusion}
In this paper, we have studied the entropy production rate in two related models of dry polar flocks, namely the Hydrodynamic Vicsek Model and the Diffusive Flocking Model \cite{Solon2015a,Marchetti2013,Dadhichi2018}. Our main results relate to the observation that the scaling of the EPR with the noise parameter $D$ changes depending on the phase behaviour of the steady-state, and that the asymptotic scaling exponent takes integer values $\geq-1$. This provides us with a handle to understand how the EPR reflects TRS violation at different orders due to small fluctuations away from the mean dynamics. In particular, truly nonequilibrium behaviour is characterised by a divergent EPR in the limit $D\rightarrow0$, and is caused by ground-state dynamics that violate detailed balance pathwise. On the other hand, in the marginal and effectively equilibrium cases where the scaling exponent is $\geq0$, the ground-state dynamics is pathwise equilibrium and entropy is produced only at the level of fluctuations. In particular, when the scaling exponent is strictly positive, the dynamics at small noise can be mapped onto equilibrium dynamics.
\begin{table}
\caption{\label{tab:scaling-sum}Scaling of the EPRs $\dS\sim D^{\chi}$, $\dS^{\pm}\sim D^{\chi^{\pm}}$ and $\dS^{\pm}_J\sim D^{\chi^{\pm}_J}$ with the noise parameter $D$ when $D\ll1$ for the isotropic, polar liquid, microphase-separated, polar cluster and crystal regimes of the HVM and DFM.}
\begin{indented}
\item[]\begin{tabular}{l|c|cccc}
 & HVM & \multicolumn{4}{c}{DFM} \\
 & $\chi$ & $\chi^+$ & $\chi^+_J$ & $\chi^-$ & $\chi^-_J$ \\
 \br
Isotropic & 1 & 0 & 0 & 0 & 0 \\
Polar liquid & 0 & 0 & -1 & 0 & 0 \\
MPS & -1 & -1 & -1 & -1 & -1 \\
Polar cluster & -1 & -1 & -1 & -1 & -1 \\
Crystal & N/A & 0 & 0 & -1 & -1
\end{tabular}
\end{indented}
\end{table}

Both models studied display a transition from an isotropic gas to a polar liquid, in addition to nonlinear polar cluster and microphase-separated phases \cite{Solon2015a,Solon2015b,Gopinath2012}. In the miscibility gap where microphase-separation occurs, high-density banded profiles that break parity travel against an isotropic background. For densities beyond the polar liquid threshold $\rho_{\ell}$ in \eref{eq:rho-ell}, the phase diagram is divided into two regions with a phase boundary that can be parameterised by the self-advection parameter $\lambda$ and local swimming velocity $w$. For small $\lambda$ and large $w$ the system is a polar liquid, and as $\lambda$ is increased or $w$ decreased this state becomes unstable to perturbations leading to the formation of polar clusters. For the HVM we were able to explicitly locate both the banded-to-flock as well as the flock-to-cluster transition lines from a linear analysis. The phase diagram of the DFM, which may be considered an extension of the HVM, contains additional structure at low densities where we find a novel crystal phase in which a stationary hexagonal lattice of high-density ridges surround low density valleys. Numerical integration of the DFM also shows that the same qualitative behaviour is retained at high densities even though the density dynamics are modified by the addition of a diffusive fluctuating current. This is, however, to be expected since the diffusive dynamics are only significant to the large scale behaviour when the advective transport is comparatively small \cite{Marchetti2013}.

Generally for systems with polar symmetry such as those considered here, the EPR may be constructed in two different ways depending on how we choose to implement time-reversal at the level of fluctuating trajectories \cite{Dadhichi2018,Shankar2018}. Specifically, we may choose whether the polar density should transform as a velocity-like odd quantity or a head-to-tail-like even one under time-reversal, which changes the physics of the model. An exception to this is presented by the HVM, which is constructed in such a way that we only have one choice. Here, the continuity equation imposes a constraint on the space of observable trajectories, i.e. those that lie in the support of the transition probability density, which excludes the time-reversed trajectory of all observable trajectories when the polar density does not flip sign. On the other hand, when the density advection is driven by independent fluctuations, we may consider both time-signatures. In addition, we may promote the current to an explicit dynamical variable and thus construct an additional EPR at this level \cite{Nardini2017}. Surprisingly, for this latter construction, we find that the additional knowledge of the current changes the EPR only when the time-signature of the current differs from that of the polar density. When it does, detailed balance at ground-state level for a homogeneously polarised system is broken by a mismatch between the density current and polar density, analogously to the way in which time-reversal symmetry may be broken on the microscopic scale by ABPs or AOUPs \cite{Shankar2018}.

For both time-signatures and models considered, as well as when explicitly tracking the density current, we find that the entropy production rate diverges in the limit $D\rightarrow0$ in the microphase-separated and polar cluster regimes. We attribute this to the observation that both bands and polar clusters lead to traveling spatially asymmetrical profiles, which engenders a discrepancy between the time-forward and reversed movies that cannot be transformed away by parity. It is not sufficient that a profile is inhomogeneous alone, however, which is exemplified by the stationary crystal phase of the DFM. Indeed, in this case we find that when the polar density does not change sign on time-reversal, the dynamics are only marginally nonequilibrium. Also, interestingly the mode of TRS violation that causes the microphase-separated and polar cluster dynamics to be truly nonequilibrium at small noise has no analog on the microscopic scale for a single active particle, and should be considered an emergent collective phenomenon. 

We also find that the polar liquid phase is marginally nonequilibrium, except in the case where we explicitly track the density current and the polar density is even under time-reversal, as noted above. When the polar density transforms like a velocity, the zero-point EPR associated with ground-state flocking vanishes due to the rotational symmetry of the dynamics. Interestingly, we may conclude from this that if we were to break rotational symmetry a priori, for example by introducing an external driving field, then flocking would in fact be truly nonequilibrium when the polar density is odd under time-reversal. This is not the case for the isotropic phase, however, which is at most marginally nonequilibrium in all cases.

We have also shown that for both the isotropic and polar liquid phases, a linearisation of the dynamics at small noise allows us access the leading order coefficient of the EPR in the marginally nonequilibrium case by evaluating steady-state averages within the linear theory. In principle, this procedure can be adapted to access coefficients at arbitrary order in an expansion in small $D$, although the algebra involved becomes exceedingly complex at higher orders. Moreover, we find that our analytical predictions agree well with simulations, confirming that the procedure is well suited to analyse the EPR at small noise. In \tref{tab:scaling-sum} we summarise the scaling of the EPRs $\dS$ of the HVM in addition to $\dS^{\pm}$ and $\dS^{\pm}_J$ of the DFM with the noise parameter $D$ for the various phases investigated.

\section*{Acknowledgements}
\label{sec:acknowledgements}
We thank Robert L. Jack, Ronojoy Adhikari, Yongjoo Baek and Elsen Tjhung for useful discussions. \O B thanks the Aker Scholarship and Cambridge Trust for a PhD studentship. \'EF acknowledges support from an ATTRACT Investigator Grant of the Luxembourg National Research Fund, an Oppenheimer Research Fellowship from the University of Cambridge, and a Junior Research Fellowship from St Catharines College. This work was funded in part by the European Research Council under the EU's Horizon 2020 Programme, Grant number 740269. MEC is funded by the Royal Society.

\appendix
\section{Linear analysis}
\label{app:lin-analysis}
In this appendix we derive and summarize the results employed in the main text from the linear theory of the HVM as well as the DFM. Specifically, we derive the linear stability conditions cited in \eref{eq:rho-ell}, \eref{eq:pc-boundary} and \eref{eq:c-stability}, in addition to expressions for the correlators used when calculating the coefficient at $\Or(D^0)$ of the entropy production rates $\dS$ and $\dS^{\pm}$. Many results that are similar to those presented here may be found elsewhere in the literature (see e.g. \cite{Toner1998,Marchetti2013}), and we therefore reiterate them here only to make the main content sufficiently self-contained.

We begin by assuming that an expansion of the fields $\rho$ and $\bi{P}$ in small $D$ as in \eref{eq:density-expansion}, \eref{eq:polarity-expansion} is valid, and that the ground-state trajectory $(\rho_0,\bi{P}_0)$ is constant and homogeneous. Substituting this into the dynamics in \eref{eq:density}-\eref{eq:polarity} we obtain a hierarchy of equations by equating terms at $\Or(D^{\alpha})$, $\alpha=0,\frac{1}{2},1,\ldots$. Since the continuity equation is linear, we obtain the trivial hierarchy
\begin{equation}
\label{eq:density-hierarchy}
\partial_t\rho_n=-w\nabla\cdot\bi{P}_n,\qquad n\geq0
\end{equation}
for the density coefficients $\rho_n$. The equation for $\bi{P}$ requires more work, however. At $\Or(D^0)$ we find that
\begin{equation}
\label{eq:zeroth-order}
a_0\bi{P}_0=0
\end{equation}
where $a_0=1-\rho_0+|\bi{P}_0|^2$. Solving this equation for the polar density gives the isotropic and polar liquid solutions $\bi{P}_0=0$ and $|\bi{P}_0|^2=\rho_0-1$ respectively. At higher orders, we find that
\begin{equation}
\label{eq:polarity-hierarchy}
\fl\partial_t\bi{P}_n+\lambda\bi{P}_0\cdot\nabla\bi{P}_n=-\funcdiff{F_L}{\bi{P}}[\rho_n,\bi{P}_n]+\bDelta^P_n\left(\{\rho_k,\bi{P}_k,\nabla\rho_k,\nabla\bi{P}_k,\ldots\}_{k<n}\right),\quad n\geq1.
\end{equation}
Here, $F_L$ is the quadratic functional defined in \eref{eq:f-lin}. The driving term $\bDelta^P_n$ at each order $n\geq2$ must be derived explicitly for each case, although it depends only on the fields $\rho_k$, $\bi{P}_k$ (and their spatial derivatives) for $k<n$. For $n=1$, $\bDelta^P_1=\bfeta_1$ is a mean zero Gaussian white noise process with covariance
\begin{equation}
\langle\eta_{1\alpha}(\bi{x},t)\eta_{1\beta}(\bi{x}',t')\rangle=2\delta_{\alpha\beta}\delta(\bi{x}-\bi{x}')\delta(t-t'),
\end{equation}
and in particular does not depend on $D$. At first nontrivial order, i.e. $n=2$, the driving term is given explicitly by
\begin{equation}
\label{eq:second-order-noise}
\fl\bDelta^P_2=-\lambda\bi{P}_1\cdot\nabla\bi{P}_1+(\rho_1-2\bi{P}_0\cdot\bi{P}_1)\bi{P}_1-|\bi{P}_1|^2\bi{P}_0+\frac{\kappa}{2}\nabla|\bi{P}_1|^2-\kappa\bi{P}_1\nabla\cdot\bi{P}_1.
\end{equation}
In particular, when $\bi{P}_0=0$, we know that $(\rho_1,\bi{P}_1)$ is in fact an equilibrium dynamics, albeit with Fourier modes driven by heat baths at different temperatures. The coupling of these via \eref{eq:second-order-noise} consequently drives the next order process $(\rho_2,\bi{P}_2)$ out of equilibrium. Moreover, since \eref{eq:polarity-hierarchy} is inhomogeneous and linear in $\rho_n$, $\bi{P}_n$, the system of equations \eref{eq:density-hierarchy}-\eref{eq:polarity-hierarchy} may in principle be solved recursively to arbitrary order. Thus, we may think of the higher order driving terms $\bDelta^P_n$ in a similar vein. Despite this, our analysis here will be restricted to $n\leq1$.

The situation is quite similar for the DFM, although the continuity equation is no longer linear and the hierarchy in \eref{eq:density-hierarchy} changes accordingly. Specifically, we find that for the DFM
\begin{equation}
\label{eq:density-hierarchy-dfm}
\partial_t\rho_n=-\nabla\cdot\left(w\bi{P}_n-\gamma^{-1}\nabla\funcdiff{F_L}{\rho}[\rho_n,\bi{P}_n]\right)+\Delta^{\rho}_n,\qquad n\geq1,
\end{equation}
where $F_L$ is now given in \eref{eq:f-dfm-lin}. Note that even though $F_L$ is modified slightly for the DFM, the hierarchy of equations in \eref{eq:polarity-hierarchy} is in fact unchanged since $\delta F_L/\delta\bi{P}$ remains the same. Again, the driving term $\Delta^{\rho}_n\equiv\Delta^{\rho}_n(\{\rho_k,\bi{P}_k,\ldots\}_{k<n})$ is in general a nonlinear function of $\rho_k$, $\bi{P}_k$ and their gradients for $k<n$. For $n=1$, $\Delta^{\rho}_1=-\nabla\cdot\bxi_1$, where $\bxi_1$ is a mean zero Gaussian white noise process with covariance
\begin{equation}
\langle\xi_{1\alpha}(\bi{x},t)\xi_{1\beta}(\bi{x}',t')\rangle=2\gamma^{-1}\delta_{\alpha\beta}\delta(\bi{x}-\bi{x}')\delta(t-t').
\end{equation}

In the following, we wish to examine both the linear stability of the constant homogeneous ground-states as well as to deduce expressions for the correlators in the linearised theory. We will perform this calculation in two parts: first we look at the isotropic state with $\bi{P}_0=0$, and subsequently the polar liquid with $|\bi{P}_0|^2=\rho_0-1$. In both cases, we consider the more general DFM and observe that predictions for the HVM may be made by considering the limit $\gamma\rightarrow\infty$.

\subsection{Isotropic ground-state}
Beginning with the isotropic state, we transform the linearised equations \eref{eq:polarity-hierarchy} and \eref{eq:density-hierarchy-dfm} for the DFM to Fourier space when $n=1$. The resulting equations are most conveniently expressed in matrix form as
\begin{equation}
\label{eq:linear-iso}
\frac{\rmd}{\rmd t}\left(\begin{array}{c}
\hat{\rho}_1 \\
\hat{P}_L \\
\hat{P}_T
\end{array}\right)=-\underbrace{\left(\begin{array}{ccc}
q^2\Gamma_{\rho}/\gamma & i\tilde{w}q & 0 \\
iw_1q & \Gamma & 0 \\
0 & 0 & \Gamma
\end{array}\right)}_{\L^{\mathrm{iso}}(q)}\left(\begin{array}{c}
\hat{\rho}_1 \\
\hat{P}_L \\
\hat{P}_T
\end{array}\right)+
\left(\begin{array}{c}
-iq\hat{\xi}_L \\
\hat{\eta}_L \\
\hat{\eta}_T
\end{array}\right).
\end{equation}
Here, $\hat{\xi}_L=\hat{\bi{q}}\cdot\hat{\bxi}_1$ is the longitudinal component of the Fourier coefficient $\hat{\bxi}_1$, while $\hat{\eta}_L=\hat{\bi{q}}\cdot\hat{\bfeta}_1$, $\hat{\eta}_T=\hat{\bi{q}}_{\perp}\cdot\hat{\bfeta}_1$ are the longitudinal and transverse components of $\hat{\bfeta}_1$ respectively. Because of the convention \eref{eq:fourier-conv} we have chosen for Fourier transforms, noise correlations in Fourier space contain an extra factor of $\V^{-1}$, i.e.
\begin{eqnarray}
\langle\hat{\xi}_L(\bi{q},t)\hat{\xi}^*_L(\bi{q}',t')\rangle=2(\gamma\V)^{-1}\delta_{\bi{q},\bi{q}'}\delta(t-t'), \\
\langle\hat{\eta}_L(\bi{q},t)\hat{\eta}^*_L(\bi{q}',t')\rangle=\langle\hat{\eta}_T(\bi{q},t)\hat{\eta}^*_T(\bi{q}',t')\rangle=2\V^{-1}\delta_{\bi{q},\bi{q}'}\delta(t-t'),
\end{eqnarray}
while $\langle\hat{\eta}_L(\bi{q},t)\hat{\eta}^*_T(\bi{q}',t')\rangle=\langle\hat{\eta}_L(\bi{q},t)\hat{\xi}^*_L(\bi{q}',t')\rangle=\langle\hat{\eta}_T(\bi{q},t)\hat{\xi}^*_L(\bi{q}',t')\rangle=0$. Note also that the damping coefficient $\Gamma(q)=1-\rho_0+q^2>0$ ensures that transverse fluctuations $\hat{\bi{P}}_T$ decay on non-hydrodynamic timescales when $\rho_0<1$. 

Observe that solutions to \eref{eq:linear-iso} are stable and decay at an exponential rate when the eigenvalues $\sigma_{\pm}$, $\sigma_T$ of the linear operator $\L^{\mathrm{iso}}$ have positive real parts. These are straightforwardly found from the characteristic equation of $\L^{\mathrm{iso}}$, and are given by
\begin{eqnarray}
\label{eq:sigmapm}
\sigma_{\pm}=\frac{1}{2}\left(\Gamma+\gamma^{-1}q^2\Gamma_{\rho}\pm\sqrt{(\Gamma+\gamma^{-1}q^2\Gamma_{\rho})^2-4q^2(\tilde{w}w_1+\gamma^{-1}\Gamma_{\rho}\Gamma)}\right), \\
\sigma_T=\Gamma.
\end{eqnarray}
\Eref{eq:sigmapm} for $\sigma_{\pm}$ may be unraveled by first understanding its behaviour at $\gamma=\infty$. In this limit, we find that
\begin{equation}
\sigma_{\pm}(\gamma\rightarrow\infty)=\frac{1}{2}\left(\Gamma\pm\sqrt{\Gamma^2-4q^2ww_1}\right).
\end{equation}
Since $w,w_1>0$ it follows that $\Re\,\sigma_{\pm}(\gamma\rightarrow\infty)>0$ if and only if $\sigma_T=\Gamma>0$. Thus, at $\gamma=\infty$, the isotropic state is linearly stable when $\rho_0<1$, while it is unstable otherwise.

From this, it is fairly straightforward to see that a similar condition holds for finite $\gamma$. Specifically, $\Re\,\sigma_{\pm}>0$ only when $\rho_0<1$. However, we see from \eref{eq:sigmapm} that in this case stability also requires that
\begin{equation}
\tilde{w}w_1+\gamma^{-1}\Gamma_{\rho}\Gamma>0.
\end{equation}
After some algebra, one finds that this latter condition holds for all $q$ if and only if
\begin{equation}
\label{eq:crystal-condition}
w_1^2<a_{\rho}+\nu_{\rho}(1-\rho_0)+4\gamma\nu_{\rho}+2\sqrt{\nu_{\rho}(1-\rho_0+2\gamma)(a_{\rho}+2\gamma\nu_{\rho})}.
\end{equation}
Thus, the phase diagram of the DFM in the region where $\rho_0<1$ as predicted by the linear theory is no longer trivial. Indeed, when condition \eref{eq:crystal-condition} is broken, there is a finite range of wave numbers $q\in[q_-,q_+]$ for which the corresponding modes $\hat{\rho}_1$, $\hat{P}_L$ grow in time, where
\begin{eqnarray}
\fl 2\nu_{\rho}q_{\pm}^2=w_1^2-\alpha_{\rho}-\nu_{\rho}(1-\rho_0) \\
\pm\sqrt{\left(w_1^2-\alpha_{\rho}-\nu_{\rho}(1-\rho_0)\right)^2-4\nu_{\rho}\left(\gamma ww_1+\alpha_{\rho}(1-\rho_0)\right)}.
\end{eqnarray}
From simulations, we find that this instability leads to the polar crystal phase reported in \sref{sec:density-flucts}.

We are also interested in calculating the equal-time correlation functions of the linear theory in order to make analytical predictions about the EPR. Since the dynamics is linear, the correlation matrix $\mathcal{C}^{\mathrm{iso}}$ in \eref{eq:corr-mat-iso} solves the algebraic Riccati equation
\begin{equation}
\label{eq:riccati}
\L^{\mathrm{iso}}\mathcal{C}^{\mathrm{iso}}+\mathcal{C}^{\mathrm{iso}}\left(\L^{\mathrm{iso}}\right)^{\dagger}=2\D,
\end{equation}
where we have defined the diffusion matrix $\D$ by
\begin{equation}
\D=\frac{1}{\V}\left(\begin{array}{ccc}
q^2/\gamma & 0 & 0 \\
0 & 1 & 0 \\
0 & 0 & 1
\end{array}\right).
\end{equation}
To derive this, one simply needs to apply the chain rule to the left-hand side of
\begin{equation}
\left\langle\frac{\rmd}{\rmd t}\left(\hat{u}^{\mathrm{iso}}_i(\bi{q},t)\left(\hat{u}^{\mathrm{iso}}_j(\bi{q}',t)\right)^*\right)\right\rangle=0,
\end{equation}
where $\hat{\bi{u}}^{\mathrm{iso}}$ is defined in \eref{eq:u-vec}. It is well known that the solution to the Riccati equation \eref{eq:riccati} may be expressed in integral form. However, we find that for our present purposes it is less cumbersome to tackle it straight on. First we observe that the linear system in \eref{eq:linear-iso} reduces to the two-dimensional coupled dynamics of $(\hat{\rho}_1,\hat{P}_L)$, in addition to the one-dimensional dynamics of $\hat{P}_T$. Thus, clearly, we may treat these separately. Beginning with the former, the Riccati equation \eref{eq:riccati} may be re-expressed as a four-by-four linear system, specifically
\begin{equation}
\fl\underbrace{\left(\begin{array}{cccc}
2q^2\Gamma_{\rho}/\gamma & -i\tilde{w}q & i\tilde{w}q & 0 \\
-iw_1q & \gamma^{-1}q^2\Gamma_{\rho}+\Gamma & 0 & i\tilde{w}q \\
iw_1q & 0 & \gamma^{-1}q^2\Gamma_{\rho}+\Gamma & -i\tilde{w}q \\
0 & iw_1q & -iw_1q & 2\Gamma
\end{array}\right)}_{\mathcal{R}^{\mathrm{iso}}(q)}\left(\begin{array}{c}
\mathcal{C}^{\mathrm{iso}}_{11} \\
\mathcal{C}^{\mathrm{iso}}_{12} \\
\mathcal{C}^{\mathrm{iso}}_{21} \\
\mathcal{C}^{\mathrm{iso}}_{22} \\
\end{array}\right) = \frac{2}{\V}\left(\begin{array}{c}
q^2/\gamma \\
0 \\
0 \\
1 \\
\end{array}\right).
\end{equation}
To find the correlators of the linear theory we therefore simply invert the matrix $\mathcal{R}^{\mathrm{iso}}$, and one may check that the solution is given by
\begin{eqnarray}
\label{eq:c11}
\mathcal{C}^{\mathrm{iso}}_{11}=\frac{1}{\V}\frac{\tilde{w}w+\gamma^{-1}\Gamma(\Gamma+\gamma^{-1}q^2\Gamma_{\rho})}{(\Gamma+\gamma^{-1}q^2\Gamma_{\rho})(\tilde{w}w_1+\gamma^{-1}\Gamma_{\rho}\Gamma)}, \\
\mathcal{C}^{\mathrm{iso}}_{12}=(C^{\mathrm{iso}}_{21})^*=\frac{1}{\V}\frac{i\gamma^{-1}q(w_1\Gamma-\tilde{w}\Gamma_{\rho})}{(\Gamma+\gamma^{-1}q^2\Gamma_{\rho})(\tilde{w}w_1+\gamma^{-1}\Gamma_{\rho}\Gamma)}, \\
\mathcal{C}^{\mathrm{iso}}_{22}=\frac{1}{\V}\frac{ww_1+\gamma^{-1}\Gamma_{\rho}(\Gamma+\gamma^{-1}q^2\Gamma_{\rho})}{(\Gamma+\gamma^{-1}q^2\Gamma_{\rho})(\tilde{w}w_1+\gamma^{-1}\Gamma_{\rho}\Gamma)}, \\
\end{eqnarray}
The final non-trivial component of $\mathcal{C}^{\mathrm{iso}}$ is found straighforwardly from the dynamics of $\hat{P}_T$, and is given by
\begin{equation}
\label{eq:c33}
\mathcal{C}^{\mathrm{iso}}_{33} = \frac{1}{\V\Gamma}.
\end{equation}

To recover the correlators in the linearised HVM, we simply take the limit $\gamma\rightarrow\infty$ in \eref{eq:c11}-\eref{eq:c33}. We obtain that
\begin{equation}
\mathcal{C}^{\mathrm{iso}}(q,\gamma\rightarrow\infty)=\frac{1}{\V\Gamma}\left(\begin{array}{ccc}
w/w_1 & 0 & 0 \\
0 & 1 & 0 \\
0 & 0 & 1
\end{array}\right).
\end{equation}
In particular, this result verifies that $\langle\hat{\rho}_1\hat{P}_L^*\rangle=0$ as advertised in \eref{eq:density-polarity-correlator}.

\subsection{Polar liquid ground-state}
The linearised dynamics about the homogenenous polar liquid phase may be treated similarly to the isotropic case considered above. Transforming \eref{eq:polarity-hierarchy} and \eref{eq:density-hierarchy-dfm} for $n=1$ and $|\bi{P}_0|^2=\rho_0-1$ to Fourier space we find that
\begin{equation}
\label{eq:linear-pl}
\frac{\rmd}{\rmd t}\left(\begin{array}{c}
\hat{\rho}_1 \\
\hat{P}_{\parallel} \\
\hat{P}_{\perp}
\end{array}\right)=-\L^{\mathrm{pl}}(\bi{q})\left(\begin{array}{c}
\hat{\rho}_1 \\
\hat{P}_{\parallel} \\
\hat{P}_{\perp}
\end{array}\right)+
\left(\begin{array}{c}
-iq\hat{\xi}_L \\
\hat{\eta}_{\parallel} \\
\hat{\eta}_{\perp}
\end{array}\right),
\end{equation}
where we have defined
\begin{equation}
\label{eq:lmat-pl}
\L^{\mathrm{pl}}(\bi{q})=\left(\begin{array}{ccc}
q^2\Gamma_{\rho}/\gamma & i\tilde{w}q_{\parallel}-q^2P_0/\gamma & i\tilde{w}q_{\perp} \\
iw_1q_{\parallel}-P_0 & \Gamma_{\parallel}+i\lambda P_0q_{\parallel} & i\kappa P_0q_{\perp} \\
iw_1q_{\perp} & -i\kappa P_0q_{\perp} & \Gamma_{\perp}+i\lambda P_0q_{\parallel}
\end{array}\right),
\end{equation}
in addition to new damping coefficients
\begin{eqnarray}
\Gamma_{\parallel}=2(\rho_0-1)+q^2, \\
\Gamma_{\perp}=q^2.
\end{eqnarray}
As before, $\hat{\xi}_L=\hat{\bi{q}}\cdot\hat{\bxi}$ is the longitudinal component of the noise $\hat{\bxi}$, while $\eta_{\parallel}$ and $\eta_{\perp}$ are respectively the parallel and perpendicular components of $\bfeta_1$ with respect to $\bi{P}_0$. Furthermore, all noise terms $\hat{\xi}_L$, $\hat{\eta}_{\parallel}$ and $\hat{\eta}_{\perp}$ are independent and of the same covariance as in the isotropic case.

Rather than solving the full cubic polynomial characteristic equation of $\L^{\mathrm{pl}}$, we will study its roots only for wave vectors that are parallel or perpendicular to $\bi{P}_0$, i.e. those for which either $q_{\perp}=0$ or $q_{\parallel}=0$ respectively. For both of these cases, we further assume that the limit $\gamma\rightarrow\infty$ has been taken. In other words, we will study the roots of the two polynomial equations 
\begin{eqnarray}
\label{eq:char-eq-parallel}
\det\left(\sigma_{\parallel}-\L^{\mathrm{pl}}(q_{\parallel},0,\gamma\rightarrow\infty)\right)=0, \\
\label{eq:char-eq-perp}
\det\left(\sigma_{\perp}-\L^{\mathrm{pl}}(0,q_{\perp},\gamma\rightarrow\infty)\right)=0,
\end{eqnarray}
and aim to deduce the conditions under which $\sigma_{\parallel}\equiv\sigma_{\parallel}(q_{\parallel})$ and $\sigma_{\perp}\equiv\sigma_{\perp}(q_{\perp})$ have positive real parts. 

\begin{figure}
\centering
\includegraphics[scale=0.6]{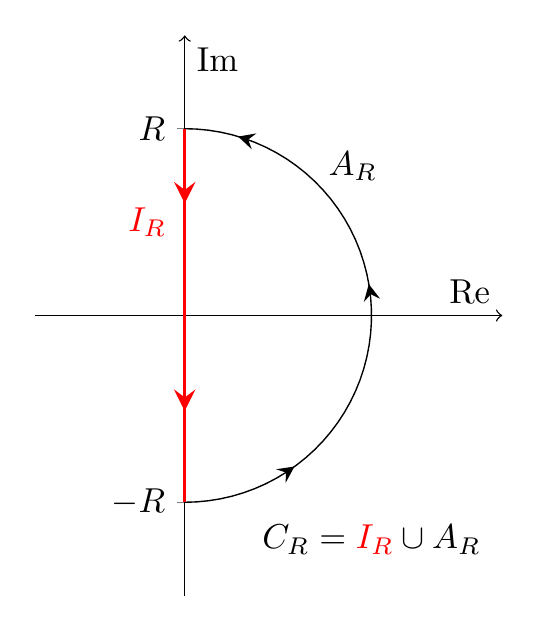}
\begin{tikzpicture}
\node[] at (0,1.9) {$\overset{g_{\parallel},g_{\perp}}{\mapsto}$};
\node[] at (0,0) {};
\end{tikzpicture}
\includegraphics[scale=0.6]{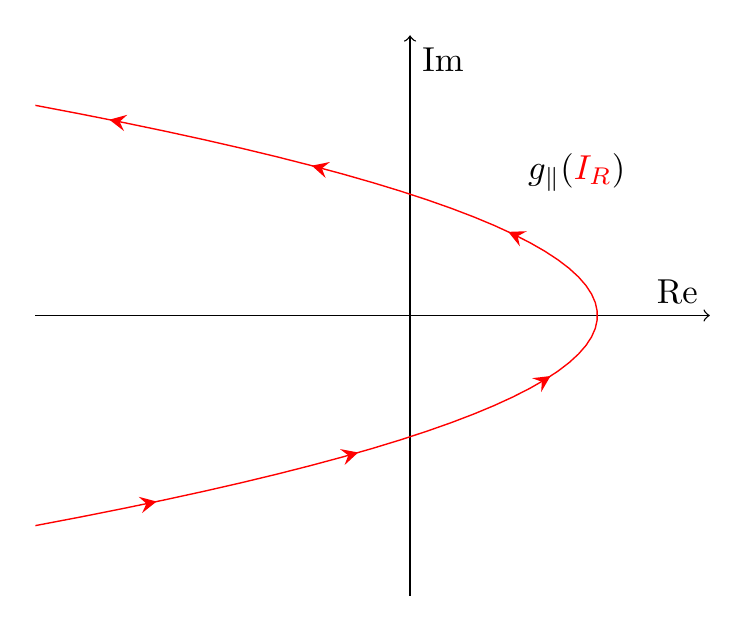}
\includegraphics[scale=0.6]{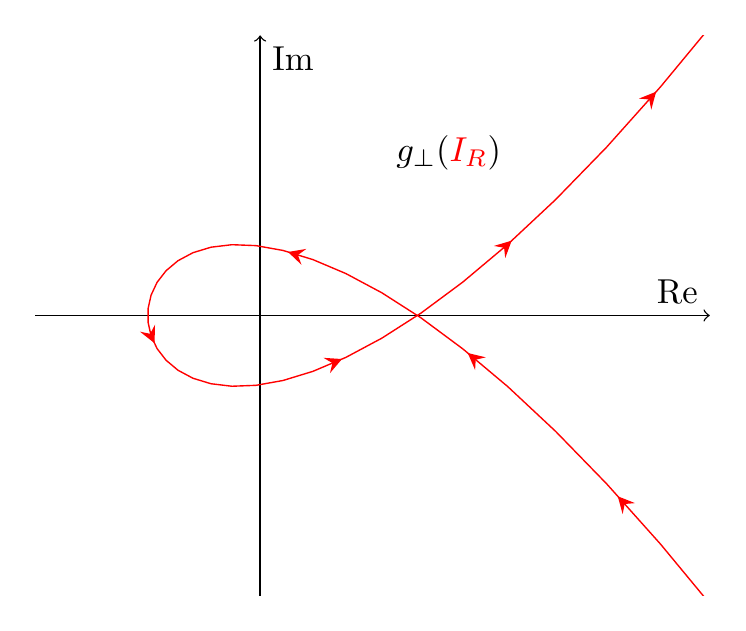}
\caption{On the semi-circle arc $A_R$, the polynomials $g_{\parallel}\sim R^2e^{2i\theta}$ and $g_{\perp}\sim R^3e^{3i\theta}$. Hence, in the limit $R\rightarrow\infty$, the number of zeros of $g_{\parallel}$ and $g_{\perp}$ in $C_R$ is fully determined by the winding number of $g_{\parallel}(I_R)$ and $g_{\perp}(I_R)$ (respectively) about the origin.}   
\label{fig:stability} 
\end{figure}

Starting with \eref{eq:char-eq-parallel}, it is straightforward to show that $\sigma_{\parallel}$ solves 
\begin{equation}
\left(\sigma_{\parallel}-\Gamma_{\perp}-i\lambda P_0q_{\parallel}\right)g_{\parallel}(\sigma_{\parallel})=0,
\end{equation}
where the polynomial $g_{\parallel}$ is given by
\begin{equation}
\label{eq:gpll}
g_{\parallel}(\sigma)=\sigma^2-(\Gamma_{\parallel}+i\lambda P_0q_{\parallel})\sigma+ww_1q_{\parallel}^2+iw P_0q_{\parallel}.
\end{equation}
In order to determine the number of roots of $g_{\parallel}$ in the halfplane $\{\Re\,\sigma>0\}$ we apply the argument principle (for an introduction to this method, see e.g. \cite{Greene2006} chapter 5, theorem 5.1.4, or \cite{Krantz2008} chapter 5). It states that the number $Z_R$ of zeros of $g_{\parallel}$ inside the semi-circle contour $C_R=I_R\cup A_R$, where
\begin{equation}
I_R=[-iR,iR],
\end{equation}
\begin{equation}
A_R=\{Re^{i\theta}:\theta\in[-\pi/2,\pi/2]\},
\end{equation}
is given by the change in the argument of $g_{\parallel}$ as we trace $C_R$ counterclockwise, i.e.
\begin{equation}
\label{eq:zr-defn}
Z_R=\frac{1}{2\pi i}\oint_{C_R}\rmd\sigma\,\frac{g_{\parallel}'(\sigma)}{g_{\parallel}(\sigma)}=\frac{\Delta\arg_{C_R}(g_{\parallel})}{2\pi i},
\end{equation}
which is illustrated in \fref{fig:stability}. Thus, by determining $Z_R$ in the limit $R\rightarrow\infty$, we may in particular deduce the number of roots of $g_{\parallel}$ with positive real part. Moreover, the choice of contour is quite convenient when dealing with polynomials such as \eref{eq:gpll}, since we know that on the arc $A_R$, $g_{\parallel}(Re^{i\theta})\sim R^2e^{2i\theta}+\Or(R)$. In other words,
\begin{eqnarray}
\Delta\arg_{A_R}(g_{\parallel})&=\log g_{\parallel}(Re^{i\pi/2})-\log g_{\parallel}(Re^{-i\pi/2}) \nonumber \\
\label{eq:delta-ar}
&=2\pi i+\Or(R^{-1}).
\end{eqnarray}
To compute $Z_R$ in the limit $R\rightarrow\infty$, we are therefore left with having to find the change in the argument of $g_{\parallel}$ along $I_R$. 

Stability clearly requires that $Z_R\rightarrow2$ as $R\rightarrow\infty$, or combining \eref{eq:zr-defn} and \eref{eq:delta-ar},
\begin{equation}
\lim_{R\rightarrow\infty}\Delta\arg_{I_R}(g_{\parallel})=2\pi i.
\end{equation} 
This occurs if and only if the image $g_{\parallel}(I_R)$ wraps around the origin once, as illustrated in \fref{fig:stability}, or equivalently that the winding number of $g_{\parallel}(I_R)$ about the origin is $1$. To investigate when this occurs, we decompose $g_{\parallel}(iy)$ where $y\in[-R,R]$ into its real and imaginary parts, thus
\begin{eqnarray}
\label{eq:gpll-real}
\Re\,g_{\parallel}(iy)=-y^2+\lambda P_0q_{\parallel} y+ww_1q_{\parallel}^2, \\
\label{eq:gpll-imag}
\Im\,g_{\parallel}(iy)=-\Gamma_{\parallel}y+wP_0q_{\parallel}.
\end{eqnarray}
Firstly, from \eref{eq:gpll-imag} we see that we must require $\Gamma_{\parallel}>0$, or the winding number of $g_{\parallel}(I_R)$ could only be $0$ or $-1$ (recall that we are tracing the line segment $I_R=[-iR,iR]$ from $iR$ to $-iR$ since $C_R$ is traced counterclockwise). This is ensured so long as $\rho_0>1$, which we assume in the following. Furthermore, from \eref{eq:gpll-real}, it follows that the quadratic $\Re\,g_{\parallel}(iy)$ has two distinct real roots for all $q_{\parallel}\neq0$, given by
\begin{equation}
y_{\pm}=\frac{q_{\parallel}}{2}\left(\lambda P_0\pm\sqrt{(\lambda P_0)^2+4ww_1}\right).
\end{equation}
Thus, we only need to require that $\Im\,g_{\parallel}(iy_+)<0$ and $\Im\,g_{\parallel}(iy_-)>0$. One may show by standard means that this occurs if and only if
\begin{equation}
\label{eq:stab-cond-flock}
\rho_0>1+\frac{1}{2}\frac{w}{\lambda+2w_1}.
\end{equation}
In conclusion therefore, it follows that all the roots of the characteristic equation of $\L^{\mathrm{pl}}$ are positive only when \eref{eq:stab-cond-flock} is satisfied.

Proceeding with \eref{eq:char-eq-perp}, i.e. the characteristic equation of $\L^{\mathrm{pl}}$ for wave vectors that are \textit{perpendicular to} $\bi{P}_0$, we apply the argument principle once more. In this case, the roots $\sigma_{\perp}$ solve the cubic polynomial equation 
\begin{equation}
g_{\perp}(\sigma_{\perp})\equiv\sigma_{\perp}^3-c_2\sigma_{\perp}^2+c_1\sigma_{\perp}-c_0=0,
\end{equation}
where the coefficients $c_2$, $c_1$ and $c_0$ are given by
\begin{eqnarray}
c_2=\Gamma_{\parallel}+\Gamma_{\perp}, \\
c_1=\Gamma_{\parallel}\Gamma_{\perp}+q_{\perp}^2(ww_1-\kappa^2P_0^2), \\
c_0=wq_{\perp}^2\left(w_1\Gamma_{\parallel}-\kappa P_0^2\right).
\end{eqnarray}
By considering the image of $C_R$ under $g_{\perp}$, we see that along the semi-circle arc $A_R$, the change in the argument of $g_{\perp}$ is $3\pi i+\Or(R^{-1})$. Thus, in this case we must require that the winding number of $g_{\perp}(I_R)$ is $\frac{3}{2}$ in order to have $Z_R\rightarrow3$. This occurs if and only if $g_{\perp}(I_R)$ wraps arounnd the origin in the way illustrated in \fref{fig:stability}. To uncover the conditions under which this occurs, we again decompose $g_{\perp}(iy)$ into its real and imaginary parts:
\begin{eqnarray}
\label{eq:gpp-real}
\Re\,g_{\perp}(iy)=c_2y^2-c_0, \\
\label{eq:gpp-imag}
\Im\,g_{\perp}(iy)=-y^3+c_1y.
\end{eqnarray}
Assuming that $\rho_0>1$, we have $c_2>0$ which is necessary in order for the winding number of $g_{\perp}(I_R)$ to be positive. Furthermore, from \eref{eq:gpp-real} we see that we must require $c_0>0$ so that the quadratic $\Re\,g_{\perp}(iy)$ has two distinct real roots. This holds if and only if 
\begin{equation}
\label{eq:stab-cond-pc}
\kappa<2w_1.
\end{equation}
Similarly, from \eref{eq:gpp-imag} we deduce that we must have $c_1>0$, or equivalently
\begin{equation}
\label{eq:extra-condition-one}
\kappa^2<2+\frac{ww_1}{\rho_0-1},
\end{equation}
so that the cubic $\Im\,g_{\perp}(iy)$ has three distinct real roots. Finally, requiring that $\Re\,g_{\perp}(\pm i\sqrt{c_1})>0$ one straightforwardly verifies that we are indeed guaranteed that $Z_R\rightarrow3$. This last condition is equivalent to having
\begin{equation}
\label{eq:extra-condition-two}
c_2c_1-c_0>0.
\end{equation}
One may show that \eref{eq:extra-condition-two} holds if and only if either
\begin{equation}
\kappa^2<3+\frac{ww_1}{2(\rho_0-1)}\qquad\mathrm{and}\qquad2P_0^2(2-\kappa^2)+w\kappa>0
\end{equation}
or
\begin{equation}
\left(2P_0^2(3-\kappa^2)+ww_1\right)^2<8P_0^2\left(2P_0^2(2-\kappa^2)+w\kappa\right)
\end{equation}
For all simulations we perform, conditions \eref{eq:extra-condition-one} and \eref{eq:extra-condition-two} are satisfied. Because of this, we will only be concerned the stability requirement in \eref{eq:stab-cond-pc}. It is also worth highlighting once more that, despite the rather involved analysis undertaken here, we have not solved the full cubic characteristic equation of $\L^{\mathrm{pl}}$ and thus have not fully identified all necessary and sufficient conditions for linear stability.
\begin{table}
\caption{\label{tab:corr-scaling}Leading order asymptotic expressions for the components $\mathcal{C}_{ij}^{\mathrm{pl}}$ of the equal-time correlation matrix $\mathcal{C}^{\mathrm{pl}}$ in the limit where $q\rightarrow\infty$, as well as when $\gamma\rightarrow\infty$ is taken first.}
\begin{indented}
\item[]\begin{tabular}{@{}l|ll}
\br
$\mathcal{C}^{\mathrm{pl}}_{ij}$ & \begin{tabular}{@{}l@{}}$\gamma\rightarrow\infty$, \\ $q\rightarrow\infty$\end{tabular} & $q\rightarrow\infty$ \\
\mr
$\mathcal{C}^{\mathrm{pl}}_{11}$ & $w/(\V w_1q^2)$ & $1/(\V \nu_{\rho} q^2)$ \\
$\mathcal{C}^{\mathrm{pl}}_{12}$ & $P_0w(\cos^2\theta+(1+\kappa w_1)\sin^2\theta)/(\V w_1q^4)$ & $iw_1\cos\theta/(\V \nu_{\rho} q^3)$ \\
$\mathcal{C}^{\mathrm{pl}}_{13}$ & $-P_0\kappa w\cos\theta\sin\theta/(\V q^4)$ & $iw_1\sin\theta/(\V\nu_{\rho} q^3)$ \\
$\mathcal{C}^{\mathrm{pl}}_{22}$ & $1/(\V q^2)$ & $1/(\V q^2)$ \\
$\mathcal{C}^{\mathrm{pl}}_{23}$ & $-iP_0\kappa\sin\theta/(\V q^3)$ & $-iP_0\kappa\sin\theta/(\V q^3)$ \\
$\mathcal{C}^{\mathrm{pl}}_{33}$ & $1/(\V q^2)$ & $1/(\V q^2)$ \\
\br 
\end{tabular}
\end{indented}
\end{table}

Lastly, we also investigate the structure of the correlators of the linearised theory about the polar liquid state. In analogy with the isotropic calculation, the matrix $\mathcal{C}^{\mathrm{pl}}$ in \eref{eq:corr-mat-pl} solves an algebraic Riccati equation as in \eref{eq:riccati}, although in this case all three modes $(\hat{\rho}_1,\hat{P}_{\parallel},\hat{P}_{\perp})$ remain coupled for general $\bi{q}$. Finding its solution can be achieved by solving the linear system
\begin{equation}
\label{eq:riccati-pl}
\mathcal{R}^{\mathrm{pl}}\bi{C}^{\mathrm{pl}}=2\bi{D},
\end{equation}
where $\bi{C}^{\mathrm{pl}}=(\mathcal{C}^{\mathrm{pl}}_{11},\mathcal{C}^{\mathrm{pl}}_{12},\ldots,\mathcal{C}^{\mathrm{pl}}_{33})^T$, $\bi{D}=(\D_{11},\D_{12},\ldots,\D_{33})^T$. We avoid explicitly writing out the nine-by-nine matrix $\mathcal{R}^{\mathrm{pl}}$ here for sake of clarity of presentation, although it may be found fairly straightforwardly from the Riccati equation. Analytical inversion of \eref{eq:riccati-pl} may be done using standard computer algebra systems that perform symbolic computations. Due to the algebraic complexity of the resulting expressions, we choose to only state the result in certain limiting cases. More specifically, we look at the asymptotic behaviour of the components $\mathcal{C}^{\mathrm{pl}}_{ij}$ in the limit $q\rightarrow\infty$ at fixed $q_{\parallel}/q_{\perp}$, as well as when $\gamma\rightarrow\infty$ is taken first. From this, we deduce in \ref{app:epr-calculation} the dependence of the EPR on the ultraviolet cutoff $\Lambda$ quoted in the main text. For the six independent components of $\mathcal{C}^{\mathrm{pl}}$ we summarize the results in \tref{tab:corr-scaling}.

\section{Structure of the EPR expansion at small noise}
\label{app:epr-half-order}
In this appendix, we aim to sketch a proof to show that we do not expect to see terms of order $D^{n/2}$ in the expansion of the EPR. We will keep our notation in this appendix fairly general, to illustrate that this is indeed something we expect generically for field theories of this type. Motivated by \eref{eq:density-hierarchy}, \eref{eq:polarity-hierarchy} and \eref{eq:density-hierarchy-dfm}, we observe that when we expand the field $\bi{u}$ in small $D$ as
\begin{equation}
\bi{u}=\sum_{n=0}^{\infty}\bi{u}_nD^{n/2},
\end{equation} 
the resulting hierarchy of equations for the $\bi{u}_n$ may in general be expressed as
\begin{equation}
\label{eq:gen-hierarchy}
\partial_t\bi{u}_n=-\L\bi{u}_n+\bDelta_n,\qquad n\geq1,
\end{equation}
while $\bi{u}_0$ is a ground-state solution to the equations of motion at $D=0$. In \eref{eq:gen-hierarchy}, $\L$ is a linear operator that depends on the $D=0$ solution, and we assume that its spectrum is strictly positive so that solutions to \eref{eq:gen-hierarchy} are stable (although this is not in general sufficient for the expansion to be valid, see e.g. \cite{Gardiner2009}). 

For the HVM and DFM, we take $\bi{u}=(\rho,\bi{P})^T$, $\bi{u}_n=(\rho_n,\bi{P}_n)^T$ and $\bDelta_n=(\Delta^{\rho}_n,\bDelta^P_n)^T$, with 
\begin{equation}
\bDelta_1=\left(-\nabla\cdot\bxi_1,\bfeta_1\right)^T.
\end{equation}
In general, when the equations of motion contain multiplicative noise, $\bDelta_1$ will also contain multiplicative factors of $\bi{u}_0$ although this does not modify our argument. Crucially, we do however assume that $\bDelta_1$ is mean-zero, Gaussian and white. The higher order terms $\bDelta_n$ can in general be expressed as functionals of $\bi{u}_k$ for $k<n$ and $\bDelta_1$, i.e.
\begin{equation}
\bDelta_n\equiv\bDelta_n[\bi{u}_0,\ldots,\bi{u}_{n-1};\bDelta_1],
\end{equation}
and in particular do not depend on $\bi{u}_n$. Moreover, each term that composes $\bDelta_n$ must \textit{preserve order}. For example, for the HVM and DFM the driving terms $\bDelta_3^P$ and $\bDelta_4^P$ can be written as linear combinations
\begin{eqnarray}
\bDelta^P_3=c_1\left((\nabla\cdot\bi{P}_2)\nabla\rho_1\right)+c_2\nabla\left(\rho_1\nabla^2\rho_2\right)+c_3\left(\bi{P}_1\nabla^2|\bi{P}_1|^2\right)+\ldots \\
\bDelta^P_4=d_1\left(\rho_3\nabla^2\bi{P}_1\right)+d_2\nabla(\nabla\cdot\bi{P}_2)^2+\ldots
\end{eqnarray}
where $c_1,c_2,c_3,\ldots$ and $d_1,d_2,\ldots$ are constants (potentially zero). More specifically, a term of $\bDelta_n$ that contains $\alpha_k$ factors of $\bi{u}_k$ for $0<k<n$ and $\beta$ factors of $\bDelta_1$ must satisfy
\begin{equation}
\beta + \sum_{k=1}^{n-1}k\alpha_k=n.
\end{equation}

Our goal in the following will be to demonstrate that the hierarchy \eref{eq:gen-hierarchy} can be solved recursively, and that in particular the solution $\bi{u}_n$ can be expressed as a linear combination of terms containing only $\bi{u}_0$ and $\bDelta_1$. Then, using the fact that each term in $\bDelta_n$ preserves order, we are able to determine whether the expectation of $\bi{u}_n$ vanishes. More specifically, we will show that we may write
\begin{equation}
\bi{u}_n\equiv\bi{u}_n[\bi{u}_0,\bDelta_1].
\end{equation} 
To see this, note first that the general solution to \eref{eq:gen-hierarchy} is given by
\begin{eqnarray}
\label{eq:formal-soln}
\bi{u}_n(\bi{x},t)&=\int_{-\infty}^{\infty}\rmd s\int_{\V}\rmd\bi{y}\,\mathcal{G}(\bi{x},\bi{y},t,s)\bDelta_n(\bi{y},s) \\
&\equiv \bi{G}[\bDelta_n],
\end{eqnarray}
where $\mathcal{G}\equiv(\mathcal{G}_{ij})$ is the Green's function of the linear operator $\partial_t+\L$, i.e. it solves
\begin{equation}
(\partial_t+\L)\mathcal{G}(\bi{x},\bi{y},t,s)=I\delta(\bi{x}-\bi{y})\delta(t-s).
\end{equation}
Moreover, $\bi{G}$ is the linear integral operator with kernel $\mathcal{G}$, which depends only on $\bi{u}_0$, and $I$ is the three-by-three identity matrix. Since $\bi{G}$ is a linear operator, it follows that $\bi{u}_n$ is a linear combination of terms that preserve order. 

The iterative solution to \eref{eq:gen-hierarchy} is now readily found by first computing
\begin{equation}
\label{eq:u1-soln}
\bi{u}_1=\bi{G}[\bDelta_1]\equiv\bi{u}_1[\bi{u}_0,\bDelta_1].
\end{equation}
Next, we solve for $\bi{u}_2$ and substitute in our solution for $\bi{u}_1$, thus
\begin{equation}
\bi{u}_2=\bi{G}[\bDelta_2]=\bi{G}_2[\bi{u}_0,\bi{u}_1[\bi{u}_0,\bDelta_1];\bDelta_1]\equiv\bi{u}_2[\bi{u}_0,\bDelta_1],
\end{equation}
where $\bi{G}_2\equiv\bi{G}\circ\bDelta_2$. Continuing iteratively, we find that
\begin{eqnarray}
\bi{u}_3&=\bi{G}_3[\bi{u}_0,\bi{u}_1[\bi{u}_0,\bDelta_1],\bi{u}_2[\bi{u}_0,\bDelta_1];\bDelta_1]\equiv\bi{u}_3[\bi{u}_0,\bDelta_1], \\
\vdots \nonumber\\
\label{eq:un-soln}
\bi{u}_n&=\bi{G}_n[\bi{u}_0,\bi{u}_1[\bi{u}_0,\bDelta_1],\ldots,\bi{u}_{n-1}[\bi{u}_0,\bDelta_1];\bDelta_1]\equiv\bi{u}_n[\bi{u}_0,\bDelta_1],
\end{eqnarray}
where $\bi{G}_k\equiv\bi{G}\circ\bDelta_k$. Clearly, each operation preserves order. Consequently, we have in fact shown that each $\bi{u}_n$ may be written as a linear combination of terms composed only of factors of $\bi{u}_0$ and $\bDelta_1$, all of order $n$. Since $\bDelta_1$ is mean-zero Gaussian and white, only even moments of its distribution can be non-vanishing by Wick's theorem. Thus, it follows that
\begin{equation}
\langle\bi{u}_n\rangle=0,\qquad n\mbox{ odd}.
\end{equation}
Similarly, expanding the EPR as
\begin{equation}
\dS[\bi{u}]=\dS_{-1}D^{-1}+\dS_{-1/2}D^{-1/2}+\dS_0+\ldots,
\end{equation}
it is fairly straightforward to see that each coefficient $\dS_{k/2}$, $k\geq-2$, must be a linear combination of terms of order $k+2$. In particular, it follows again by Wick's theorem that
\begin{equation}
\dS_{k/2}=0\qquad k\mbox{ odd}.
\end{equation}

\section{Calculation of the EPR}
\label{app:epr-calculation}
Here we derive explicitly the expression \eref{eq:epr} for the entropy production rate $\dS$ of the HVM, as well as \eref{eq:eprp} and \eref{eq:eprm} for $\dS^{\pm}$ of the DFM. In addition, we deduce the small noise expansion of the EPRs about the constant homogeneous isotropic and polar liquid ground-states quoted in the main text. For clarity, we choose to consider the HVM and the DFM separately. In particular, in contrast with our treatment above in \ref{app:lin-analysis}, the results obtained in the former model cannot in general be computed from the latter by sending $\gamma\rightarrow\infty$.

\subsection{Hydrodynamic Vicsek Model}
Starting from the definitions \eref{eq:epr-defn} for the EPR $\dS$ and \eref{eq:path-prob} for the path transition probability density $\P$ of the HVM, one finds that we may equivalently express $\dS$ in terms of the $\T$-antisymmetric part of the Freidlin-Wentzell action $\A$, i.e.
\begin{equation}
\label{eq:epr-action-defn}
\dS=\lim_{\tau\rightarrow\infty}\frac{\cev{\A}-\A}{2D\tau},
\end{equation}
where $\cev{\A}=\A\circ\T$ as before. By applying $\T$ to $\A$ as given in \eref{eq:fw-action}, we find that the action $\cev{\A}$ for the time-reversed ensemble may be expressed more explicitly by
\begin{equation}
\fl\cev{\A}=\frac{1}{4}\int_{-\tau}^{\tau}\rmd t\int_{\V}\rmd\bi{x}\,\left|\partial_t\bi{P}+\lambda\bi{P}\cdot\nabla\bi{P}-\funcdiff{F^S}{\bi{P}}+\funcdiff{F^A}{\bi{P}}\right|^2\quad\mbox{if}\quad\partial_t\rho+w\nabla\cdot\bi{P}=0,
\end{equation}
and $\cev{\A}=\infty$ otherwise. In particular, from \eref{eq:epr-action-defn} one straightforwardly deduces that
\begin{equation}
\label{eq:epr-1}
\dS=\lim_{\tau\rightarrow\infty}\frac{-1}{2D\tau}\int_{-\tau}^{\tau}\rmd t\int_{\V}\rmd\bi{x}\,\left(\odot\partial_t\bi{P}+\lambda\bi{P}\cdot\nabla\bi{P}+\funcdiff{F^A}{\bi{P}}\right)\cdot\funcdiff{F^S}{\bi{P}}.
\end{equation}
In \eref{eq:epr-1} -- and throughout this appendix -- we indicate by $\odot$ that the product with $\partial_t\bi{P}$ should be interpreted in the Stratonovich sense, i.e. employ mid-point discretisation in time.

To transform this into the expression given in \eref{eq:epr}, we simply have to observe that some terms in \eref{eq:epr-1} are integrable. Specifically, we find that we may write
\begin{equation}
\label{eq:epr-2}
\fl\dS=\lim_{\tau\rightarrow\infty}\left[\frac{-\Delta\mathcal{I}}{2D\tau}+\frac{1}{2D\tau}\int_{-\tau}^{\tau}\rmd t\int_{\V}\rmd\bi{x}\left(\rho\bi{P}\odot\partial_t\bi{P}-\left(\lambda\bi{P}\cdot\nabla\bi{P}+\funcdiff{F^A}{\bi{P}}\right)\cdot\funcdiff{F^S}{\bi{P}}\right)\right],
\end{equation}
where we have defined $\mathcal{I}$ to be the functional
\begin{equation}
\mathcal{I}[\bi{P}]=\int_{\V}\rmd\bi{x}\,\left(\frac{1}{2}|\bi{P}|^2+\frac{1}{2}(\nabla_{\alpha}P_{\beta})^2+\frac{1}{4}|\bi{P}|^4\right),
\end{equation}
i.e. as the part of $F^S$ which does not explicitly depend on the density $\rho$. Since we assume that the moments of $\bi{P}$ and its higher order spatial derivatives are finite in steady-state, it follows that $\Delta\mathcal{I}/\tau\rightarrow0$ as $\tau\rightarrow\infty$ so that the first term in \eref{eq:epr-2} may safely be ignored. In a similar vein, an integration by parts allows us to write the integral over the first term appearing in the integrand in \eref{eq:epr-2} as
\begin{equation}
\label{eq:epr-3}
\fl\int_{-\tau}^{\tau}\rmd t\int_{\V}\rmd\bi{x}\,\rho\odot\partial_t|\bi{P}|^2=\left.\int_{\V}\rmd\bi{x}\,\rho|\bi{P}|^2\right|_{-\tau}^{\tau}+w\int_{-\tau}^{\tau}\rmd t\int_{\V}\rmd\bi{x}\,(\nabla\cdot\bi{P})|\bi{P}|^2,
\end{equation} 
where we have used the continuity equation $\partial_t\rho=-w\nabla\cdot\bi{P}$. Again, after dividing by $4D\tau$, the first term on the right-hand side of \eref{eq:epr-3} goes away in the limit $\tau\rightarrow\infty$ since it grows sublinearly in $\tau$. Thus, assuming that we may replace temporal averages by averages over noise realizations, we finally arrive at the expression
\begin{equation}
\label{eq:epr-appendix}
\dS=D^{-1}\int_{\V}\rmd\bi{x}\,\left\langle\frac{w}{2}|\bi{P}|^2(\nabla\cdot\bi{P})-\left(\lambda\bi{P}\cdot\nabla\bi{P}+\funcdiff{F^A}{\bi{P}}\right)\cdot\funcdiff{F^S}{\bi{P}}\right\rangle
\end{equation}
reported in the main text.

Next, we investigate the expansion of \eref{eq:epr-appendix} about the constant homogeneous ground-states. To do this, we substitute the expansions \eref{eq:density-expansion} and \eref{eq:polarity-expansion} into \eref{eq:epr-appendix} and collect terms at equal order in $D$. It is fairly straightforward to check that there are no contributions to $\dS$ at orders $D^{-1}$ or $D^{-1/2}$. At order $D^0$ we find after some fairly tedious algebra that
\begin{eqnarray}
\label{eq:epr-zero-appendix}
\fl\dS_0=\bi{P}_0\cdot\int_{\V}\rmd\bi{x}\,\bigg\langle w\left(\nabla\cdot\bi{P}_1\right)\bi{P}_1+\rho_1\left((2w_1-\kappa)\nabla\left(\bi{P}_0\cdot\bi{P}_1\right)+\lambda\left(\bi{P}_0\cdot\nabla\right)\bi{P}_1\right) \nonumber \\
+2\kappa\left(\nabla\cdot\bi{P}_1\right)\left(\nabla^2\bi{P}_1-|\bi{P}_0|^2\bi{P}_1\right)\bigg\rangle
\end{eqnarray}
In order to arrive at this expression we have only assumed that $\bi{P}_0$ satisfies the zeroth order equation \eref{eq:zeroth-order}, and used \eref{eq:density-hierarchy} repeatedly. If we further assume that $\bi{P}_0$ is a polarised solution with $|\bi{P}_0|>0$, we may write $\bi{P}_0=(P_0,0)$ without loss of generality, from which \eref{eq:epr-zero} follows immediately after integrating out total derivatives. 

After transforming \eref{eq:epr-zero-appendix} to Fourier space, we obtain \eref{eq:epr-zero-fourier} as stated in the main text. Using our results from \ref{app:lin-analysis}, we may now investigate the scaling of this expression with $\Lambda\rightarrow\infty$. To determine this, we note that the scaling of the sum in this limit is determined by the corresponding integral in $q$-space, i.e.
\begin{equation}
\dS_0/\V=\sum_{|\bi{q}|\leq\Lambda}\Tr(\dot{\sigma}^{\mathrm{pl}}\,C^{\mathrm{pl}})\sim\frac{\V}{(2\pi)^2}\int_0^{2\pi}\rmd\theta\int_0^{\Lambda}\rmd q\,q\Tr(\dot{\sigma}^{\mathrm{pl}}\,C^{\mathrm{pl}}),
\end{equation}
where $\dot{\sigma}^{\mathrm{pl}}$ is given by \eref{eq:sigma-pl}. In particular, direct substitution from \tref{tab:corr-scaling} and subsequently performing the integrals over $q$ and $\theta$ yields
\begin{equation}
\dS_0/\V\sim \frac{P_0^2\kappa^2}{4\pi}\Lambda^2.
\end{equation}

\subsection{Diffusive flocking model}
In analogy with the above calculation, we may compute the EPRs $\dS^{\pm}$ of the DFM from the $\T^{\pm}$-antisymmetric part of the Freidlin-Wentzell action $\adf$, specifically
\begin{equation}
\dS^{\pm}=\lim_{\tau\rightarrow\infty}\frac{\adfb^{\pm}-\adf}{2D\tau},
\end{equation}
where we have defined $\adfb^{\pm}=\adf\circ\T^{\pm}$. Writing out the actions for the time-reversed ensembles explicitly, we find that
\begin{equation}
\fl\adfb^+=\frac{1}{4}\int_{-\tau}^{\tau}\rmd t\int_{\V}\rmd\bi{x}\,\left[\gamma\left|\nabla^{-1}\left(\partial_t\rho-\nabla\cdot\bi{J}_d\right)\right|^2+\left|\partial_t\bi{P}-\lambda\bi{P}\cdot\nabla\bi{P}-\funcdiff{F}{\bi{P}}\right|^2\right],
\end{equation}
and
\begin{eqnarray}
\fl\adfb^-=\frac{1}{4}\int_{-\tau}^{\tau}\rmd t\int_{\V}\rmd\bi{x}\,\Bigg[\gamma\left|\nabla^{-1}\left(\partial_t\rho-\nabla\cdot\bi{J}_d^S+\nabla\cdot\bi{J}_d^A\right)\right|^2 \nonumber\\
+\left|\partial_t\bi{P}+\lambda\bi{P}\cdot\nabla\bi{P}-\funcdiff{F^S}{\bi{P}}+\funcdiff{F^A}{\bi{P}}\right|^2\Bigg].
\end{eqnarray}
Thus, proceeding as above we straighforwardly find that
\begin{eqnarray}
\fl\dS^+=\lim_{\tau\rightarrow\infty}\Bigg[\frac{-\Delta F}{2D\tau} \nonumber \\
\label{eq:eprs-1}
+\frac{1}{2D\tau}\int_{-\tau}^{\tau}\rmd t\int_{\V}\rmd\bi{x}\,\left(\gamma w(\nabla\cdot\bi{P})\nabla^{-2}\odot\partial_t\rho-\lambda\left[(\bi{P}\cdot\nabla)\bi{P}\right]\odot\partial_t\bi{P}\right)\Bigg]
\end{eqnarray}
and
\begin{eqnarray}
\fl\dS^-=\lim_{\tau\rightarrow\infty}\Bigg[\frac{-\Delta F^S}{2D\tau} \nonumber\\
+\frac{1}{2D\tau}\int_{-\tau}^{\tau}\rmd t\int_{\V}\rmd\bi{x}\,\left(\bi{J}_d^A\cdot\nabla\mu^S-\left(\lambda\bi{P}\cdot\nabla\bi{P}+\funcdiff{F^A}{\bi{P}}\right)\cdot\funcdiff{F^S}{\bi{P}}\right)\Bigg].
\end{eqnarray}
The expression for $\dS^-$ immediately reduces to \eref{eq:eprm} provided in the main text upon replacing the temporal average by an average over noise realizations and noting again that $\Delta F^S/\tau\rightarrow0$ as $\tau\rightarrow\infty$. On the other hand, \eref{eq:eprs-1} requires a bit more massaging. Firstly, we observe that the equal-time expectation
\begin{equation}
\left\langle(\nabla\cdot\bi{P})\nabla^{-2}\nabla\odot\bxi\right\rangle=0,
\end{equation}
since here the Stratonovich product coincides with the corresponding Ito product (i.e. there is no spurious drift term). It follows that
\begin{equation}
\left\langle(\nabla\cdot\bi{P})\nabla^{-2}\odot\partial_t\rho\right\rangle=-w\left\langle(\nabla\cdot\bi{P})\nabla^{-2}\nabla\cdot\bi{P}\right\rangle+\gamma^{-1}\left\langle(\nabla\cdot\bi{P})\mu\right\rangle.
\end{equation}
To treat the product with $\odot\partial_t\bi{P}$ in \eref{eq:eprs-1} we must explicitly compute the spurious drift. We will show that, in fact, the spurious drift is a total derivative and therefore does not contribute to the EPR $\dS^+$. To do this, we consider a finite discretisation of the process in Fourier space with $|\bi{q}|\leq\Lambda$, which is consistent with our numerical scheme. By applying standard stochastic calculus, we then obtain
\begin{eqnarray}
\fl\int_{\V}\rmd\bi{x}\,\left\langle[(\bi{P}\cdot\nabla)\bi{P}]\odot\bfeta\right\rangle&=\V\sum_{|\bi{q}|\leq\Lambda}\sum_{|\bi{k}|\leq\Lambda}ik_{\beta}\left\langle P_{\beta}(\bi{q}-\bi{k})P_{\alpha}(\bi{k})\odot\eta_{\alpha}(-\bi{q})\right\rangle \nonumber \\
&=D\sum_{|\bi{q}|\leq\Lambda}\sum_{|\bi{k}|\leq\Lambda}ik_{\beta}\left\langle\left(\delta_{\bi{k},\bi{0}}P_{\beta}(\bi{k})+2\delta_{\bi{q},\bi{k}}P_{\beta}(\bi{q}-\bi{k})\right)\right\rangle \nonumber \\
&=0
\end{eqnarray}
Here, the second equality follows from the transformation rule between Ito and Stratonovich processes \cite{Gardiner2009}, i.e.
\begin{equation}
\left\langle h(\bi{P}(\bi{q}_1),\ldots,\bi{P}(\bi{q}_n))\odot{\eta}_{\alpha}(\bi{k})\right\rangle=D\sum_{m=1}^n\left\langle\frac{\partial h}{\partial P_{\beta}(\bi{q}_m)}\right\rangle\delta_{\bi{q}_m,-\bi{k}}\delta_{\alpha\beta},
\end{equation}
and the fact that $\delta_{\alpha\alpha}=2$. To see why the final equality holds, note also that
\begin{equation}
\sum_{|\bi{k}|\leq\Lambda}k_{\alpha}=0
\end{equation}
since the sum is finite. From this, and taking $\tau\rightarrow\infty$ in \eref{eq:eprs-1}, the result \eref{eq:eprp} reported in the main text follows immediately.

For the DFM, we may additionally calculate the EPRs $\dS^{\pm}_J$ at the level of the fluctuating density current $\bi{J}$. At this level, the actions $\adfb^{J,\pm}$ for the time-reversed ensembles may be expressed as
\begin{equation}
\adfb^{J,+}=\frac{1}{4}\int_{-\tau}^{\tau}\rmd t\int_{\V}\rmd\bi{x}\left[\gamma\left|\bi{J}+\bi{J}_d\right|^2+\left|\partial_t\bi{P}-\lambda\bi{P}\cdot\nabla\bi{P}-\funcdiff{F}{\bi{P}}\right|^2\right]
\end{equation}
and
\begin{equation}
\fl\adfb^{J,-}=\frac{1}{4}\int_{-\tau}^{\tau}\rmd t\int_{\V}\rmd\bi{x}\left[\gamma\left|\bi{J}+\bi{J}_d^S-\bi{J}_d^A\right|^2+\left|\partial_t\bi{P}+\lambda\bi{P}\cdot\nabla\bi{P}-\funcdiff{F^S}{\bi{P}}+\funcdiff{F^A}{\bi{P}}\right|^2\right]
\end{equation}
if $\partial_t\rho+\nabla\cdot\bi{J}=0$ and $\adfb^{J,\pm}=\infty$ otherwise. By direct substitution we then find that
\begin{equation}
\label{eq:eprjp-app}
\dS^+_J=\frac{1}{2D\tau}\int_{-\tau}^{\tau}\rmd t\int_{\V}\rmd\bi{x}\left(\gamma\bi{J}_d\odot\bi{J}-\left(\lambda\bi{P}\cdot\nabla\bi{P}+\funcdiff{F}{\bi{P}}\right)\odot\partial_t\bi{P}\right),
\end{equation}
in addition to
\begin{eqnarray}
\fl\dS^-_J=\frac{1}{2D\tau}\int_{-\tau}^{\tau}\rmd t\int_{\V}\rmd\bi{x}\Bigg(\gamma\left(\odot\bi{J}-\bi{J}_d^A\right)\cdot\bi{J}_d^S \nonumber \\
\label{eq:eprjm-app}
-\left(\odot\partial_t\bi{P}+\lambda\bi{P}\cdot\nabla\bi{P}+\funcdiff{F^A}{\bi{P}}\right)\cdot\funcdiff{F^S}{\bi{P}}\Bigg),
\end{eqnarray}
Now, it is fairly easy to see that
\begin{eqnarray}
\int_{\V}\rmd\bi{x}\,\left\langle\bi{J}_d\odot\bi{J}\right\rangle&=\int_{\V}\rmd\bi{x}\,\left\langle w\bi{P}\odot\bi{J}-\gamma^{-1}\odot\bi{J}\cdot\nabla\mu\right\rangle \nonumber \\
&=\int_{\V}\rmd\bi{x}\,\left\langle w^2|\bi{P}|^2-\gamma^{-1}w\bi{P}\cdot\nabla\mu-\gamma^{-1}\funcdiff{F}{\rho}\odot\partial_t\rho\right\rangle
\end{eqnarray}
where the second equality follows from an integration by parts and the fact that $\langle\bi{P}\odot\bi{J}\rangle=\langle\bi{P}\cdot\bi{J}_d\rangle$. Substituting this back into \eref{eq:eprjp-app} gives the desired result for $\dS_J^+$, stated in \eref{eq:eprjp}. Similarly, we have that
\begin{equation}
\int_{\V}\rmd\bi{x}\,\left\langle\left(\odot\bi{J}-\bi{J}_d^A\right)\cdot\bi{J}_d^S\right\rangle=-\int_{\V}\rmd\bi{x}\,\left\langle\gamma^{-1}\funcdiff{F^S}{\rho}\odot\partial_t\rho+\bi{J}_d^A\cdot\bi{J}_d^S\right\rangle,
\end{equation}
from which the fact that $\dS^-_J=\dS^-$ follows upon substitution back into \eref{eq:eprjm-app}.
\begin{table}
\caption{\label{tab:epr-mat}Independent components of the Hermitian bilinear EPR coupling matrices $\dot{\sigma}^{\pm,\mathrm{pl}}$.}
\begin{indented}
\item[]\begin{tabular}{@{}l|ll}
\br
$(i,j)$ & $\dot{\sigma}^{+,\mathrm{pl}}_{ij}$ & $\dot{\sigma}^{-,\mathrm{pl}}_{ij}$ \\
\mr
$(1,1)$ & 0 & 0 \\
$(1,2)$ & $\frac{i}{2}\left(w\Gamma_{\rho}-P_0^2\lambda-iP_0\lambda w_1q_{\parallel}\right)q_{\parallel}$ & $\frac{i}{2}\left(w_1\Gamma_{\parallel}-\tilde{w}\Gamma_{\rho}+P_0^2\lambda\right)q_{\parallel}$ \\
$(1,3)$ & $\frac{i}{2}\left(w\Gamma_{\rho}-iP_0\lambda w_1q_{\parallel}\right)q_{\perp}$ & $\frac{i}{2}\left(w_1\Gamma_{\perp}-\tilde{w}\Gamma_{\rho}+P_0^2\kappa\right)q_{\perp}$ \\
$(2,2)$ & $\left(\gamma w\tilde{w}+P_0^2\lambda^2q^2\right)q_{\parallel}^2/q^2$ & $0$ \\
$(2,3)$ & $-\frac{i}{2}w\left(P_0q^2+2i\gamma\tilde{w}q_{\parallel}\right)q_{\perp}/q^2$ & $\frac{i}{2}P_0\left(\tilde{w}-\kappa\left(\Gamma_{\parallel}+\Gamma_{\perp}\right)\right)q_{\perp}$ \\
$(3,3)$ & $\left(\gamma w\tilde{w}+P_0^2\lambda^2q^2\left(\frac{q_{\parallel}}{q_{\perp}}\right)^2\right)q_{\perp}^2/q^2$ & $0$ \\
\br 
\end{tabular}
\end{indented}
\end{table}

Again we may linearize the expressions \eref{eq:eprp} and \eref{eq:eprm} about the homogeneous isotropic and polar liquid states by substituting in an expansion of the form \eref{eq:density-expansion}, \eref{eq:polarity-expansion}. Treating this as above for the HVM, we find that
\begin{eqnarray}
\fl\dS^+_0=\int_{\V}\rmd\bi{x}\,\Bigg\langle\gamma w^2|K\bi{P}_1|^2+w\left(\nabla\cdot\bi{P}_1\right)\left(a_{\rho}\rho_1-\nu_{\rho}\nabla^2\rho_1\right)-P_0wP_{\parallel}\left(\partial_{\perp}P_{\perp}\right) \nonumber \\
-ww_1\left(\nabla\cdot\bi{P}_1\right)^2+P_0^2\lambda^2|\partial_{\parallel}\bi{P}_1|^2-P_0^2\lambda\rho_1\left(\partial_{\parallel}P_{\parallel}\right) \nonumber \\
\label{eq:eprp-0-appendix} 
+ P_0\lambda w_1\left(\partial_{\parallel}\rho_1\right)\left(\nabla\cdot\bi{P}_1\right)\Bigg\rangle, \\
\fl\dS^-_0=\int_{\V}\rmd\bi{x}\,\Bigg\langle P_0P_{\parallel}\partial_{\perp}\left(w P_{\perp}+\gamma^{-1}w_1\nabla^2P_{\perp}\right) \nonumber \\ 
-\left(a_{\rho}\rho_1-\nu_{\rho}\nabla^2\rho_1\right)\nabla\cdot\left(w\bi{P}_1+\gamma^{-1}w_1\nabla^2\bi{P}_1\right) \nonumber \\
\label{eq:eprm-0-appendix}
+P_0^2\left(2w_1-\kappa+\lambda\right)\rho_1\partial_{\parallel}P_{\parallel}-2P_0\kappa\left(\partial_{\perp}P_{\perp}\right)\left(P_0^2P_{\parallel}-\nabla^2P_{\parallel}\right)\Bigg\rangle,
\end{eqnarray}
where both expressions hold for general $P_0\geq0$. We may equivalently express \eref{eq:eprp-0-appendix} and \eref{eq:eprm-0-appendix} in Fourier space, and for the constant homogeneous ground-states we obtain expressions analogous to those which we encountered for the HVM \eref{eq:epr-zero-fourier}. Specifically, we find that for the isotropic and polar liquid states,
\begin{equation}
\label{eq:eprpm-iso-app}
\dS_0^{\pm}/\V=\sum_{|\bi{q}|\leq\Lambda}\Tr\left(\dot{\sigma}^{\pm,\mathrm{iso}}C^{\mathrm{iso}}\right),
\end{equation}
and
\begin{equation}
\label{eq:eprpm-pl-app}
\dS_0^{\pm}/\V=\sum_{|\bi{q}|\leq\Lambda}\Tr\left(\dot{\sigma}^{\pm,\mathrm{pl}}C^{\mathrm{pl}}\right),
\end{equation}
respectively. Furthermore, we may choose to write the sum such that $\dot{\sigma}^{\pm,\mathrm{iso}}$ and $\dot{\sigma}^{\pm,\mathrm{pl}}$ are Hermitian. Taking $P_0=0$ in \eref{eq:eprp-0-appendix} and \eref{eq:eprm-0-appendix} and transforming to Fourier space we find that $\dot{\sigma}^{\pm,\mathrm{iso}}$ are given by \eref{eq:sigmap-iso} and \eref{eq:sigmam-iso} as advertised. For $\sigma^{\pm,\mathrm{pl}}$ we list the six independent components of each matrix in \tref{tab:epr-mat}. Finally, from \eref{eq:eprpm-pl-app} in addition to tables \ref{tab:corr-scaling} and \ref{tab:epr-mat}, we straightforwardly deduce that in the polar liquid phase, $\dS_0^+/\V\sim P_0^2\lambda^2\Lambda^2/(4\pi)$ and $\dS_0^-/\V\sim w_1^2\Lambda^4/(8\pi\gamma)$, while the exact results in the isotropic phase are presented in the main text.

\section*{Bibliography}
\bibliography{v8}

\providecommand{\newblock}{}
\begin{thebibliography}{10}
\expandafter\ifx\csname url\endcsname\relax
  \def\url#1{{\tt #1}}\fi
\expandafter\ifx\csname urlprefix\endcsname\relax\def\urlprefix{URL }\fi
\providecommand{\eprint}[2][]{\url{#2}}

\bibitem{Sekimoto2010}
Sekimoto K 2010 {\em Stochastic Energetics\/} 1st ed ({\em Lecture Notes in
  Physics\/} vol 799) (Springer-Verlag Berlin Heidelberg)

\bibitem{Kubo1991}
Kubo R, Toda M and Hashitsume N 1991 {\em Statistical Physics II:
  Nonequilibrium Statistical Mechanics\/} 2nd ed ({\em Springer Series in
  Solid-State Sciences\/} vol~31) (Springer-Verlag Berlin Heidelberg)

\bibitem{Zwanzig2001}
Zwanzig R 2001 {\em Nonequilibrium Statistical Mechanics\/} (Oxford University
  Press)

\bibitem{Lebowitz1998}
Lebowitz J~L and Spohn H 1998 {\em J. Stat. Phys.\/} {\bf 95}(1) 333--365

\bibitem{Seifert2012}
Seifert U 2012 {\em Rep. Prog. Phys.\/} {\bf 75}(12) 126001

\bibitem{Wittkowski2014}
Wittkowski R, Tiribocchi A, Stenhammar J, Allen R~J, Marenduzzo D and Cates M~E
  2014 {\em Nat. Commun.\/} {\bf 5}(1) 4351

\bibitem{Tailleur2008}
Tailleur J and Cates M~E 2008 {\em Phys. Rev. Lett.\/} {\bf 100}(21) 218103

\bibitem{Solon2015}
Solon A~P, Cates M~E and Tailleur J 2015 {\em Eur. Phys. J. Spec. Top.\/} {\bf
  224}(7) 1231--1262

\bibitem{Bertin2006}
Bertin E, Droz M and Gr\'egoire G 2006 {\em Phys. Rev. E\/} {\bf 74}(2) 022101

\bibitem{Peshkov2014}
Peshkov A, Bertin E, Ginelli F and Chat\'e H 2014 {\em Eur. Phys. J. Spec.
  Top.\/} {\bf 223}(7) 1315--1344

\bibitem{Chate2008}
Chat\'e H, Ginelli F, Gr\'egoire G and Raynaud F 2008 {\em Phys. Rev. E\/} {\bf
  77}(4) 046113

\bibitem{Mahault2018}
Mahault B, Patelli A and Chat\'e H 2018 {\em J. Stat. Mech.\/} {\bf 2018}(9)
  093202

\bibitem{Fodor2020}
Fodor E, Nemoto T and Vaikuntanathan S 2020 {\em New J. Phys.\/} {\bf 22}(1)
  013052

\bibitem{Chate2020}
Chat\'e H 2020 {\em Annu. Rev. Condens. Matter Phys.\/} {\bf 11}(1) 189--212

\bibitem{Fodor2016}
Fodor E, Nardini C, Cates M~E, Tailleur J, Visco P and van Wijland F 2016 {\em
  Phys. Rev. Lett.\/} {\bf 117}(3) 038103

\bibitem{Toner1998}
Toner J and Tu Y 1998 {\em Phys. Rev. E\/} {\bf 58}(4) 4828--4858

\bibitem{Toner2005}
Toner J, Tu Y and Ramaswamy S 2005 {\em Ann. Phys.\/} {\bf 318}(1) 170--244

\bibitem{Toner2012}
Toner J 2012 {\em Phys. Rev. Lett.\/} {\bf 108}(8) 088102

\bibitem{Marchetti2013}
Marchetti M~C, Joanny J~F, Ramaswamy S, Liverpool T~B, Prost J, Rao M and Simha
  R~A 2013 {\em Rev. Mod. Phys.\/} {\bf 85}(3) 1143--1189

\bibitem{Thompson2011}
Thompson A~G, Tailleur J, Cates M~E and Blythe R~A 2011 {\em J. Stat. Mech.\/}
  {\bf 2011}(02) P02029

\bibitem{Cates2015}
Cates M~E and Tailleur J 2015 {\em Annu. Rev. Condens. Matter Phys.\/} {\bf
  6}(1) 219--244

\bibitem{Barre2014}
Barr\'e J, Ch\'etrite R, Muratori M and Peruani F 2015 {\em J. Stat. Phys.\/}
  {\bf 158}(3) 589--600

\bibitem{Vicsek1995}
Vicsek T, Czir\'ok A, Ben-Jacob E, Cohen I and Shochet O 1995 {\em Phys. Rev.
  Lett.\/} {\bf 75}(6) 1226--1229

\bibitem{Nemoto2019}
Nemoto T, Fodor E, Cates M~E, Jack R~L and Tailleur J 2019 {\em Phys. Rev. E\/}
  {\bf 99}(2) 022605

\bibitem{Ramaswamy2010}
Ramaswamy S 2010 {\em Annu. Rev. Condens. Matter Phys.\/} {\bf 1}(1) 323--345

\bibitem{Li2020}
Li Y~I and Cates M~E 2020 {\em J. Stat. Mech.\/} {\bf 2020}(5) 053206

\bibitem{Grafke2017}
Grafke T, Cates M~E and Vanden-Eijnden E 2017 {\em Phys. Rev. Lett.\/} {\bf
  119}(18) 188003

\bibitem{Farrell2012}
Farrell F~D~C, Marchetti M~C, Marenduzzo D and Tailleur J 2012 {\em Phys. Rev.
  Lett.\/} {\bf 108}(24) 248101

\bibitem{Deseigne2010}
Deseigne J, Dauchot O and Chat{\'{e}} H 2010 {\em Phys. Rev. Lett.\/} {\bf
  105}(9) 098001

\bibitem{Kumar2014}
Kumar N, Soni H, Ramaswamy S and Sood A~K 2014 {\em Nat. Commun.\/} {\bf 5}(1)
  4688

\bibitem{Deblais2018}
Deblais A, Barois T, Guerin T, Delville P~H, Vaudaine R, Lintuvuori J~S, Boudet
  J~F, Baret J~C and Kellay H 2018 {\em Phys. Rev. Lett.\/} {\bf 120}(18)
  188002

\bibitem{Palacci2013}
Palacci J, Sacanna S, Steinberg A~P, Pine D~J and Chaikin P~M 2013 {\em
  Science\/} {\bf 339}(6122) 936--940

\bibitem{Nardini2017}
Nardini C, Fodor E, Tjhung E, van Wijland F, Tailleur J and Cates M~E 2017 {\em
  Phys. Rev. X\/} {\bf 7}(2) 021007

\bibitem{Ganguly2013}
Ganguly C and Chaudhuri D 2013 {\em Phys. Rev. E\/} {\bf 88}(3) 032102

\bibitem{Dadhichi2018}
Dadhichi L~P, Maitra A and Ramaswamy S 2018 {\em J. Stat. Mech.\/} {\bf
  2018}(12) 123201

\bibitem{Shim2016}
Shim P~S, Chun H~M and Noh J~D 2016 {\em Phys. Rev. E\/} {\bf 93}(1) 012113

\bibitem{Crosato2019}
Crosato E, Prokopenko M and Spinney R~E 2019 {\em Phys. Rev. E\/} {\bf 100}(4)
  042613

\bibitem{Dabelow2019}
Dabelow L, Bo S and Eichhorn R 2019 {\em Phys. Rev. X\/} {\bf 9}(2) 021009

\bibitem{Markovich2020}
Markovich T, Fodor E, Tjhung E and Cates M~E 2020  (\textit{Preprint}
  \eprint{2008.06735})

\bibitem{Caballero2020}
Caballero F and Cates M~E 2020 {\em Phys. Rev. Lett.\/} {\bf 124}(24) 240604

\bibitem{Bertin2009}
Bertin E, Droz M and Gr\'egoire G 2009 {\em J. Phys. A: Math. Theor.\/} {\bf
  42}(44) 445001

\bibitem{Dean1996}
Dean D~S 1996 {\em J. Phys. A: Math. Gen.\/} {\bf 29}(24) L613--L617

\bibitem{Solon2015a}
Solon A~P, Chat\'e H and Tailleur J 2015 {\em Phys. Rev. Lett.\/} {\bf 114}(6)
  068101

\bibitem{Ginelli2016}
Ginelli F 2016 {\em Eur. Phys. J. Spec. Top.\/} {\bf 225}(11-12) 2099--2117

\bibitem{OLaighleis2018}
Laighl\'eis E~O, Evans M~R and Blythe R~A 2018 {\em Phys. Rev. E\/} {\bf 98}(6)
  062127

\bibitem{Castellana2016}
Castellana M, Bialek W, Cavagna A and Giardina I 2016 {\em Phys. Rev. E\/} {\bf
  93}(5) 052416

\bibitem{Peshkov2012}
Peshkov A, Ngo S, Bertin E, Chat\'e H and Ginelli F 2012 {\em Phys. Rev.
  Lett.\/} {\bf 109}(9) 098101

\bibitem{Patelli2019}
Patelli A, Djafer-Cherif I, Aranson I~S, Bertin E and Chat\'e H 2019 {\em Phys.
  Rev. Lett.\/} {\bf 123}(25) 258001

\bibitem{Solon2015b}
Solon A~P, Caussin J~B, Bartolo D, Chat\'e H and Tailleur J 2015 {\em Phys.
  Rev. E\/} {\bf 92}(6) 062111

\bibitem{Hohenberg1977}
Hohenberg P~C and Halperin B~I 1977 {\em Rev. Mod. Phys.\/} {\bf 49}(3)
  435--479

\bibitem{Gopinath2012}
Gopinath A, Hagan M~F, Marchetti M~C and Baskaran A 2012 {\em Phys. Rev. E\/}
  {\bf 85}(6) 061903

\bibitem{Kloeden1992}
Kloeden P~E and Platen E 1992 {\em Numerical Solution of Stochastic
  Differential Equations\/} (Berlin, Heidelberg: Springer Berlin Heidelberg)

\bibitem{Canuto2006}
Canuto C, Hussaini M~Y, Quarteroni A and Zang T~A 2006 {\em Spectral Methods:
  Fundamentals in Single Domains\/} Scientific Computation (Berlin, Heidelberg:
  Springer Berlin Heidelberg)

\bibitem{Touchette2009}
Touchette H 2009 {\em Phys. Rep.\/} {\bf 478}(1-3) 1--69

\bibitem{Shankar2018}
Shankar S and Marchetti M~C 2018 {\em Phys. Rev. E\/} {\bf 98}(2) 020604

\bibitem{Greene2006}
Greene R~E and Krantz S~G 2006 {\em Function Theory of One Complex Variable\/}
  (American Mathematical Society)

\bibitem{Krantz2008}
Krantz S~G 2008 {\em A Guide to Complex Variables\/} (Mathematical Association
  of America)

\bibitem{Gardiner2009}
Gardiner C 2009 {\em Stochastic Methods: A Handbook for the Natural and Social
  Sciences\/} 4th ed Springer Series in Synergetics (Berlin, Heidelberg:
  Springer-Verlag Berlin Heidelberg)

\end{thebibliography}

\end{document}